\documentclass[10pt]{article}
\usepackage[utf8]{inputenc}
\usepackage{setspace}
\usepackage[margin=1in]{geometry}
\usepackage{amsmath}
\usepackage{amsthm}
\usepackage{amssymb}
\usepackage{bbm}
\usepackage{cleveref}
\usepackage{breakcites}
\usepackage{tikz}
\usepackage{dcolumn}
\usepackage{booktabs}
\usepackage{colortbl}
\usepackage{mathrsfs}
\usepackage{rotating}
\usepackage{graphicx} 
\usepackage{algorithm}
\usepackage{algorithmicx}
\usepackage{algpseudocode} 
\usepackage{caption,subcaption}
\usepackage{bm}
\usepackage{multirow,array}

\DeclareMathOperator{\bg}{\textbf{g}}

\newtheorem{theorem}{Theorem}

\newtheorem{corollary}[theorem]{Corollary}

\theoremstyle{definition}

\newcommand{\ra}[1]{\renewcommand{\arraystretch}{#1}}

\title{Control and spread of contagion in networks with global effects}
\author{
  John Higgins\footnote{University of Wisconsin, Madison, WI, USA}\\
  \texttt{jfhiggins@wisc.edu}
  \and
  Tarun Sabarwal\footnote{University of Kansas, Lawrence, KS, USA}\\
  \texttt{sabarwal@ku.edu} 
}

\begin{document}

\thispagestyle{empty}

\maketitle 

\begin{center}
First draft: April 2021 
\end{center} 

\begin{center}
\large{Abstract} 
\end{center}
\renewcommand{\baselinestretch}{1} \small \normalsize
\noindent 
We study proliferation of an action in binary action network coordination games that are generalized to include global effects. This captures important aspects of proliferation of a particular action or narrative in online social networks, providing a basis to understand their impact on societal outcomes. Our model naturally captures complementarities among starting sets, network resilience, and global effects, and highlights interdependence in channels through which contagion spreads. We present new, natural, computationally tractable, and efficient algorithms to define and compute equilibrium objects that facilitate the general study of contagion in networks and prove their theoretical properties. Our algorithms are easy to implement and help to quantify relationships previously inaccessible due to computational intractability. Using these algorithms, we study the spread of contagion in scale-free networks with 1,000 players using millions of Monte Carlo simulations. Our analysis provides quantitative and qualitative insight into the design of policies to control or spread contagion in networks. The scope of application is enlarged given the many other situations across different fields that may be modeled using this framework.


\vspace{\fill}

\pagebreak

\renewcommand{\baselinestretch}{1.5} \small \normalsize
\section{Introduction}

Proliferation of contagion in networks is an increasingly central problem with systemic consequences for society. 
For example, political divisiveness, anti-vaccination rhetoric, climate change skepticism, conspiracy theories, and stereotyping on social networks contribute to influence broad societal outcomes \cite{vosoughi2018}. 
Proliferation of divisive politics on social media exacerbates a fractious political climate, intensifying divisions across party lines \cite{grinberg}. 
Attacks on election integrity undermine trust in the election process \cite{frenkel_2020}. 
Repeated assaults on the science behind disease transmission contribute to significant long term harm to public health, especially during the COVID-19 pandemic \cite{brennen2020types}. 
Anti-vaccine rhetoric contributes to a decline in vaccination rates, posing another risk to public health \cite{broniatowski}. 
Persistent promotion of distrust in scientific evidence prevents adoption of measures to mitigate impacts of climate change.  
Crimes against minorities (ethnic, immigrant, or religious) may be precipitated by posting and spreading incendiary rhetoric on social networks. 
Mental health of vulnerable communities may be worsened by social network pressures \cite{wells2021}. 
These problems are exacerbated by the rapid and widespread diffusion of information in online social networks \cite{varol2017} and \cite{shao2018}. 
These activities can cause systemic upheaval, as evidenced by the mob attack on the U.S. Capitol on January 6, 2021, and in similar events in other countries as described in the 2021 Nobel peace prize lecture by \cite{ressa_2021}.
Collectively, such events are a significant and growing threat to both democratic institutions and society at large. 

Proliferation of an alternative action can also be beneficial for society. For example, proliferation of welfare-improving innovations such as more secure transactions, health or vaccine adoption, environmentally beneficial technology, and collective action for citizen safety can influence broad societal outcomes.

Even though different campaigns to proliferate an action or narrative in society have different goals, their structural characteristics are similar. An alternative narrative is promoted by a small group of participants who rely on social network dynamics for its proliferation. This is achieved both by person-to-person spread through individual connectivity in the network and by global effects based on similar activity by others in different parts of the network. The objective of the promoting group is typically achieved with partial proliferation of the narrative in the network. Once the objective is achieved, the underlying truth or falsity of the narrative may not matter; it may be undermined or simply swept away. 

We study these characteristics of proliferation of an action in a network using a binary action network coordination game. As coordination incentives arise in many socio-economic scenarios with interdependent decision-making, our model and analysis can be applied to many additional situations, including regime change, technology adoption, bank runs, currency crises, run on groceries in a pandemic, marketing new products, segregation and desegregation, success of social platforms, agglomeration in urban economics, and others.

Consider a network in which each person has a preference to coordinate with their friends (or network neighbors) on the choice of two actions. Suppose this preference dictates that I find it beneficial to choose an action if a sufficiently large fraction of my friends also choose that action. This causes me to choose an action based on the choices of some of my friends, which may cause some of my other friends to choose the same action, and in turn, cause additional friends of friends to choose the same action, propagating a chain reaction and causing contagion. 

Theoretical properties of contagion arising from individual decisions in coordination games on networks in the manner above have been studied in the existing literature. A central result of interest is the following. Suppose there are finitely many players in a connected network. Each player can take one of two actions, 0 or 1, and each player is initially playing 0. Suppose an initial subset of players $S$ is exogenously infected to play 1 and we want to characterize when best response dynamics starting from $S$ lead to all players in the network playing 1. As a parameter of the game, suppose that action 1 is a best response for a player, if, and only if, fraction $q \in [0, 1]$ of the player's neighbors play 1. (The parameter $q$ is derived analytically from an underlying tradeoff between benefit of coordination $b$ and cost of miscoordination $c$ to yield $q = \frac{c}{b+c}$. In this sense, it may be viewed as relative cost of miscoordination as well as a measure of network resilience to contagion. The higher is $q$, the harder it is for a player to play $1$.) An adaptation of \cite{morris2000} in \cite{jackson2008} shows that \textit{contagion occurs from $S$ to the entire network if, and only if, the complement of $S$ is uniformly no more than $(1-q)$-cohesive}. Recall that a set $A$ of players is $(1-q)$-cohesive if each player in $A$ has at least fraction $1-q$ of their neighbors in $A$. Set $A$ is uniformly no more than $(1-q)$-cohesive if every nonempty subset of $A$ is at most $(1-q)$-cohesive.  

In order to apply this result, we need to check that every subset of the complement of $S$ is at most $(1-q)$-cohesive. Such a task is exponentially expensive, making its application infeasible beyond very small networks. 

In addition to the computational obstacle, the existing result applies only in coordination games with local network effects, that is, where a player's payoff depends solely on the actions of their neighbors (and neighbors have uniform unit weights). It does not apply when a player's best response depends on actions of both their neighbors and the rest of the network. This dependence arises naturally in the types of phenomena we want to study. 

We solve both problems in this paper. We extend the existing theoretical model of a network coordination game to include a new, flexible, and tractable formulation of global effects, which capture the notion that each player's decision depends not just on decisions of their neighbors but also on an aggregate based on decisions taken by others in the network. We provide new, natural, tractable, and efficient algorithms that can be applied to an arbitrary network coordination game with heterogeneous local and global effects and arbitrary starting set $S$ to compute equilibrium depth of contagion starting from $S$ and $q$, and also compute all $q$ for which contagion spreads from $S$ to the entire network in equilibrium. 

Global effects are included as an additive component to a player's payoff, making them transparent and tractable while still allowing for natural interactions with local effects in the best response and facilitating computation of equilibrium. These effects are otherwise flexible in that we only require global effects to be weakly increasing. There are no restrictions related to continuity, convexity, concavity, differentiability, and so on, and we allow for heterogeneity in global effects for different players. Moreover, we allow for heterogeneity in local effects as well (different weights for different neighbors of a player, asymmetric weights for different players, and unidirectional links). We also include players who may be exogenously infected to play $1$. Our formulation subsumes the existing formulation with local effects (and uniform unit weights on neighbors) as a natural special case. Our formulation allows a natural parametric flexibility to study the differential impact of global effects on each player's decision and the corresponding equilibrium. 

Our algorithms are natural and tractable because they are based on best response dynamics. They are efficient because we reduce the number of subsets checked to no more than the number of players in the complement of $S$, reducing the computational burden from exponential to linear. This is better than the polynomial or quadratic complexity of best response algorithms in the computer science literature. We present examples to show that in general, the linearity bound cannot be improved upon using best response dynamics. We achieve this efficiency by innovatively leveraging and extending ideas from the general theory of complementarities.

Algorithm 1 formalizes best response dynamics for our generalized model. Starting from an arbitrary set $S$ of players with incentive to play $1$ at $q \in [0,1]$, it delivers an equilibrium object $C(S,q)$, which is the smallest Nash equilibrium (of players playing $1$) that contains $S$. We prove several properties of Algorithm 1, some that are adaptations and extensions of similar properties in the literature and others that are new results such as a bootstrapping property. 

Algorithm 2 is new and leverages the bootstrapping property of Algorithm 1 to derive a finite sequence of strictly increasing Nash equilibria $C(S, q_n)$ that converges to the full network and a corresponding strictly decreasing sequence of $(q_n)$ that converges to $q^*$, which we prove is the largest $q$ for which contagion occurs from $S$ to the entire network. This provides a new and computationally tractable characterization of the set of all $q$ for which contagion occurs from $S$ to the entire network. Bootstrapping implies that for our combined Algorithm 3 (Algorithm 2 with nested Algorithm 1) to run, the maximum number of nonempty subsets of the complement of $S$ that are checked is no more than the number of players in the complement of $S$. Examples show that this bound cannot be reduced in general. 

Our algorithms provide new and computationally tractable characterizations of several concepts from the existing literature. This includes characterizing when contagion occurs from $S$ to the entire network, characterizing when a set $S$ is $q$-cohesive and its complement is $(1-q)$-cohesive, characterizing when the complement of $S$ is uniformly no more than $(1-q)$-cohesive, and characterizing when conventions $0$ and $1$ coexist in equilibrium. 

We use our algorithm outputs to define and compute several equilibrium objects that facilitate the general study of contagion in networks. In addition to $q^*$, outputs of Algorithm 2 can be used to compute the (equilibrium) contagion depth function, $\delta(S,q)$, for every $q \in [0,1]$. This measures the equilibrium depth of proliferation of an action in the population starting from a given set $S$ at $q$. The difference between contagion depth $\delta(S,q)$ and size of its starting set $S$ provides an (equilibrium) definition of virality, measuring the incremental proliferation of an action beyond the starting set $S$, in equilibrium. We show how to (numerically) invert $\delta(S,q)$ to compute the answer to another important question: Given a depth of contagion $\delta \in [0,1]$ and $q \in [0,1]$, what is the smallest starting set size to achieve this depth in equilibrium? Our results make it feasible to compute comparative statics of these equilibrium objects.

Our algorithms and results are useful for a broader audience in the sense that they hold for all network coordination games with otherwise arbitrarily specified and heterogeneous local and global effects and the algorithms may be run with computing power that is accessible to most practitioners. Given the wide range of potential applications of our formulation, practitioners can flexibly study interactions among infinitely many different configurations of networks, local effects, and global effects, and also study tradeoffs among connectivity, network resilience, global impact and other variables of interest. Pseudocode for the algorithms is included to aid implementation.

We apply our theoretical results to study the spread of contagion in the class of Barab\'{a}si-Albert scale-free networks \cite{barabasi1999emergence} using Monte Carlo methods.  Scale-free networks are used widely in modeling social networks and naturally capture the idea of preferential attachment. Given a connectivity parameter $m$ (a positive integer signifying the number of existing nodes to which each new node is connected), the B-A procedure generates a scale-free network of the desired size. 

For each parameter $m \in \{ 5, 10, 20 \}$, we generate 40 scale-free networks of 1,000 players each using the B-A procedure. For each network, we randomly sample 50 starting sets $S$ each of sizes 10 through 990 (at increments of 10). Global effects are parameterized by $\alpha \in \{0, 0.5, 1\}$ with $\alpha = 0$ signifying the model without global effects, $\alpha = 1$ for the model with global effects, and $\alpha = 0.5$ for the case of intermediate impact of global effects. For each combination $(m, \alpha)$ and each starting set $S$, we use our algorithms to compute the contagion threshold and contagion depth function. This yields 1,782,000 complete runs of our algorithms. Robustness checks with additional variations in parameters are conducted with several million additional runs of our algorithms. 

Several features of contagion on networks emerge in the data. The broad patterns are as follows. Details are included later in the paper. As expected, increasing starting set size, or decreasing network resilience, or increasing global impact increases both equilibrium depth of contagion and the interval of $q$ for which contagion occurs to the whole network. Importantly, there is a natural complementarity among several parameters of interest. 

The larger is the starting set size the greater is the impact of global effects on the contagion threshold and depth of contagion. A large starting set directly influences more people to play $1$ and also increases the global impact because more people are playing $1$ in the aggregate. This complementarity naturally captures a snowball effect of contagion resulting from increasing returns. Indeed, after a point, global impact overrides local obstacles to switch to $1$. In low or moderate connectivity networks ($m=5,10$) with global effects ($\alpha=1$), if 40 percent of the network is infected (has an incentive to play 1 at $q=1$), there is no hope of stopping contagion to the entire network, regardless of local resilience of the remaining network. Increasing connectivity ($m$) makes it harder to spread contagion from a given starting set, because with more neighbors, a given starting set $S$ flips a smaller fraction of neighbors. 

In the absence of global effects, it is harder to spread contagion at higher resilience $q$. With global effects, contagion spreads deeper and using smaller starting sets. Complementarity between global effects and network resilience (local effects) implies that the equilibrium impact of global effects is higher when network resilience is higher. This increases the range of starting set sizes that lead to full network contagion and brings contagion depth closer to a type of singularity (also termed a \textit{tipping point} in the literature); a narrow interval of starting set sizes below which there is no contagion and above which there is full contagion. In most cases, the length of this interval is about 6-7 percentage points (in terms of proportion of number of players in the network).

For inverse depth of contagion, the broad pattern that emerges is that on average, in order to spread contagion to a given depth $\delta \in [0,1]$ of the network, smaller starting sets are needed if network resilience is lower, or global effects are higher, or connectivity is lower, or a combination of these. Different combinations of these parameters have differing impacts on the starting set size needed to achieve a given depth of contagion.

Our analysis provides insight into the design of policies to control or spread contagion in networks. Limiting starting set size, making the network more resilient, and reducing cross-network interaction (global effects) each reduces contagion by itself. Taking the reverse actions increases spread of contagion. Complementarities among these factors have important effects on depth of contagion.  

Global effects amplify the impact of a larger starting set, and therefore, in networks where global interaction is a common feature (or promoted by network owners), it is easier to curtail contagion by nipping it in the bud. Waiting for things to play out will add to the likelihood of considerably larger spread of contagion. 

Complementarity between global effects and network resilience (local effects) implies that the equilibrium impact of global effects is higher when network resilience is higher. In other words, it is especially in networks where we believe it is hard to spread contagion because local network resilience is high that we need greater vigilance against global effects (as they have a greater impact on virality). 

This complementarity brings the spread of contagion closer to a type of singularity or tipping point, and therefore, in more resilient and connected networks, an ability to control global effects or global interaction may be the difference between localized contagion around the starting set or widespread contagion throughout the network. 

More broadly, our research provides a framework to analyze the effects of different policies in a situation-dependent manner. The scope of application is enlarged given the many other situations across different fields and disciplines that may be modeled using this framework.  

The paper proceeds as follows. Section 2 presents the related literature. Section 3 presents the model of network coordination games with local and global effects, section 4 presents the contagion algorithms, proves their theoretical properties, and presents an illustrative example. Section 5 is devoted to a Monte Carlo study of scale-free networks using results from section 4. Section 6 provides broad principles to guide policies to control or spread contagion.

\section{Related literature} 

There is a vast literature that studies contagion using different models and tools. Some of the related literature is as follows. 

This paper belongs to the literature on contagion models in network coordination games. In particular, we study and extend the model of contagion on networks proposed by \cite{morris2000} and its adaptation to finitely many players in \cite{jackson2008} to include heterogeneity and global effects. An early reference is \cite{schelling1978sorting}. \cite{young2011} uses a coordination game on a network to model social innovation and the development of social norms, and provides theoretical results regarding speed of diffusion using perturbed best response dynamics that depend on neighbor actions. \cite{chwe} defines the concept of a minimal dependence network and uses it to characterize when coordination is possible in networks. \cite{jackson2019behavioral} investigate behavioral communities where people in some parts of the network adopt a given behavior while people in other parts of the network do not. \cite{oyama2015contagion} extend \cite{morris2000} to include the bilingual game and supermodular games using countably infinite players, local and symmetric interactions, and sequential best response dynamics. \cite{board2021} study learning dynamics based on adoption decisions of neighbors. 

Comprehensive frameworks for network games are presented in \cite{galeotti2010network}, \cite{jackson2015}, \cite{bramoulle15}, and \cite{jackson2017}, and book length treatments are available in \cite{jackson2008} and  \cite{goyal2009connections}. Applications to financial networks and contagion are studied in \cite{elliott2014financial}, to public goods in \cite{elliott2019network}, to innovation in \cite{dasaratha2019innovation}, and person-to-person spread in diffusion games is studied in \cite{sadler2020diffusion}. Models of learning in social networks are studied in \cite{golub2010naive} and an extensive survey is available in \cite{golublearning}. \cite{xu2018social} studies econometric estimation for particular network games with a unique equilibrium. 
\cite{bich2020} study existence of pairwise stable networks.Optimal lockdowns for the Covid-19 pandemic are studied in \cite{gallic2022optimal}, optimal pandemic interventions in a network are studied in \cite{freiberger2022chasing}, and a model with transmission delays is studied in \cite{hritonenko2022model}.

Applications of models of contagion on networks have been stymied by the difficulty of characterizing equilibrium analytically.  \cite{jackson2006diffusion}, \cite{jackson2007diffusion} address this by studying diffusion properties using the more analytically tractable mean-field analysis. They also use a Bayesian game formulation where agents are unsure of the larger network in which they reside and use symmetric Bayesian equilibrium as the equilibrium concept. The model in this paper does not require these additional features and uses computationally tractable algorithms to characterize different equilibrium objects. 

Mean-field models are an example of the more general class of models based on population-level arguments, which have proved to be tractable and yielded valuable insights. These models require same treatment of players with the same degree and cost structure, regardless of their position in the network. An early model in the sociology literature is given in \cite{granovetter1978threshold}. More recent examples are in \cite{watts2002simple}, \cite{dodds2004universal}, and \cite{wiedermann2020network}. Applications using experiments include \cite{centola2007complex} and \cite{centola2010spread}. These models rule out situations in which a player closer to a starting set is infected, but a player with the same characteristics further away is not infected. Our model allows for these situations in equilibrium and is based more explicitly on additional micro-foundations, including allowing for heterogeneity in local network topology and starting set location. 

Population-based models are related to another literature that is more common in epidemiology based on diffusion models of the form Susceptible-Infected-xxx, which includes Susceptible-Infected-Susceptible (SIS), Susceptible-Infected-Removed (SIR) and their many additional variants. These models help to quantify important aspects of disease transmission. They typically do not have an explicit network structure and are based on random matching derived from exogenously specified probabilities of being in particular states. In particular, a common homogeneous mixing assumption implies that each individual may have contact with any other individual, and therefore, even one infected individual eventually leads to infection of any individual in the community. This is the not case in the model studied here. 

There is another literature in physics and computer science. Bootstrap percolation models in physics (for example, \cite{adler1991bootstrap}) typically use node activation based on number of neighbors rather than fraction of neighbors. Work in computer science, for example, \cite{blume2011}, \cite{kempe2003}, and \cite{mossel2010}, identifies starting sets (or seeds) that maximize contagion in networks and proves some of their algorithmic properties. These build on ideas from viral marketing. These classes of models don't have global effects. Our work may be viewed as extending the foundation for these models in the decentralized and interdependent decision-making framework of game theory and Nash equilibrium as formulated in \cite{morris2000} to include global effects and provide computational techniques to characterize equilibrium. More recently, \cite{kobayashi2021dynamics} use a message-passing approach in locally tree-like networks to more accurately replicate the dynamics of diffusion as compared to mean-field methods. \cite{akbarpour2020} show that in the class SIR-type diffusion models, optimal seeding of a given size is dominated by random seeding of a slightly larger size, using undirected communication. 

There are additional non-network based game theory models that study contagion. Global games models of contagion include the model proposed by \cite{carlsson1993global}, and its developments in \cite{morris2003global}, \cite{frankel2003equilibrium}, \cite{hoffmann2019global}, and several others. Applications to signaling in global games, dynamic global games, and coordinating diffusion risk are studied in \cite{angeletos2006signaling}, \cite{angeletos2007dynamic}, and \cite{basak2020diffusing}. This literature focuses more on equilibrium selection by embedding a given game in a larger global game using uncertainty in observables and investigating how an equilibrium in dominant strategies in particular states may cause contagion to one of multiple equilibria in other states. These models typically use iterated deletion of strictly dominated strategies as a selection mechanism and do not impose a network structure. \cite{leister2021} use this approach to study binary choice adoption in networks.

\section{Network coordination games with local and global effects}

We follow \cite{morris2000} and its adaptation in \cite{jackson2008} to define a binary action network coordination game as follows. 

Consider a canonical $2 \times 2$ coordination game in which each player chooses action $0$ or $1$. Action $0$ may be viewed as a baseline action (or established belief) and $1$ as a new action or deviation from baseline. For example, playing $0$ may entail abiding by some societal norm, adhering to a traditional policy, or believing an established narrative. Playing $1$ entails a departure from an established societal norm, adopting a new standard or policy, or choosing to believe an alternative narrative.
The payoff from coordinating on the baseline action is normalized to $0$ for each player. The benefit from coordinating on action $1$ is $b \geq 0$ for each player. The cost of deviating to $1$ when the other player stays at baseline is $c \geq 0$ for the deviating player. By miscoordinating with a friend, one breaks with existing social norms or accepted policy and risks conflict and estrangement from those playing $0$. We assume that at least one of $b$ or $c$ is positive, or equivalently, $b+c>0$. Payoffs are summarized as follows. The game has two strict Nash equilibria: $(0,0)$ and $(1,1)$. 
\begin{table}[!htbp]
    \centering
    \setlength{\extrarowheight}{2pt}
    \begin{tabular}{cc|c|c|}
      & \multicolumn{1}{c}{} & \multicolumn{2}{c}{}\\
      & \multicolumn{1}{c}{} & \multicolumn{1}{c}{$1$}  & \multicolumn{1}{c}{$0$} \\\cline{3-4}
      \multirow{2}*{}  & $1$ & $b,b$ & $-c,0$ \\\cline{3-4}
      & $0$ & $0,-c$ & $0,0$ \\\cline{3-4}
    \end{tabular}
  \end{table}

A \textit{network} is a pair $(\mathcal{I}, \textbf{g})$, where $\mathcal{I} = \{1,\ldots, I \}$ is a finite set of players (or nodes) in the network and $\textbf{g}$ specifies which players are connected to each other. 
For each player $i$ in network $(\mathcal{I}, \textbf{g})$, the neighbors of $i$ are $N_i = \{j \in \mathcal{I}\backslash \{i\} \mid ij \in \bg\}$ and the degree of $i$ is the number of its neighbors, denoted $d_i = \lvert N_i \rvert$. A network $(\mathcal{I}, \bg)$ is \textit{connected}, if for every pair of players $i$ and $j$, there is a sequence of indices $(i_1 = i, i_2,\ldots, i_{k-1}, i_{k} = j)$ such that $i_{\ell} i_{\ell+1} \in \bg$ for every $\ell = 1, \ldots, k-1$. 

In a network, incentives for each player to coordinate with their neighbors are viewed in terms of the $2 \times 2$ coordination game above. Each person's payoff depends on the choice of their network neighbors. If player $i$ plays $1$ and neighbor $j \in N_i$ also plays $1$, player $i$ garners some benefit due to solidarity, conformity effects, or other positive effects of coordination. Or, seeing a neighbor or friend deviate to $1$ can provide justification for one's own deviation. This is modeled as benefit $b \ge 0$ that player $i$ receives for each neighbor $j$ with whom $i$ coordinates on playing $1$. Playing $1$ has a drawback if neighbor $j$ plays $0$. Miscoordinating with a neighbor is modeled as cost $c \ge 0$ that player $i$ playing $1$ incurs for each neighbor $j$ who plays $0$. For convenience, payoff to a player from playing baseline action $0$ is normalized to be $0$ so that $b$ and $c$ represent benefit and cost of deviation from baseline. 

We allow for heterogeneity in the importance of different neighbors by including weight $w_{i,j} \geq 0$ that player $i$ puts on relevance of neighbor $j$'s action. Thus, weighted benefit to $i$ from coordination on $1$ with neighbor $j$ is $w_{i,j}b$ and weighted cost to $i$ from playing $1$ when neighbor $j$ plays $0$ is $w_{i,j}c$, for each $j \in N_i$. Let $w_i = \sum_{j \in N_i} w_{i,j}$ and for non-trivial results, assume that $w_i > 0$. In addition to heterogeneity in neighbor effects, this formulation allows for asymmetric weights, that is, for $i\neq j$, $w_{i,j}$ is not necessarily equal to $w_{j,i}$, and it also allows for unidirectional links, that is, for $i\neq j$, $w_{i,j}$ may be zero but $w_{j,i}$ may be positive.  

A \textit{\textbf{network coordination game with local effects}} is a tuple $\Gamma = (\mathcal{I}, \bg, (A_i, u_i)_{i \in \mathcal{I}})$, where $(\mathcal{I}, \bg)$ is a connected network, the action space of player $i \in \mathcal{I}$ is $A_i = \{0,1\}$, the joint action space is $A = A_1 \times \cdots \times A_I$, and player $i$ payoff, $u_i: A \rightarrow \mathbb{R}$, is given by $u_i(a_i, a_{-i}) = \sum_{j \in N_i} w_{i,j} \hat{u}_i(a_i, a_j)$, where $\hat{u}_i(a_i, a_j)$ is player $i$ payoff from playing the $2 \times 2$ coordination game with neighbor $j$ and $w_{i,j}$ is weight placed by player $i$ on neighbor $j$. 

In a network coordination game with (heterogeneous) local effects, $1$ is a best response of player $i$, if, and only if, $u_i(1, a_{-i}) \ge u_i(0, a_{-i})$, which holds exactly when $\frac{1}{w_i}\sum_{j \in N_i} w_{i,j}a_j \ge \frac{c}{b+c}$. Let $s_i = \sum_{j \in N_i} w_{i,j}a_j$ be the weighted number of neighbors of $i$ who play $1$ and let $q=\frac{c}{b+c} \in [0,1]$ be a reduced form parameter summarizing relative cost of miscoordination. In other words, \textbf{$1$ is a best response of player $i$, if, and only if, $\frac{s_i}{w_i} \geq q$}. In the special case when each neighbor has uniform unit weight, that is, $w_{i,j} = 1$, this condition specializes to $\frac{s_i}{d_i} \geq q$, that is, the fraction of player $i$'s neighbors who are playing $1$ is at least $q$. This is the condition in \cite{morris2000} and \cite{jackson2008}. 

The model does not ascribe special relevance to either action $0$ or $1$. That can depend on the particular application. We follow the literature in assuming that coordinating on $1$ gives payoff $b$ and coordinating on $0$ gives payoff $0$. Relabeling can reverse these payoffs, for example, by setting the payoff from playing $1$ to $0$, and the payoff from playing $0$ depends on whether the opponent is playing $0$ (in which case payoff is $b$) or $1$ (in which case payoff is $-c$). The threshold to play $1$ is then given in terms of $1-q$ and the entire analysis to proliferate action $1$ can be carried out analogously. We follow the formulation in the literature to build transparently on existing work.

The parameter $q$ may be viewed as an index of network resilience to contagion. When player $i$ plays $1$, $b$ indexes intensity of benefit of coordination with neighbors who play $1$, $c$ indexes intensity of cost of miscoordination with neighbors who play $0$, and $q = \frac{c}{b+c}$ measures the relative cost of miscoordination with neighbors who play $0$. When $q$ is low, miscoordination cost is relatively low, and it is easier for player $i$ to switch to $1$ (easier for $\frac{s_i}{w_i} \geq q$ to hold). When $q$ is high, miscoordination cost is relatively high and it is harder for player $i$ to switch to $1$. In other words, in a network coordination game with low $q$, the network is less resilient to contagion and in one with high $q$, the network is more resilient to contagion.  

An important feature of spread of contagion on networks is the role of global effects. In its simplest form, global effects capture the notion that player $i$ may take action $1$ when a large number of other players take action $1$ even if they are not network neighbors of $i$. For example, in social networks, this may be achieved by publishing trending topics, or the numbers of likes or retweets, and these are visible to every player. 

We extend the network coordination game to include global effects as follows. Given player $i$, and a profile $a_{-i}$ of actions of other players in the network, let $p_i(a_{-i}) = \frac{1}{I-d_i-1}\sum_{j \not\in N_i \cup \{i\}} a_j$ be the fraction of the network (not including $i$ or neighbors of $i$) that plays $1$. Player $i$ gets some additional benefit from playing $1$ that depends positively on this aggregate proportion. This is formalized by a (weakly) increasing function $\phi_i: [0, 1] \to \mathbb{R}_+$ such that payoff to player $i$ from playing $1$ is given by $u_i(1, a_{-i}) = \sum_{j \in N_i} w_{i,j}\hat{u}_i(1, a_j) + \phi_i(p_i(a_{-i}))$ with the natural normalization that $\phi_i(0) = 0$. When convenient, we suppress the notation $a_{-i}$ in $p_i(a_{-i})$. 

In a \textbf{\textit{network coordination game with local and global effects}} payoff of each player $i$ is given by $u_i(1, a_{-i}) = \sum_{j \in N_i} w_{i,j} \hat{u}_i(1, a_j)+ \phi_i(p_i(a_{-i}))$, and $u_i(0, a_{-i}) = 0$, where $\phi_i: [0, 1] \to \mathbb{R}_+$ is a (weakly) increasing function with $\phi_i(0) = 0$. Notice that a network coordination game with local and global effects is a game with strategic complements, because payoff of each player satisfies increasing differences: $u_i(1, a_{-i}) - u_i(0, a_{-i})$ is (weakly) increasing in $a_{-i}$. Indeed, this holds for a (weakly) increasing function of the form $\phi_i(a_{-i})$, without composing with $p_i(a_{-i})$, and our results hold for this generalization. We use $p_i(a_{-i})$ to aid intuition for global effects. Moreover, it follows immediately that a network coordination game with local effects is the special case when each $\phi_i$ is identically zero. Our formulation is more general than assuming a weight $w_{i,j} \geq 0$ for each non-neighbor $j$ of $i$. 

With heterogeneous local and global effects, \textbf{$1$ is a best response of player $i$, if, and only if, $u_i(1, a_{-i}) \ge u_i(0, a_{-i})$, which holds exactly when $\frac{s_i}{w_i} \geq q - \frac{\phi_i(p_i)}{w_i(b+c)}$}, where $s_i$ and $q$ are as above. As $\phi_i$ is (weakly) increasing, the threshold (weighted) fraction of neighbors who play $1$ before $i$ switches to $1$ decreases in $p_i$. In other words, with global effects, a smaller fraction of a given player's neighbors need to play $1$ as compared to the case with local effects only. In this sense, global effects reduce network resilience to contagion. 

To ensure that the threshold fraction $q - \frac{\phi_i(p_i)}{w_i(b+c)}$ is nonnegative (that is, global effects alone do not make $1$ strictly dominant for player $i$), it is necessary and sufficient that $\phi_i(p_i) \leq cw_i$. We impose this constraint going forward. There are no other restrictions on $\phi_i$ in terms of continuity, differentiability, concavity, convexity, and so on. 

When $\phi_i$ is identically zero, the condition $\frac{s_i}{w_i} + \frac{\phi_i(p_i)}{w_i(b+c)} \geq q$, specializes to $\frac{s_i}{w_i} \geq q$, and with equal and unit weights to $\frac{s_i}{d_i} \geq q$ as in the earlier literature. With global effects, there is an additional contribution of $b$ and $c$ in a player's decision, as shown in the term $\frac{\phi_i(p_i)}{w_i(b+c)}$. In this case, considering a reduced form $q$ alone is not sufficient to analyze the model, because there is an indeterminacy among the infinitely many $b$ and $c$ that yield the same $q$. Without loss of generality, we adopt a normalization to address this by fixing $c$ to be a positive constant throughout the paper, and considering the one-to-one mapping $b \mapsto q = \frac{c}{b+c}$. Intuitively, this normalization fixes miscoordination cost at some baseline value $c > 0$ (for example, $c=1$) and coordination benefit $b$ is measured relative to baseline cost. 

Our model includes complementarities in several parameters in a natural manner. The condition $u_i(1, a_{-i}) \ge u_i(0, a_{-i})$ can be written equivalently as $\frac{s_i(a_{-i})}{w_i} + \frac{\phi_i(p_i(a_{-i}))}{w_ic}q \geq q$ and further as $\frac{cs_i(a_{-i})}{cw_i - \phi_i(p_i(a_{-i}))} - q \geq 0$. These formulations show that in addition to increasing differences in $(a_i, a_{-i})$, the payoff function has single crossing property in $(a_i, \phi_i)$ capturing a natural complementarity with overall global effects. Considering $-q$ as inverse network resilience or network ``vulnerability" (to local effects), the payoff function has single crossing property in $(a_i, -q)$ capturing a complementarity with inverse network resilience or network vulnerability. A cross partial of payoff with respect to $\phi_i$ and $q$ ($c$ is fixed) yields $\frac{1}{w_ic}$ showing a complementarity between network resilience and global effects. The greater is network resilience the larger is the impact of global effect to switch a given player to play $1$. These relationships emerge explicitly in Monte Carlo simulations of equilibrium objects. 

As usual, a profile of actions $a^* = (a^*_1,\ldots, a^*_I)$ is a (pure strategy) \textbf{\textit{Nash equilibrium at $q$}}, if for every player $i$, $a^*_i$ is a best response to $a^*_{-i}$. 

Following \cite{morris2000}, a profile of actions $a = (a_1,\ldots,a_I)$ is identified naturally with the subset of players who play $1$, that is, with $E = \{ i \in \mathcal{I} \mid a_i=1 \}$, and conversely, each subset $E$ of $\mathcal{I}$ is identified with the profile in which player $i$ plays $1$, if, and only if, $i \in E$. Thus, if $E \subseteq \mathcal{I}$ is the set of players playing $1$, we may write the (weighted) number of neighbors of $i$ who play $1$ as $s_i = \sum_{j\in N_i \cap E} w_{i,j}$ and the (weighted) fraction of neighbors of $i$ who play $1$ as $\frac{s_i}{w_i}$. The fraction of the rest of the network (excluding $i$ and neighbors of $i$) that plays $1$ is computed as  
$p_i = \frac{\lvert E \setminus (N_i \cup \{ i \}) \rvert}{I-d_i-1}$, where, as usual, $\vert A \rvert$ denotes cardinality of set $A$. When needed, the dependence of $s_i$ and $p_i$ on $E$ is denoted $s_i(E)$ and $p_i(E)$, respectively. If $E=\emptyset$, then $s_i(E) = p_i(E) = 0$. 

Also following \cite{morris2000}, we assume that in case of indifference between 0 and 1, a player plays 1. It follows that in our formulation of a network coordination game with heterogeneous local and global effects, \textbf{set $E \subseteq \mathcal{I}$ is a Nash equilibrium at $q$, if, and only if, for every $i \in E$, $\frac{s_i(E)}{w_i} +  \frac{\phi_i(p_i(E))}{w_i(b+c)} \geq q$ and for every $i \in \mathcal{I}\backslash E$, $\frac{s_i(E)}{w_i} +  \frac{\phi_i(p_i(E))}{w_i(b+c)} < q$}. 

We include players who may be exogenously infected to play $1$ as follows. \textbf{Player $i$ is exogenously infected to play $1$}, if $1$ is strictly dominant for player $i$. We may formalize this by setting $u_i(1,a_{-i})=1$ and $u_i(0,a_{-i})=0$ for such a player. The set of players exogenously infected to play $1$ is denoted $\mathcal{D} \subseteq \mathcal{I}$. We allow for the case that $\mathcal{D} = \emptyset$. Equilibrium is characterized as follows. A \textbf{set $E \subseteq \mathcal{I}$ is a Nash equilibrium at $q$, if, and only if, $\mathcal{D} \subseteq E$ and for every $i \in E\backslash \mathcal{D}$ $\frac{s_i(E)}{w_i} +  \frac{\phi_i(p_i(E))}{w_i(b+c)} \geq q$, and for every $i \in \mathcal{I} \backslash E$, $\frac{s_i(E)}{w_i} + \frac{\phi_i(p_i(E))}{w_i(b+c)} < q$}. This characterization is helpful in the analysis below. 

Both conditions for player $i$ to play $1$ are summarized by saying that for a given set $S$ of players, \textbf{\textit{player $i \in S$ has an incentive to play $1$ at $q$}}, if either $i \in \mathcal{D}$ or $\frac{s_i(S)}{w_i} +  \frac{\phi_i(p_i(S))}{w_i(b+c)} \geq q$. Moreover, \textbf{\textit{$S$ is a set of players with incentive to play $1$ at $q$}}, if every player $i \in S$ has an incentive to play $1$ at $q$. Notice that if $S \subseteq \mathcal{D}$, then for every $q\in [0,1]$, $S$ is a set of players with incentive to play $1$ at $q$. 

For a computationally tractable version of the model that we use in Monte Carlo simulations, we assume that each neighbor has the same unit weight, that is, $w_{i,j}=1$ for every $j \in N_i$, and suppose $\phi_i(p_i) = \alpha cd_i p_i$, where $\alpha \in [0, 1]$ is a parameter capturing the intensity of global effect on payoffs. This captures the different impacts succinctly. The function $\phi_i$ is (weakly) increasing in $p_i$ (and strictly increasing when $\alpha > 0$). The term $cd_i$ is the upper bound on $\phi_i$ to keep the threshold nonnegative. The term $\alpha$ captures intensity of global effect; when $\alpha=0$, there are no global effects on payoffs and $\alpha=1$ indexes the full global effect. The formulation is computationally tractable in the sense that \textit{$1$ is a best response of player $i$, if, and only if, $\frac{s_i}{d_i} \geq q(1 - \alpha p_i)$}. Both $\alpha$ and $p_i$ serve to lower the switching threshold in an intuitive and analytically tractable manner. There is a complementarity among $\alpha$, $p_i$, and $q$; the higher is one, the greater is the impact of the others to lower the threshold to play $1$.

\section{Contagion algorithms}

A central problem in the study of contagion on networks is to understand the proliferation of an action in the network. Suppose there is an initial set $S$ of players who have an incentive to play $1$ (either based on the configuration of play given by $S$ or due to an exogenous infection). Suppose player $i \not\in S$ is playing $0$. If a sufficiently large proportion of $i$'s neighbors play $1$, player $i$ finds it optimal to play $1$ as well. This may further induce some of player $i$'s other neighbors who were playing $0$ to play $1$, spreading action $1$ further along. This is the underlying dynamic for proliferation based on local network effects. Moreover, as more players in the network play $1$, this may induce a player to play $1$ even if fewer of its neighbors are playing $1$. This is the impact on proliferation based on global network effects. Best response dynamics iterates this process, including local and global effects simultaneously.  

Several questions are of interest. Starting from an arbitrary set of players $S$ playing $1$ at $q$, how far will action $1$ proliferate in the network? What characteristics does this limit set have? Can we compute this limit? How does this limit depend on $S$ and $q$? Under what conditions is the limit set the entire network? Our algorithms help to answer these questions. 

For network coordination games with local effects only and with uniform unit weight on neighbors, \cite{morris2000} characterizes equilibrium behavior in terms of cohesiveness of subsets of players in a network, as follows. For $r \in [0,1]$, subset $S$ of $\mathcal{I}$ is \textit{$r$-cohesive}, if every $i \in S$ has at least proportion $r$ of its neighbors in $S$. Equivalently, $S$ is $r$-cohesive, if, and only if, $\min_{i \in S} \frac{ s_i}{d_i} \geq r$, where $s_i = \lvert N_i \cap S \rvert$ is the number of neighbors of $i$ in $S$. The cohesiveness of $S$ is the largest $r$ such that $S$ is $r$-cohesive. Set $S$ is \textit{uniformly no more than $r$-cohesive}, if every nonempty subset of $S$ is no more than $r$-cohesive.  \cite{jackson2008} adapts \cite{morris2000} to characterize when contagion occurs from $S$ to $\mathcal{I}$ at $q$ in terms of uniform cohesion. Both results are as follows.

\begin{theorem} \label{jackson}
Let $\Gamma$ be a network coordination game with local effects only and uniform unit weights on neighbors, and $q \in [0,1]$. 
\begin{enumerate} 
\item {\cite{morris2000} A nonempty set $E \subset \mathcal{I}$ is a Nash equilibrium at $q$, if, and only if, $E$ is $q$-cohesive and $\mathcal{I}\backslash E$ is $(1-q)$-cohesive.} 
\item {\cite{jackson2008} Contagion occurs from $S$ to $\mathcal{I}$ at $q$, if, and only if, $\mathcal{I} \backslash S$ is uniformly no more than $(1-q)$-cohesive.}
\end{enumerate} 
\end{theorem} 

\begin{proof} Confer Propositions 9.7 and 9.8 in \cite{jackson2008}. 
\end{proof}

Both statements provide helpful insights about the role of mutual connectivity of sets of neighbors, but are hard to implement computationally. 

Statement (1) has been used to provide intuition about when different actions may coexist in equilibrium, sometimes termed \textit{coexistence of conventions} in a population. Statement (1) says that conventions coexist in equilibrium at $q$ exactly when connectivity within one set of players is at level $q$ and connectivity within its complement is at level $1-q$. In other words, two complementary sets with different actions coexist in equilibrium if players in each one of them are appropriately self-connected. 

Statement (2) characterizes when contagion occurs from $S$ to the entire network but involves checking cohesiveness of each subset of the complement of $S$. This involves checking $2^{\lvert \mathcal{I} \backslash S\rvert}$ sets, which is also computationally infeasible as network size increases. The problem is exacerbated if we also want to understand how the analysis changes with changes in $S$, $q$, virality, and network connectivity.  

Both statements in \Cref{jackson} apply to network coordination games with local effects only and with uniform unit weights on neighbors. Neither statement is necessarily true with global network effects, because in that case, actions of non-neighbors of $i$ also influence the choice of player $i$, whereas cohesion conditions are local conditions about mutual connectivity. 

We consider the more general case of network coordination games with heterogeneous local and global effects, and we present algorithms to compute the extent of proliferation of action $1$ starting from an arbitrary subset $S$ of players playing $1$ at an arbitrary threshold $q = \frac{c}{b+c} \in [0,1]$. We go further to compute the largest $q$ for which contagion occurs from $S$ to the entire network. 

\vspace{0.2in}
\noindent 
\textbf{Algorithm 1 (Depth of contagion starting from $(S,q)$)} 
\begin{enumerate} 
\item Let $S$ be a set of players with incentive to play $1$ at $q =\frac{c}{b+c} \in [0, 1]$. 

\item Let $C_0 = S \cup \mathcal{D}$ and $B_0 = \mathcal{I} \backslash C_0$. 

\item For $t \geq 0$, recursively define \\
$F_{t+1} = \left\{i \in B_t \mid \frac{s_i(C_t)}{w_i} + \frac{\phi_i(p_i(C_t))}{w_i(b+c)} \geq q \right\}$,\\ 
$C_{t+1} = C_t \cup F_{t+1}$, and\\ 
$B_{t+1} = \mathcal{I}\backslash C_{t+1}$. 

\item Let $T$ be the first $t$ such that $C_{t+1} = C_{t}$. Let $C(S,q) = C_T$. 
\end{enumerate} 

Algorithm 1 formalizes best response dynamics starting from an initial set $S$ of players with an incentive to play $1$ at $q$. We typically have in mind a situation in which the starting set is the set of exogenously infected players or contains it ($S = \mathcal{D}$ or $S \supseteq \mathcal{D}$) but we allow for the case that there may be exogenously infected players not in $S$, that is, $\mathcal{D} \backslash S \neq \emptyset$. Of course, these players play $1$ and affect best response dynamics, and therefore, they are included in the contagion set at the beginning of the algorithm by setting $C_0 = S \cup \mathcal{D}$. At each step $t+1$, the algorithm identifies players not in the contagion set $C_t$ who have an incentive to play 1, denoted $F_{t+1}$, and includes them in $C_{t+1}$. The algorithm stops when there are no more players left to flip, or the first $t$ such that $F_{t+1} = \emptyset$. The ending contagion set is $C(S,q)$. When convenient, we denote stage $t$ dependence of the algorithm on the starting parameters $(S,q)$ by $F_t = F_t(S,q)$, $C_t = C_t(S,q),$ and $B_t = B_t(S,q)$. 

By construction, the sequence $(C_t)_{t=0}^T$ is a strictly increasing sequence of sets starting at $C_0 = S \cup \mathcal{D}$ and ending at $C(S,q)$. As the number of players is finite, Algorithm 1 stops in finitely many iterations, with an upper bound of $\vert \mathcal{I}\backslash S \vert$ on the number of subsets of the complement of $S$ that are evaluated. 

Algorithm 1 allows for $S=\emptyset$ and $\mathcal{D} = \emptyset$. In this case, if $q > 0$, then $F_1 = \emptyset$, the algorithm stops in one step and $C(S,q)=\emptyset$, which is a Nash equilibrium. If $q=0$, then $F_1=\mathcal{I}$, the algorithm stops after two steps and $C(S,q) = \mathcal{I}$, which is also a Nash equilibrium. Indeed, if $q=0$, then for every $S$ and $\mathcal{D}$, Algorithm 1 stops after at most two steps and $C(S,q) = \mathcal{I}$. 

Comparing Algorithm 1 with best response dynamics in games with strategic complements (GSC), recall that in GSC, convergence of best response dynamics and its limit properties are typically shown with the smallest or largest profile of actions as the starting point, and moreover, it is possible for best response dynamics to decrease. 
In our case, best response dynamics can start at any point in the strategy space, because $S$ is an arbitrary subset of players, and moreover, by construction, best response dynamics increase (weakly) in all cases. This is made possible by our assumption that $S$ is a set of players with incentive to play $1$ at $q$. Furthermore, we prove that best response dynamic has a very useful bootstrapping property (formalized below). To the best of our knowledge, this property is not present in the previous literature. 

\begin{theorem}[Properties of Algorithm 1] \label{combined} 
Let $\Gamma$ be a network coordination game with local and global effects. 
\leavevmode
\begin{enumerate}
    \item Subset $E$ of $\mathcal{I}$ is a Nash equilibrium at $q$, if, and only if, $C(E,q) = E$. 
    \item For every $(S, q)$, $C(S, q)$ is a Nash equilibrium at $q$. 
    \item For every $q$, if $S \subseteq S'$, then $\forall t,$ $C_t(S,q) \subseteq C_t(S',q)$, and therefore, $C(S,q) \subseteq C(S',q)$.
    \item For every $S$, if $q' \leq q$, then $\forall t,$ $C_t(S,q) \subseteq C_t(S,q')$, and therefore, $C(S,q) \subseteq C(S,q')$.
    \item For every $(S, q)$, $C(S, q)$ is the smallest Nash equilibrium at $q$ that contains $S$. 
    \item Consider $(S,q)$ and let $A = C(S,q)$. If $q' \leq q$, then $C(A, q') = C(S,q')$.
\end{enumerate}
\end{theorem}
\begin{proof}
Statement (1) is proved as follows. If $E$ is a Nash equilibrium at $q$, then $\mathcal{D} \subseteq E$, for every $i \in E\backslash \mathcal{D}$, $\frac{s_i(E)}{w_i} + \frac{\phi_i(p_i(E))}{w_i(b+c)} \geq q$, and for every $i \not\in E$, $\frac{s_i(E)}{w_i} + \frac{\phi_i(p_i(E))}{w_i(b+c)} < q$. Therefore, Algorithm 1 terminates in one step ($T=0$) to yield $C(E,q) = E$. 
In the other direction, suppose $C(E,q) = E$. By construction of Algorithm 1, $E \cup \mathcal{D} = C_0 \subseteq C(E,q) = E$, whence $\mathcal{D} \subseteq E$. Moreover, $i \in E\backslash \mathcal{D} \subseteq C(E,q)$ implies  $\frac{s_i(E)}{w_i} + \frac{\phi_i(p_i(E))}{w_i(b+c)} \geq q$. Finally, $i \not\in E$ implies $i \not\in C(E,q)$, whence Algorithm 1 implies $\frac{s_i(E)}{w_i} + \frac{\phi_i(p_i(E))}{w_i(b+c)} < q$. Therefore, $E$ is a Nash equilibrium.

Statement (2) is proved as follows. If the algorithm starts at $(S,q)$ and stops at $E = C(S,q)$, then restarting it at $E$ with no change in $q$ will terminate it in one step yielding $C(E, q) = E$. Statement (1) now implies that $C(S,q)$ is a Nash equilibrium at $q$. 

Statement (3) is proved as follows. Suppose $S \subseteq S'$. We use induction on $t$. For $t = 0$, $C_0(S,q) = S \cup \mathcal{D} \subseteq S' \cup \mathcal{D} = C_0(S', q)$. Suppose $C_t(S,q) \subseteq C_t(S',q)$. Let $A = C_t(S,q) $, $A' = C_t(S',q)$, and $E = A' \backslash A$. Then $A \cup E = A'$ and $A^c = E \cup (A')^c$, where superscript $c$ denotes complement of a set. Notice that each of $A$ and $A'$ is a set of players with incentive to play $1$ at $q$ and contains $\mathcal{D}$. We show that $F_1(A,q) \subseteq F_1(A',q) \cup E$ as follows. Suppose $i \in F_1(A,q)$. Then $i\not\in A$ and $\frac{s_i(A)}{w_i} + \frac{\phi_i(p_i(A))}{w_i(b+c)} \geq q$. Moreover, $i\not\in A$ implies $i \in E \cup (A')^c$. If $i \in E$, then $i\in F_1(A'q) \cup E$. Otherwise, $i \not\in A'$ and also $q \leq \frac{s_i(A)}{w_i} + \frac{\phi_i(p_i(A))}{w_i(b+c)} \leq \frac{s_i(A')}{w_i} + \frac{\phi_i(p_i(A'))}{w_i(b+c)}$, where the last inequality follows from $A \subseteq A'$. Therefore, $i \in F_1(A'q) \subseteq F_1(A'q) \cup E$. It follows that $C_{t+1}(S,q) = C_0(A,q) \cup F_1(A,q) \subseteq C_0(A,q) \cup F_1(A',q) \cup E = C_0(A',q) \cup F_1(A',q) = C_{t+1}(S',q)$.

Statement (4) is proved as follows. Suppose $q' \leq q$. We use induction on $t$. For $t = 0$, $C_0(S,q) = S = C_0(S,q')$. Suppose $C_t(S,q) \subseteq C_t(S,q')$. Let $A = C_t(S,q)$ and $A' = C_t(S,q')$ and consider $(A,q)$ and $(A,q')$. Then $F_1(A,q) \subseteq F_1(A,q') \Rightarrow C_1(A,q) \subseteq C_1(A,q')$. Therefore, $C_{t+1}(S,q) = C_1(A,q) \subseteq C_1(A,q') \subseteq C_1(A',q') = C_{t+1}(S,q')$, where the second inclusion follows from statement (3). 

Statement (5) is proved as follows. If $E$ is a Nash equilibrium at $q$ that contains $S$, then $S \subseteq E$ implies $C(S,q) \subseteq C(E,q) = E$, where the inclusion follows from statement (3) and equality from statement (1). 

Statement (6) is proved as follows. Note that $S \subseteq A \Rightarrow C(S,q') \subseteq C(A,q')$ by statement (3). Let $A' = C(S,q')$. Statement (4) implies that $A = C(S,q) \subseteq C(S,q') = A'$. Applying statement (3) we have that $C(A, q') \subseteq C(A', q') = C(S,q')$, where the last equality follows from statement (2). 
\end{proof}

For a fixed $q$, Algorithm 1 may be viewed as a mapping $S \mapsto C(S,q)$, from an initial set $S$ of infected players to a (weakly) larger set $C(S,q)$ of all players who are infected by players in $S$ using both local network effects and global network effects. 

Statements (1), (2), and (5) are adaptations of results from the theory of GSC. Our model has some different-looking features as compared to a standard GSC, in terms of using subsets of players to formulate profiles of actions, including local and global effects for an arbitrary network, including exogenously infected players, and requiring a notion of when a set of players has an incentive to play $1$ at $q$. A benefit of our formulation is that it preserves strategic complementarities. We give direct proofs of these statements for completeness. Statement (1) shows that \textit{the set of Nash equilibria at $q$ is equal to the set of fixed points of Algorithm 1}. Statements (2) and (5) show that regardless of the starting configuration $S$ of players with an incentive to play $1$ at $q$, Algorithm 1 expands this set to deliver the smallest Nash equilibrium at $q$ containing $S$. 

Statements (3) and (4) prove comparative statics properties of best response dynamics in our model and its implications for comparative statics of equilibrium outcomes generated by Algorithm 1. If starting set increases, the number of players who play 1 goes up (weakly) at each step in the best response dynamics and in the resulting Nash equilibrium. If network resilience parameter $q$ goes down, the number of players who end up playing 1 goes up (weakly) at each step in the best response dynamics and in the resulting Nash equilibrium. These two statements generalize the corresponding statements in \cite{morris2000} to include global network effects. 

Statement (6) is an important bootstrapping property of Algorithm 1. If we start at $(S, q)$ and end at $C(S, q) \neq \mathcal{I}$, then statement (4) says that we can reduce $q$ to $q'$ and end up with a (weakly) larger contagion set $C(S, q')$. Statement (6) says that instead of running the entire analysis again by starting at $(S, q')$, we can start the algorithm at $C(S, q)$ and $q'$ and end up with the same result as starting at $(S, q')$. This bootstrapping from the end of one run of the algorithm to a continuation of the next to reach the same final outcome is an important step in providing a greatly improved algorithm to compute the contagion threshold using Algorithm 2. It is useful to compute the equilibrium depth of contagion for every $q$ as well. To the best of our knowledge, this property is not present in the existing literature.

Algorithm 1 depends on the global effect functions $(\phi_i)_{i=1}^I$ as well. Comparative statics with respect to these functions and the corresponding complementarity between local and global effects follows immediately as follows. Let $\phi = (\phi_i)_{i=1}^I$ and define $\hat{\phi} \leq \tilde{\phi}$ to mean that for every $i$, $\hat{\phi}_i \leq \tilde{\phi}_i$ pointwise. Including $\phi$ in the contagion set notation, let $C_t(S,q, \phi)$ denote the step $t$ contagion set and $C(S,q, \phi)$ the stopping set for Algorithm 1. 

\begin{theorem}\label{theo:global}
Let $\Gamma$ be a network coordination game with local and global effects.   
\begin{enumerate} 
\item For every $(S, q)$, if $\hat{\phi} \leq \tilde{\phi}$, then $\forall t,$ $C_t(S,q, \hat{\phi}) \subseteq C_t(S,q, \tilde{\phi})$, and therefore, $C(S,q, \hat{\phi}) \subseteq C(S,q, \tilde{\phi})$. 
\item For every $S$, if $\tilde{q} \leq \hat{q}$ and $\hat{\phi} \leq \tilde{\phi}$, then $C(S, \hat{q}, \hat{\phi}) \subseteq C(S, \tilde{q}, \hat{\phi})$, $C(S, \hat{q}, \tilde{\phi}) \subseteq C(S, \tilde{q}, \tilde{\phi})$, and $C(S, \tilde{q}, \hat{\phi}) \subseteq C(S, \tilde{q}, \tilde{\phi})$. 
\end{enumerate} 
\end{theorem} 

\begin{proof} Statement (1) follows immediately from 
$ \frac{s_i(C_t)}{w_i} + \frac{\tilde{\phi}_i(p_i(C_t))}{w_i(b+c)} \geq \frac{s_i(C_t)}{w_i} + \frac{\hat{\phi}_i(p_i(C_t))}{w_i(b+c)}$. 
Statement (2) follows from statement (1) in this theorem and statement (4) of \Cref{combined}. 
\end{proof} 

Statement (2) of \Cref{theo:global} formalizes a complementarity between local and global effects. Reducing $q$ (strengthening contagion due to local effect) increases contagion both at $\hat{\phi}$ and at $\tilde{\phi}$, but the final effect is larger at the higher global effect $\tilde{\phi}$, that is, $C(S, \tilde{q}, \hat{\phi}) \subseteq C(S, \tilde{q}, \tilde{\phi})$. 

The output of Algorithm 1, $C(S,q)$, is a useful equilibrium object. We use it to define the \textbf{\textit{(equilibrium) depth of contagion from $S$ at $q$}} as  
$\delta(S, q) = \frac{\lvert C(S, q)\rvert}{\lvert \mathcal{I}\rvert}$, measuring the fraction of the network to which action $1$ spreads in equilibrium starting from a given set $S$ and network resilience $q$. $C(S,q)$ provides an equilibrium definition of virality as well. In the existing literature, virality typically signifies that an action spreads far from its starting set. The notion of far is typically specified as an \textit{ad hoc} number of hops from the initial set. There is no notion of how far an action spreads \textit{in equilibrium}. In our model, the difference between $C(S,q)$ and the starting set $S$ is a natural definition. In other words, the \textbf{\textit{(equilibrium) virality of $S$ at $q$}} is $v(S,q) = \delta(S, q) - \frac{\lvert S \rvert}{\lvert \mathcal{I}\rvert} \geq 0$, measuring how far action $1$ proliferates from starting set $S$ at $q$ in equilibrium. Although we scale both notions by the size of the network as a natural normalization, the definitions can be applied equivalently in terms of number of players as well. 

Properties of Algorithm 1 in \Cref{combined} can be used to prove new characterizations of earlier results for network coordination games with local effects and with uniform unit weights.  
\begin{corollary}\label{corr:alg1}
Let $\Gamma$ be a network coordination game with local effects only and uniform unit weights on neighbors, and $q \in [0,1]$. 
\begin{enumerate} 
\item {A nonempty set $E \subset \mathcal{I}$ is $q$-cohesive and $\mathcal{I}\backslash E$ is $(1-q)$-cohesive, if, and only if, $E$ is a fixed point of Algorithm 1.} 
\item {For every nonempty set $S$ of players with incentive to play $1$ at $q$, $C(S,q)$ is the smallest set $E$ containing $S$ such that $E$ is $q$-cohesive and $\mathcal{I}\backslash E$ is $(1-q)$-cohesive.} 
\end{enumerate} 
\end{corollary} 
\begin{proof}
Statement (1) follows from statement (1) of \Cref{jackson} and statement (1) of \Cref{combined}. Statement (2) follows from statement (1) of \Cref{jackson} and statement (5) of \Cref{combined}. 
\end{proof}

Statement (2) in \Cref{corr:alg1} shows that Algorithm 1 can be used to sets $E$ that are $q$-cohesive and have a $(1-q)$-cohesive complement. Indeed, to find such a set $E$, it is sufficient to run Algorithm 1 with any nonempty set $S$ of players with incentive to play $1$ at $q$. The output of Algorithm 1 will be the smallest such $E$ that contains $S$. 

Algorithm 1 shows that if we start from an arbitrary set $S$ of players playing $1$ and an arbitrary threshold $q$, then action $1$ proliferates to the set $C(S,q)$ of players in $\mathcal{I}$. It is also of interest to know when action $1$ proliferates to the entire network. Algorithm 2 answers this question. 

\vspace{0.1in} 
\noindent 
\textbf{Algorithm 2 (Full network contagion starting from $S$)} 
\begin{enumerate} 

\item Let $S$ be a set of players with incentive to play $1$ at $q=1$. 

\item Define $A_0 = S$ and $q_0 = 1$ (and therefore, $b_0 =0$). 

\item For $n \geq 0$, define $(A_{n+1}, q_{n+1})$ recursively as follows:
\begin{enumerate}
    \item Apply Algorithm 1 with $(A_n,q_n)$ to determine $A_{n+1} = C(A_n, q_n)$.
    \item If $A_{n+1} \neq \mathcal{I}$, let $q_{n+1} = \max_{i \notin A_{n+1}} \frac{c s_i(A_{n+1})}{c w_i - \phi_i(p_i(A_{n+1}))}$. Let $b_{n+1} = \frac{c(1 - q_{n+1})}{q_{n+1}}$.
\end{enumerate}

\item Let $N$ be the first $n$ such that $A_{n+1} = \mathcal{I}$. Let $q^* = q_N$.
\end{enumerate}

In any network coordination game with local and global effects, Algorithm 2 starts with an initial set $S$ of players who have an incentive to play 1 at $q=1$. This is the case when $S = \mathcal{D}$ (or $S \subseteq \mathcal{D}$). Algorithm 2 recursively searches for the next lower $q$ that allows the contagion set resulting from Algorithm 1 to keep expanding. To motivate statement 3(b), notice that $\frac{s_i}{w_i} + \frac{\phi_i(p_i)}{w_i(b + c)} = \frac{s_i}{w_i} + \frac{\phi_i(p_i)}{cw_i }q$, and therefore, the threshold to switch $\frac{s_i}{w_i} + \frac{\phi_i(p_i)}{w_i(b + c)} = q$ can be written as $\frac{c s_i}{c w_i - \phi_i(p_i)} = q$. Moreover, $i \notin C(A_n, q_n)$ implies $\phi_{i}(p_{i}) < cw_{i}$ so that the denominator is positive.

The algorithm starts at the top, with $q_0 =1$, and applies Algorithm 1 to determine $C(S, 1)$. If contagion has not spread to the entire network with $C(S, 1)$, the algorithm looks at the threshold computations of the remaining players and selects one with the highest threshold and sets $q_1$ to be this threshold. This $q_1$ is the threshold closest to $1$ that is \textit{just enough} to give this ``marginal'' player an incentive to play $1$. Algorithm 1 is run again to see the full ramification of lowering the threshold to $q_1$. If contagion has not yet spread to the entire network, we follow the same process to lower $q$ just enough to switch the next marginal player and apply Algorithm 1 again. Algorithm 2 stops at the first time when contagion spreads to the entire network. As the number of players is finite, the algorithm stops in finitely many iterations (there are never more than the number of players in $\mathcal{I}\backslash S$, and there may be considerably fewer depending on spread of contagion in Algorithm 1 at each step). The resulting $q^*$ is the contagion threshold for the network with starting set $S$. Our simulations provide information about the prevalence and size of the starting set of exogenously infected players (and other parameters) and their effect on contagion. These are discussed in more detail below. 
Theoretical properties of Algorithm 2 are as follows.

\begin{theorem}[Properties of Algorithm 2] \label{algtheorem} 
Let $\Gamma$ be a network coordination game with local and global effects and $S$ be as in Algorithm 2.
\leavevmode 
\begin{enumerate} 
\item $(A_n)_{n=1}^{N+1}$ is a strictly increasing sequence of Nash equilibria, starting at $C(S, 1)$ and ending at $\mathcal{I}$, and for every $n \in \{ 1,\ldots, N+1 \}$, $A_n$ is a Nash equilibrium at $q_{n-1}$. 

\item $(q_n)_{n=0}^{N}$ is a strictly decreasing sequence in $[0,1]$, starting at $1$ and ending at $q^*$. 
\item For every $n \in \{0,\ldots,N\}$, $A_{n+1} = C(S,q_n)$, and therefore, $C(S,q^*) = \mathcal{I}$. 
\item For every $q \in [0,1]$, $C(S,q) = \mathcal{I}$, if, and only if, $q \leq q^*$.  
\end{enumerate} 
\end{theorem}

\begin{proof} 
For statement (1), for every $n=1,\ldots, N+1$, $A_n$ is a Nash equilibrium at $q_{n-1}$ follows from statement (2) of \Cref{combined}, as does $A_1=C(S, 1)$, and $A_{N+1} = \mathcal{I}$ follows from Algorithm 2. The sequence is (weakly) increasing follows from the fact that each term in the sequence is the output of Algorithm 1 applied to the immediately preceding term in the sequence. To see that it is strictly increasing, fix $n \in \{1,\ldots,N\}$ and notice that $A_n \neq \mathcal{I}$. Let $i_0 \notin A_n$ be such that $q_{n} = \frac{c s_{i_0}}{c w_{i_0} - \phi_{i_0}(p_{i_0})} = \max_{i \notin A_n} \frac{c s_i}{c w_i - \phi_i(p_i)}$. Notice that $i \notin A_n$ implies $\phi_{i}(p_{i}) < cw_{i}$ so that the denominator is positive. Rearranging terms yields $\frac{s_{i_0}}{w_{i_0}} + \frac{\phi_{i_0}(p_{i_0})}{w_{i_0}(b_{n}+c)} = q_{n}$, and therefore, $i_0 \in A_{n+1}$.

To prove statement (2), notice that for each $n \in \{0,\ldots, N-1\}$, $C(A_n, q_n) = A_{n+1} \neq \mathcal{I}$. Therefore, for every $i\not\in C(A_n, q_n)$, it must be that $\frac{s_i}{w_i} + \frac{\phi_i(p_i)}{w_i(b_{n} + c)} < q_n$, or equivalently, $\frac{c s_i}{c w_i - \phi_i(p_i)} < q_n$, whence $q_{n+1} = \max_{i \notin C(A_n, q_n)} \frac{c s_i}{c w_i - \phi_i(p_i)} < q_n$.  

Statement (3) is proved using induction on $n$. We know that $A_1 = C(S,q_0)$. Suppose $A_n = C(S, q_{n-1})$. Using statement (6) of \Cref{combined}, it follows that $A_{n+1} = C(A_n, q_{n}) = C(S, q_{n})$.

To prove statement (4), in one direction, suppose $q \leq q^*$. Then $\mathcal{I} = C(S,q^*) \subseteq C(S,q) \subseteq \mathcal{I}$, where the equality follows from statement (3) and the first inclusion follows from statement (4) of \Cref{combined}. In the other direction, to show the contrapositive, suppose $q^* < q$. Then $\exists n \in \{1, \ldots, N\}$ such that $q \in (q_n,q_{n-1}]$. We show that $C(S,q)=C(S,q_{n-1})$, as follows. In one direction, $q \leq q_{n-1} \Rightarrow C(S,q_{n-1}) \subseteq C(S,q)$. In the other direction, using induction on $t$, we show that for every $t$, $C_t(S,q) \subseteq C(S,q_{n-1})$, whence $C(S,q) \subseteq C(S,q_{n-1})$. For $t=0$, $C_0(S,q) = S \subseteq C(S,q_{n-1})$. Suppose $C_t(S,q) \subseteq C(S,q_{n-1})$. Fix arbitrary $i_0 \in C_{t+1}(S,q) = C_t(S,q) \cup F_{t+1}(S,q)$. Then $i_0 \in C_{t}(S,q) \Rightarrow C(S,q_{n-1})$ by the inductive hypothesis. Suppose $i_0 \in F_{t+1}(S,q)$. Then $i_0 \not\in C_t(S,q)$ and $\frac{c s_{i_0}(C_t(S,q))}{c w_{i_0} - \phi_{i_0}(p_{i_0}(C_t(S,q)))} \geq q$. If $i_0 \not\in C(S,q_{n-1})$, then 
\[ q \leq \frac{c s_{i_0}(C_t(S,q))}{c w_{i_0} - \phi_{i_0}(p_{i_0}(C_t(S,q)))} \leq \frac{c s_{i_0}(C(S,q_{n-1}))}{c w_{i_0} - \phi_{i_0}(p_{i_0}(C(S,q_{n-1})))} \leq \max_{i \notin C(S, q_{n-1})} \frac{c s_i(C(S, q_{n-1}))}{c w_i - \phi_i(p_i(C(S, q_{n-1})))} = q_n, \] 
where the second inequality follows from the inductive hypothesis. This contradicts $q \in (q_n, q_{n-1}]$. Therefore, $i_0 \in C(S,q_{n-1})$, whence $C_{t+1}(S,q) \subseteq C(S,q_{n-1})$, as desired. This shows that $C(S,q) = C(S,q_{n-1}) \neq \mathcal{I}$, as desired.  
\end{proof} 

Statements (1) and (2) of \Cref{algtheorem} show that Algorithm 2 outputs a strictly increasing sequence of Nash equilibria that starts from $C(S,1)$ and expands to the entire network. At each step, the expansion at the end of that step is the smallest Nash equilibrium extension of the beginning set for that step. Moreover, at the end of each step, Algorithm 2 determines the smallest decrease in $q$ that is needed to continue further proliferation of action $1$ in the network. In this sense, for an arbitrary starting set $S$, Algorithm 2 provides an expanding equilibrium-by-equilibrium proliferation of action $1$ to the entire network, using a minimal decrease in $q$ to expand contagion from one equilibrium to the next. 

Statement (3) equates the equilibrium-by-equilibrium expansion up to step $n+1$ with the one step contagion from initial $S$ using threshold $q_n$, using the bootstrapping feature of Algorithm 1 listed in statement (6) of \Cref{combined}. One direction of statement (4) is immediately implied by statement (3), that is,  for every threshold $q \le q^*$, contagion occurs from $S$ to $\mathcal{I}$ at $q$. The design of the algorithm helps to prove the other direction in statement (4) as well.  

\begin{figure} 
    \centering
    \begin{subfigure}[b]{0.4\textwidth}
    \centering
    \includegraphics[scale=0.1]{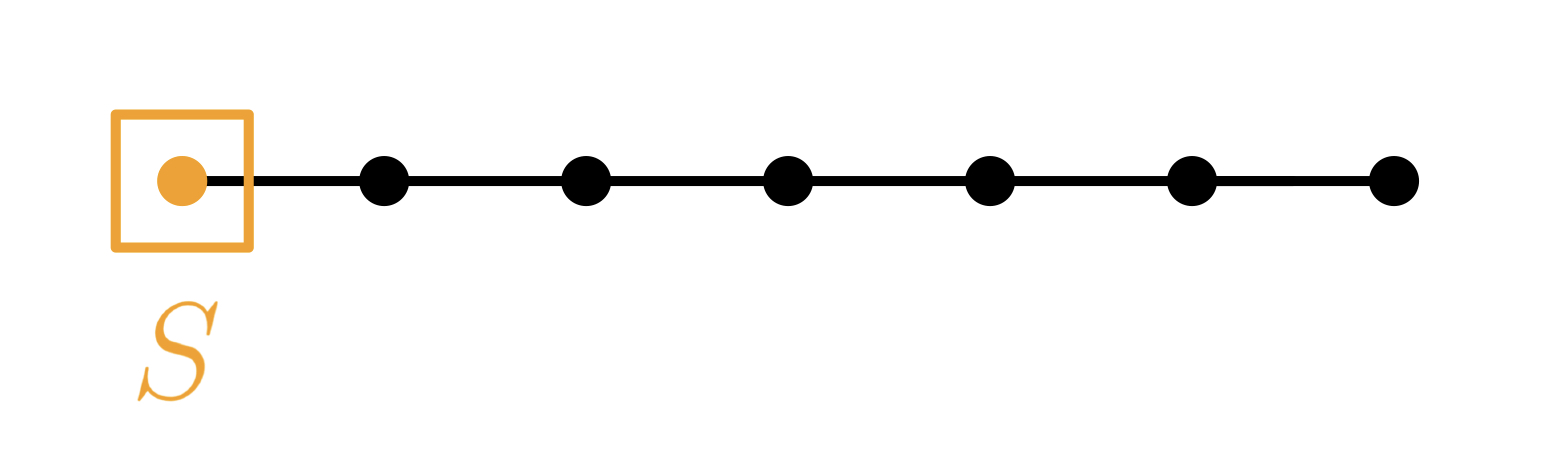}
    \caption{Example 1 ($\alpha = 0$). Algorithm 3 checks 6 subsets; $q^*=0.50$.}
    \end{subfigure}
    \hfil
    \begin{subfigure}[b]{0.4\textwidth}
    \centering
    \includegraphics[scale=0.1]{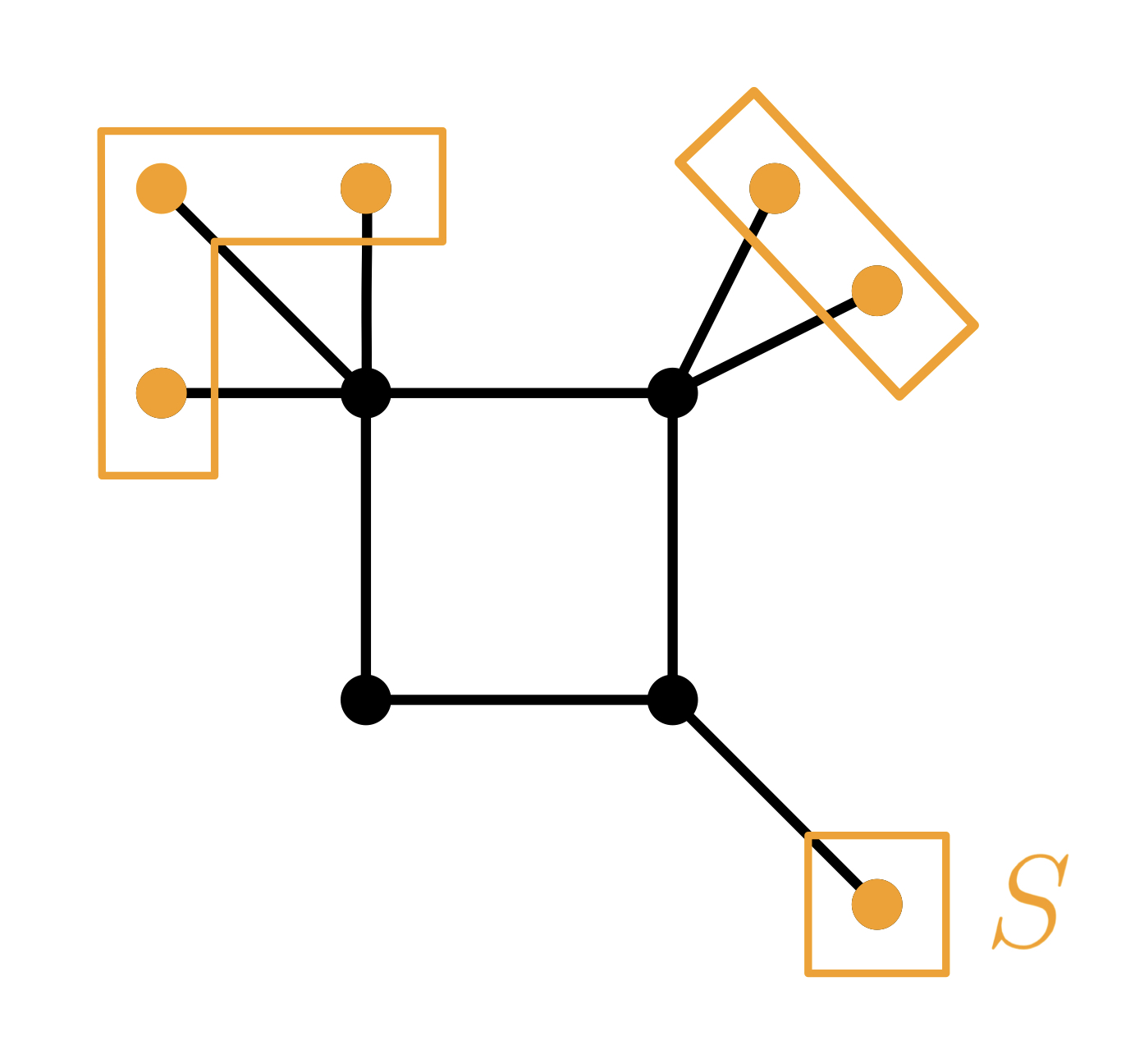}
    \caption{Example 2 ($\alpha = 0.25$). Algorithm 3 checks 4 subsets;  $q^*=0.7875$.}
    \end{subfigure}
    \caption{Two networks where upper bound of Algorithm 3 is attained. Orange nodes are exogenously infected to play 1, black nodes initially play zero. }
    \label{fig:algbounds}
\end{figure}

An important consequence of our construction of Algorithm 2 with nested Algorithm 1 is that the total number of nonempty subsets of the complement of $S$ that are checked is no more than $\vert \mathcal{I}\backslash S \vert$. This can be seen in Algorithm 3 below and in \Cref{alg3theorem}. Examples in \Cref{fig:algbounds} show that this bound cannot be lowered in general. Pseudocode for the combined algorithm is provided to aid implementation.

\vspace{0.1in} 
\noindent 
\textbf{Algorithm 3 (Algorithm 2 with nested Algorithm 1)} 
\begin{enumerate} 

\item Let $S$ be a set of players with incentive to play $1$ at $q=1$. 

\item Let $A_0=S \cup \mathcal{D}$ and $q_0=1$. 

\item For $n \geq 0$, define $(A_{n+1}, q_{n+1})$ recursively as follows:
\begin{enumerate}
    \item For $t=0$, let $C_{n,t} = A_n$ and $B_{n,t} = \mathcal{I}\backslash C_{n,t}$
    \item For $t \ge 0$, recursively define \\
        $F_{n, t+1} = \left\{i \in B_{n, t} \mid \frac{s_i(C_{n,t})}{w_i} + \frac{\phi_i(p_i(C_{n,t}))}{w_i(b+c)} \geq q \right\}$,\\ 
        $C_{n,t+1} = C_{n,t} \cup F_{n,t+1}$, and\\ 
        $B_{n,t+1} = \mathcal{I}\backslash C_{n,t+1}$. 
    \item Let $T(n)$ be first $t$ such that $C_{n,t+1} = C_{n,t}$. 
    \item Let $A_{n+1} = C_{n,T(n)}$. Let $q_{n+1} = \left\{ \begin{array}{lll} 
            \max_{i \notin A_{n+1}} \frac{c s_i(A_{n+1})}{c w_i - \phi_i(p_i(A_{n+1}))} &\mbox{if} &C_{n,T(n)} \neq \mathcal{I} \\ 
            q_n &\mbox{if} &C_{n,T(n)} = \mathcal{I}. 
            \end{array} \right. $ 
    \end{enumerate} 
\item Let $N$ be first $n$ such that $q_{n+1}=q_n$. Let $q^* = q_N$. 
\end{enumerate} 

Algorithm 3 generates a sequence of sets $\{ ( C_{n,t} )_{t=0}^{T(n)} \}_{n=0}^N$ starting from $C_{0,0} = A_0 = S \cup \mathcal{D}$ and ending at $\mathcal{I}$. 
There are redundant sets in this sequence at transition points between different runs, given by $C_{n, T(n)} = C_{n+1, 0}$. Removing these sets and including the starting set, let $\{ S, ( C_{n,t} )_{t=1}^{T(n)} \}_{n=0}^N$ be the sequence of sets given by $S, C_{0, 1}, \ldots, C_{0, T(0)}, C_{1, 1}, \ldots, C_{1, T(1)} , C_{2, 1}, \ldots, C_{N-1, T(N-1)}, C_{N, 1}, \ldots, C_{N, T(N)}.$ 

\begin{theorem}[Properties of Algorithm 3] \label{alg3theorem} 
Let $\Gamma$ be a network coordination game with local and global effects and $S$ be as in Algorithm 3.

\begin{enumerate} 
\item For $n=0$, $(S, C_{0,t})_{t=1}^{T(0)}$ is a strictly increasing sequence of sets and for $n>0$, $(C_{n,t})_{t=0}^{T(n)}$ is a strictly increasing sequence of sets. 
\item Sequence $\{ S, ( C_{n,t} )_{t=1}^{T(n)} \}_{n=0}^N$ is a strictly increasing sequence of sets starting at $S$ and ending at $\mathcal{I}$. 
\item Subsequence $\{ ( C_{n,T(n)} ) \}_{n=0}^N$ is a strictly increasing sequence of Nash equilibria starting at $C(S,1)$ and ending at $\mathcal{I}$, and for every $n \in \{ 0,\ldots, N-1 \}$, $C_{n, T(n)}$ is a Nash equilibrium at every $q \in (q_{n+1}, q_n]$, and $C_{N, T(N)} = \mathcal{I}$ is a Nash equilibrium at every $q \in (0, q_N]$.
\item The total number of sets in the sequence $\{ S, ( C_{n,t} )_{t=1}^{T(n)} \}_{n=0}^N$ is never more than $\vert \mathcal{I} \backslash S \vert + 1$. 
\item The total number of nonempty subsets of the complement of $S$ checked in Algorithm 3 is never more than $\vert \mathcal{I} \backslash S \vert$. 
\end{enumerate} 
\end{theorem}

\begin{proof} 
For statement (1), for $n=0$, $(S, C_{0,t})_{t=0}^{T(0)}$ is a strictly increasing sequence of sets generated by the first iteration of Algorithm 1 with starting set $S$ and $q=1$. For $n>0$, $(C_{n,t})_{t=0}^{T(n)}$ is a strictly increasing sequence of sets generated by the $(n+1)$-th iteration of Algorithm 1 with starting set $A_{n}$ and $q=q_{n}$. 
For statement (2), notice that the sequence $\{ S, ( C_{n,t} )_{t=1}^{T(n)} \}_{n=0}^N$ removes the redundant sets $C_{n, T(n)} = C_{n+1, 0}$ in the collection of all sets $C_{n,t}$ generated by Algorithm 3. Statement (2) now follows from statement (1) above and statement (3) of \Cref{algtheorem}, which uses the bootstrapping property of Algorithm 1. 
For statement (3), $\{ ( C_{n,T(n)} ) \}_{n=0}^N$ is a strictly increasing sequence of Nash equilibria starting at $C(S,1)$ and ending at $\mathcal{I}$ follows from statement (1) in \Cref{algtheorem}, and the remaining statement follows from statement (4) in \Cref{algtheorem} and its proof. 
Statement (4) follows from statement (2) by noting that the bijective sequence of complements given by $\{ \mathcal{I} \backslash S, ( B_{n,t} )_{t=1}^{T(n)} \}_{n=0}^N$ is strictly decreasing, starts from $\mathcal{I}\backslash S$, and ends at the empty set, and therefore, cannot have more than $\vert \mathcal{I} \backslash S \vert + 1$ terms. Statement (5) follows from statement (4), because the last term in the sequence $\{ \mathcal{I} \backslash S, ( B_{n,t} )_{t=1}^{T(n)} \}_{n=0}^N$ is the empty set. 
\end{proof} 

Statement (5) of \Cref{alg3theorem} shows that in our formulation, an upper bound to the number of subsets checked to compute $q^*$ is the number of players in $\mathcal{I} \backslash S$ as compared to the number of subsets of $\mathcal{I} \backslash S$, an exponential improvement. As shown in \Cref{fig:algbounds}, this bound cannot be lowered in general. Subfigure (a) is an example with no global effects ($\alpha=0$) with $7$ players and starting set size of $1$. The total number of nonempty subsets checked by Algorithm 3 is 6, which is the number of players in $\mathcal{I} \backslash S$, yielding $q^* = 0.50$. Subfigure (b) is an example with global effects ($\alpha = 0.25$) with $10$ players and starting set size of $6$ players. The total number of subsets checked by Algorithm 3 is 4, which is the number of players in $\mathcal{I} \backslash S$, yielding $q^* = 0.7875$.

\setcounter{algorithm}{2}
\begin{algorithm}[htb]
    \caption{Pseudocode for Algorithm 2 with nested Algorithm 1}
    \label{alg:thresh}
\begin{algorithmic}[1] 
        \State Generate network $\mathcal{I}$
        \State Define starting set $A$
        \State Create dictionary Depth
        \Function{prop}{$i, S$}
            \State $H \gets \mathcal{I} \backslash (N_i \cup \{i\})$
            \State $p$ $\gets 0$
            \State $n$ $\gets \lvert H\rvert$
            \For{$j \in H$}:
                \If{$j \in S$}:
                    \State $p \gets p +1$
                \EndIf
            \EndFor
            \State return $\frac{p}{n}$
        \EndFunction
        \State $q \gets 1$
        \State $b \gets 0$ 
        \State $c \gets 1$
        \While{$\lvert A\rvert < \lvert \mathcal{I}\rvert$}: 
            \State $F \gets \emptyset$
            \State $B \gets \mathcal{I} \backslash A$
            \State $B' \gets \emptyset$
            \While{$B \neq B'$}: 
                \For{$i$ in $B$}:
                    \State $p_i \gets $ \Call{prop}{$i, A$}
                    \If{$\frac{s_i}{w_i} + \frac{\phi_i(p_i)}{w_i(b+c)} \geq q$}
                        \State $F \gets F \cup \{i\}$
                    \EndIf
                \EndFor
                \State $A \gets A \cup F$
                \State $B' \gets B$
                \State $B \gets \mathcal{I} \backslash A$
            \EndWhile
            \If{$F = \emptyset$}
                \If{$A = \mathcal{I}$}    
                    \State Depth($q$) $\gets$ 1 
                    \State return $q$
                \Else 
                    \State Depth($q$) $\gets \frac{\lvert A\rvert }{\lvert \mathcal{I}\rvert}$
                    \For{$i \in B$}:
                        \State $p_i \gets $ \Call{prop}{$i, A$}
                    \EndFor
                    \State $q \gets \max_{i \in B} \frac{c s_i}{c w_i - \phi_i(p_i)}$
                    \State $b \gets \frac{c(1-q)}{q}$
                \EndIf
            \EndIf
        \EndWhile
    \end{algorithmic}
\end{algorithm}

Statement (4) of \Cref{algtheorem} proves that a special characteristic of the $q^*$ computed by Algorithm 2 is that it is the largest $q$ for which $C(S,q) = \mathcal{I}$. Following \cite{morris2000}, we term this the contagion threshold. Let $\Gamma$ be a network coordination game with local and global effects and consider a set $S$ of players with incentive to play $1$. The \textit{\textbf{contagion threshold for $S$}} is  $q^* = \sup \{ q \in [0,1] \mid C(S, q) = \mathcal{I} \}$. As $0$ is trivially in the set over which the supremum is taken, the supremum is well-defined. Algorithm 2 shows that the supremum is in the set itself. 

\begin{corollary}\label{corr:alg2-threshold}
Let $\Gamma$ be a network coordination game with local and global effects and $S$ be as in Algorithm 2.
\begin{enumerate}
    \item The value $q^*$ from Algorithm 2 is the contagion threshold for $S$. 
    \item The set of all $q$ for which $C(S,q) = \mathcal{I}$ is the interval $[0,q^*]$. 
\end{enumerate}
\end{corollary} 

\begin{proof} Both statements follow from statement (4) of \Cref{algtheorem}. \end{proof}

The interval $[0,q^*]$ gives a full characterization of when contagion occurs from $S$ to the entire network. In other words, for a given starting set $S$, the most resilient network that can be fully infected is the one with relative cost of miscoordination given by $q^*$. Networks with resilience less than or equal to $q^*$ will be fully infected while those with higher resilience will not be fully infected. 
Algorithms 1 and 2 provide computable solutions to the problem of determining when different conventions (in the sense that some players play $0$ and others play $1$) coexist in equilibrium. A subset \textbf{\textit{$E \subseteq \mathcal{I}$ is an equilibrium with coexisting conventions at $q$}}, if $E$ is a Nash equilibrium at $q$, $E\neq \emptyset$, and $E \neq \mathcal{I}$. 

\begin{corollary}\label{corr:conventions}
Let $\Gamma$ be a network coordination game with local and global effects.\\ 
For every nonempty set $S$ of players with incentive to play $1$ at $q$, 
\begin{enumerate}
    \item $S$ can be extended to an equilibrium with coexisting conventions at $q$, if, and only if, $C(S,q) \neq \mathcal{I}$. In this case, $C(S,q)$ is the smallest equilibrium with coexisting conventions at $q$ that contains $S$. 
    \item $S$ can be extended to an equilibrium with coexisting conventions at $q$, if, and only if, $q > q^*$. 
\end{enumerate}
\end{corollary} 

\begin{proof} For statement (1), if $S$ can be extended to an equilibrium $E$ with coexisting conventions at $q$, then \Cref{combined} implies that $C(S,q) \subseteq E \neq \mathcal{I}$. In the other direction, $\emptyset \neq S \subseteq C(S,q) \neq \mathcal{I}$ implies that $C(S,q)$ is an extension of $S$ to an equilibrium with coexisting conventions at $q$. $C(S,q)$ is the smallest equilibrium with coexisting conventions at $q$ that contains $S$ follows from \Cref{combined} as well. Statement (2) follows from statement (1) combined with statement (4) of \Cref{algtheorem}. \end{proof}

\Cref{corr:conventions} shows that Algorithms 1 and 2 are convenient tools to find equilibria with coexisting conventions. Starting with any nonempty set $S$ of players with incentive to play $1$ at $q$, we just need to check that the ending set in Algorithm 1 is not the full network, that is, $C(S,q) \neq \mathcal{I}$. This is made easier if we have $q^*$ from Algorithm 2, because $q > q^*$ is an easy and equivalent condition for $C(S,q) \neq \mathcal{I}$. 

In addition to determining the threshold $q^*$ guaranteeing contagion to the entire network, Algorithm 2 can be used to compute how far contagion progresses if it does not proliferate to the entire network. This is helpful to understand proliferation of contagion to a fraction of the network. As mentioned in the introduction, the objective of the promoters of an alternative narrative may be achieved with partial proliferation of the narrative in the network. For example, convincing a fraction of the population to distrust election results may be sufficient to undermine election integrity. Convincing a fraction of the population to doubt climate change or the role of guns in civil society or the efficacy of vaccines may be sufficient to delay or stop progress that may be beneficial for society. It is, therefore, useful to understand not only full contagion, but also partial contagion to a fraction of the network.

Algorithm 2 yields a decreasing sequence of thresholds $q_n$ and a corresponding increasing sequence of contagion sets $C(S, q_n)$ that shows how far action $1$ can spread at $q_n$. Combined with the fact that for each $q \in (q_n, q_{n-1}]$, $C(S, q) = C(S, q_{n-1})$ (as shown in proof of statement (4) of \Cref{algtheorem}), we define the \textbf{\textit{contagion depth function from $S$}} as follows: 
\[\delta(S, q) = \begin{cases} 1 &\text{ if } q \leq q^*, \ \text{and} \\ 
\frac{\lvert C(S, q_n)\rvert}{\lvert \mathcal{I}\rvert} &\text{ if } q_{n+1} < q \leq q_n, \ \text{for} \ n \in \{0,\ldots,N-1\}. 
\end{cases}\]
For a fixed starting set $S$, $\delta$ measures the proportion of the network that is infected as a function of network resilience parameter $q \in [0,1]$. For a fixed $S$, it is a step function that starts at $1$ for $q \leq q^*$ (low network resilience leads to full contagion) and strictly decreases in a step-wise manner to $\lvert C(S, q_n)\rvert/\lvert \mathcal{I} \rvert$ over each interval $(q_{n+1}, q_{n}]$ as network resilience increases. The contagion depth function is an equilibrium notion. For each $S$ and $q$, it provides the fraction of the network infected by the smallest Nash equilibrium that contains $S$.  

The contagion depth function allows us to analyze scenarios in which an action may spread to a significant proportion of a network but not the entire network. These scenarios are important in light of the aforementioned cases of partial contagion causing significant societal disruption. As it is important to understand both full contagion and partial contagion to some proportion of a network, we will show in the next section how to invert this relation to understand what starting set size is needed to reach a given depth of contagion for a given $q$. 

Special cases of properties of Algorithm 2 in \Cref{algtheorem} can be used to prove new characterizations of earlier results for network coordination games with local network effects and uniform unit weights.

\begin{corollary}\label{corr:alg2}
Let $\Gamma$ be a network coordination game with local effects only and uniform unit weights on neighbors and $S$ and $q^*$ be as in Algorithm 2.\\ 
For every $q \in [0,1]$, 
the complement of $S$ is uniformly no more than $(1-q)$-cohesive, if, and only if, $q \le q^*$.
\end{corollary} 
\begin{proof}
Follows immediately from statement (2) of \Cref{jackson} and statement (4) of \Cref{algtheorem}. 
\end{proof}

For network coordination games with local effects only and uniform unit weights, this corollary shows how Algorithm 2 helps to compute when the complement of $S$ is uniformly no more than $(1-q)$-cohesive even for significantly large networks which were not amenable to such analysis earlier.

\begin{figure}[!htb]
    \centering
    \begin{subfigure}{\textwidth}
    \begin{subfigure}{0.24\textwidth}
    \includegraphics[width=\linewidth]{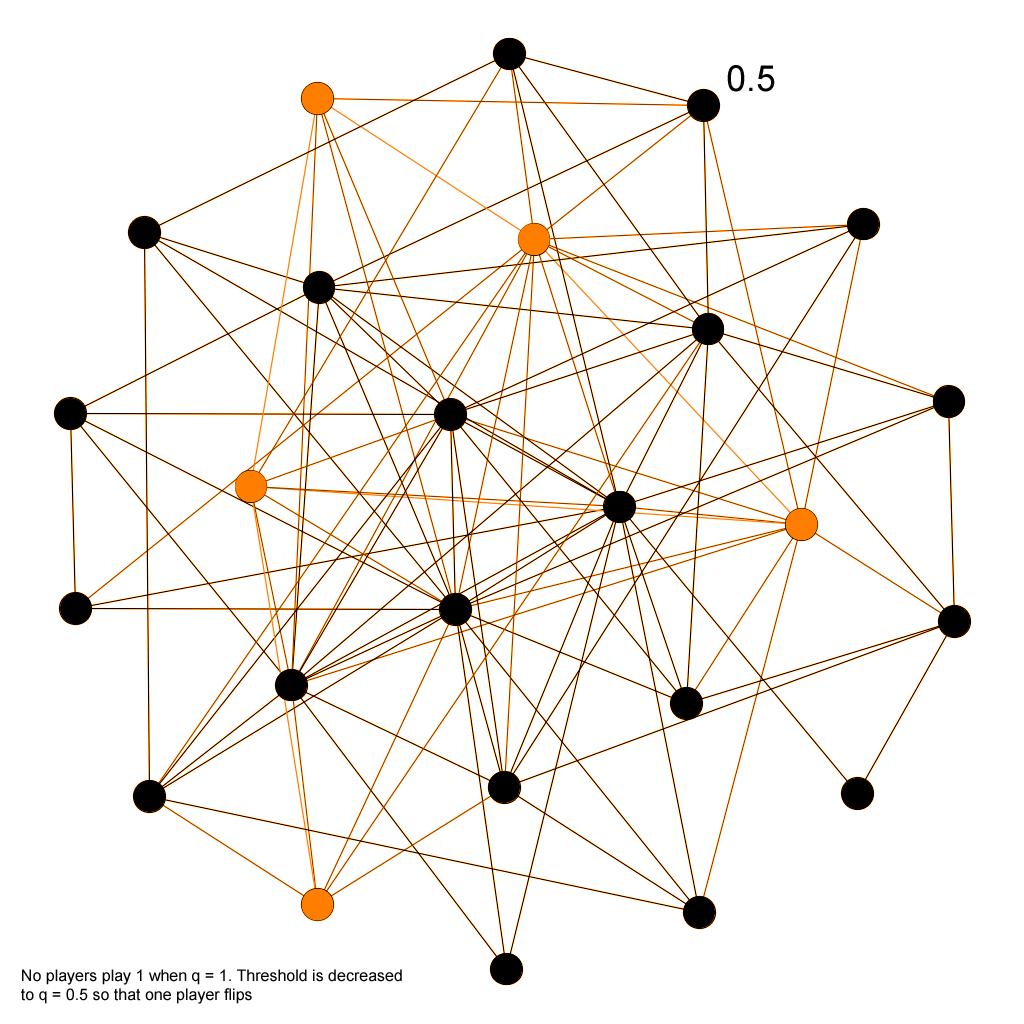}
    \end{subfigure}
    \begin{subfigure}{0.24\textwidth}
    \includegraphics[width=\linewidth]{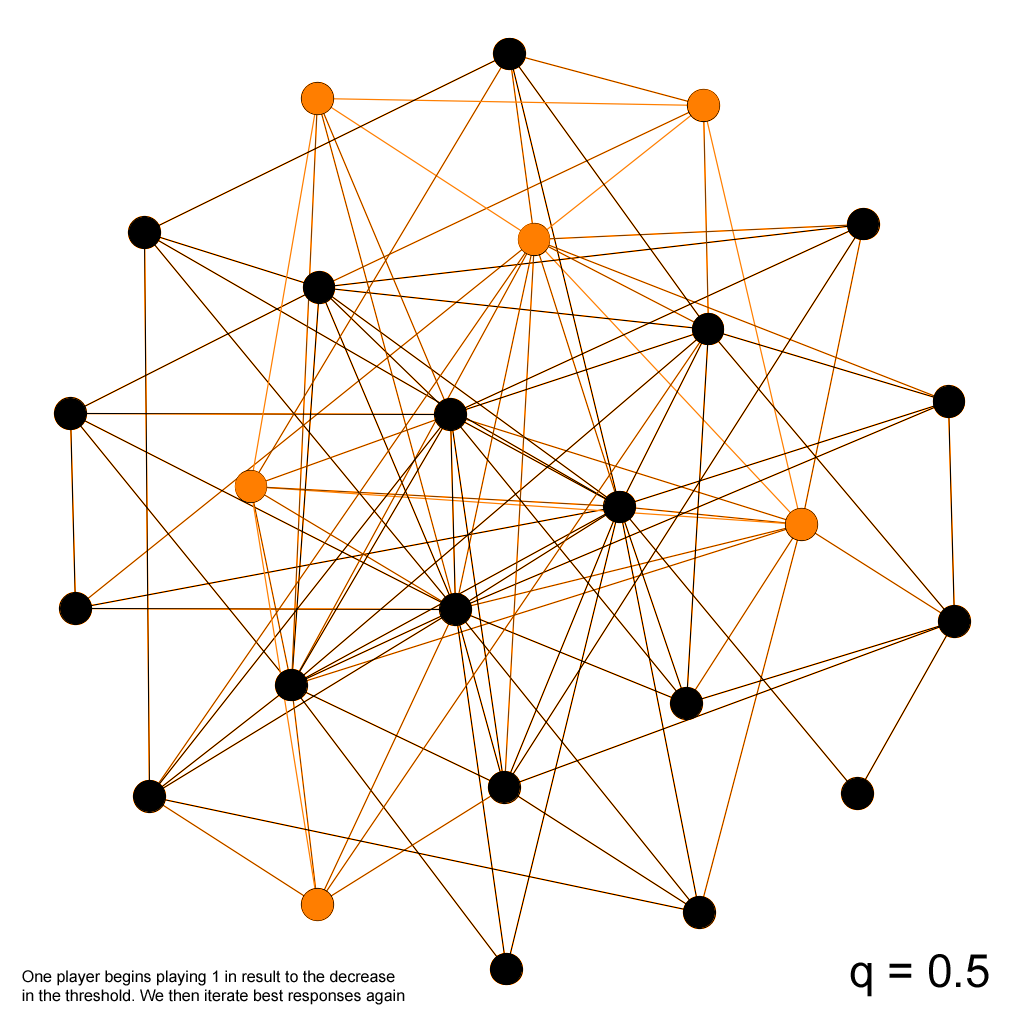}
    \end{subfigure}
    \begin{subfigure}{0.24\textwidth}
    \includegraphics[width=\linewidth]{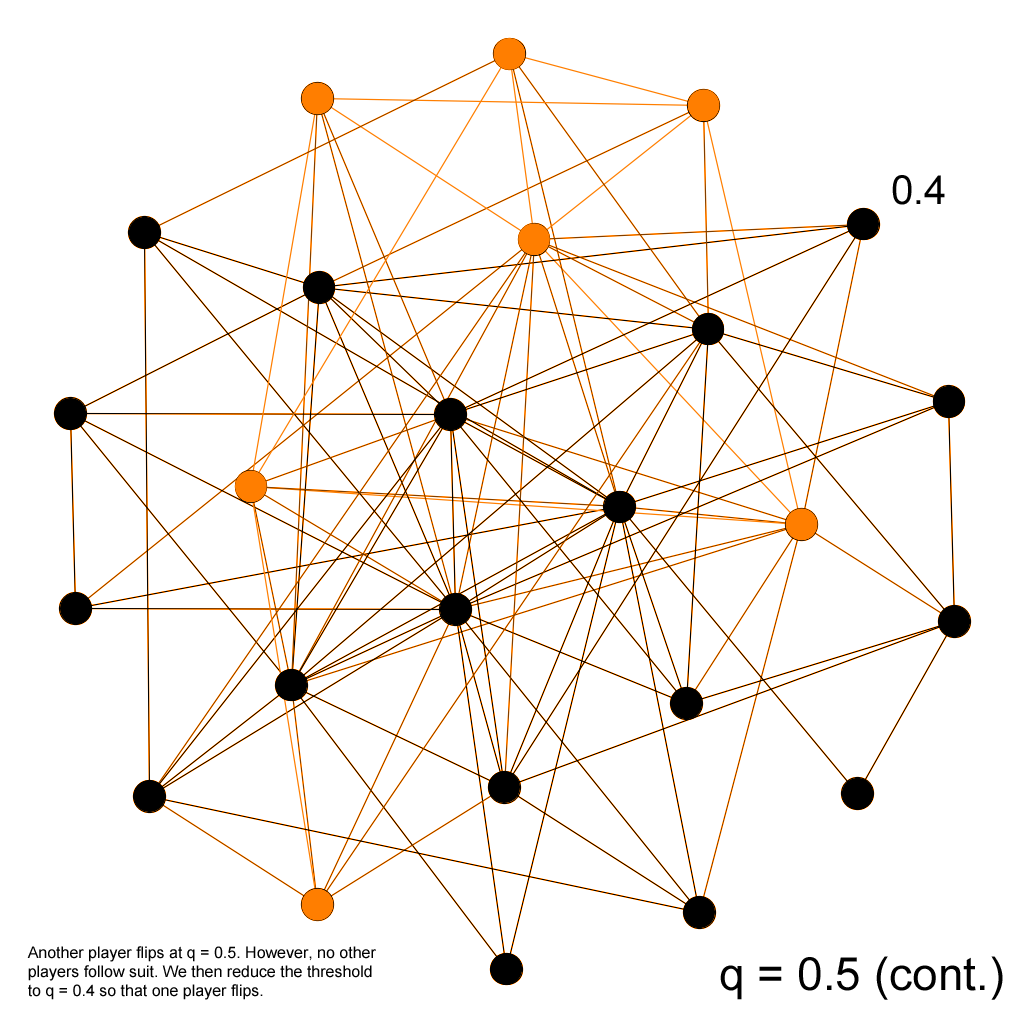}
    \end{subfigure}
    \begin{subfigure}{0.24\textwidth}
    \includegraphics[width=\linewidth]{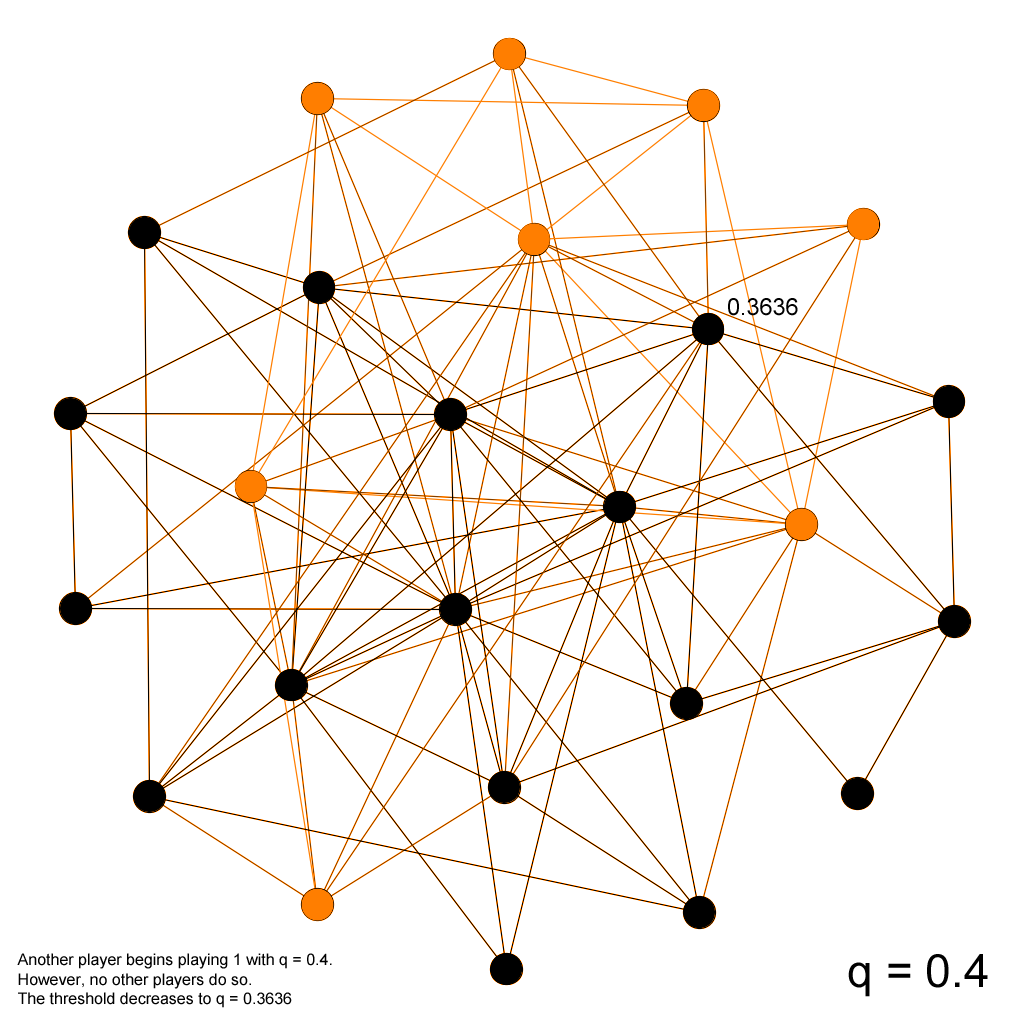}
    \end{subfigure}
    \medskip
    \begin{subfigure}{0.24\textwidth}
    \includegraphics[width=\linewidth]{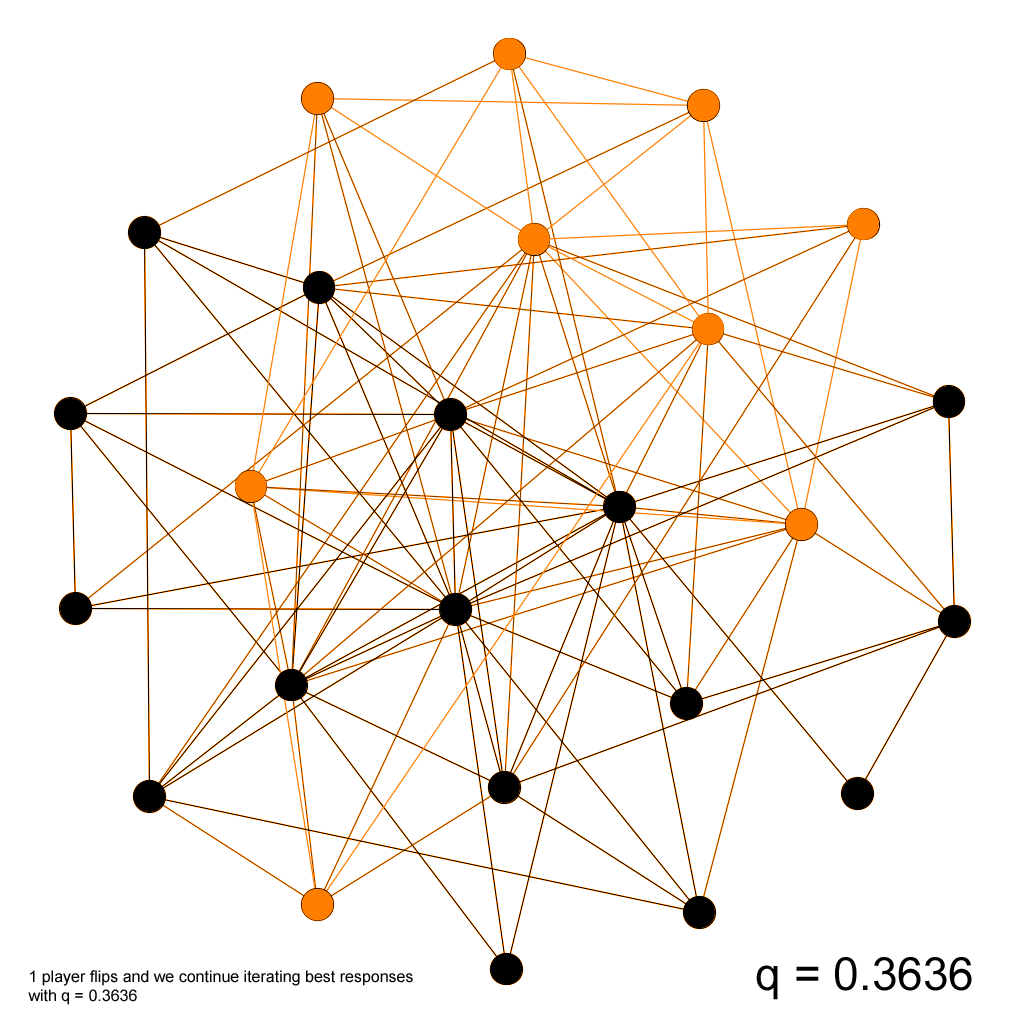}
    \end{subfigure}
    \begin{subfigure}{0.24\textwidth}
    \includegraphics[width=\linewidth]{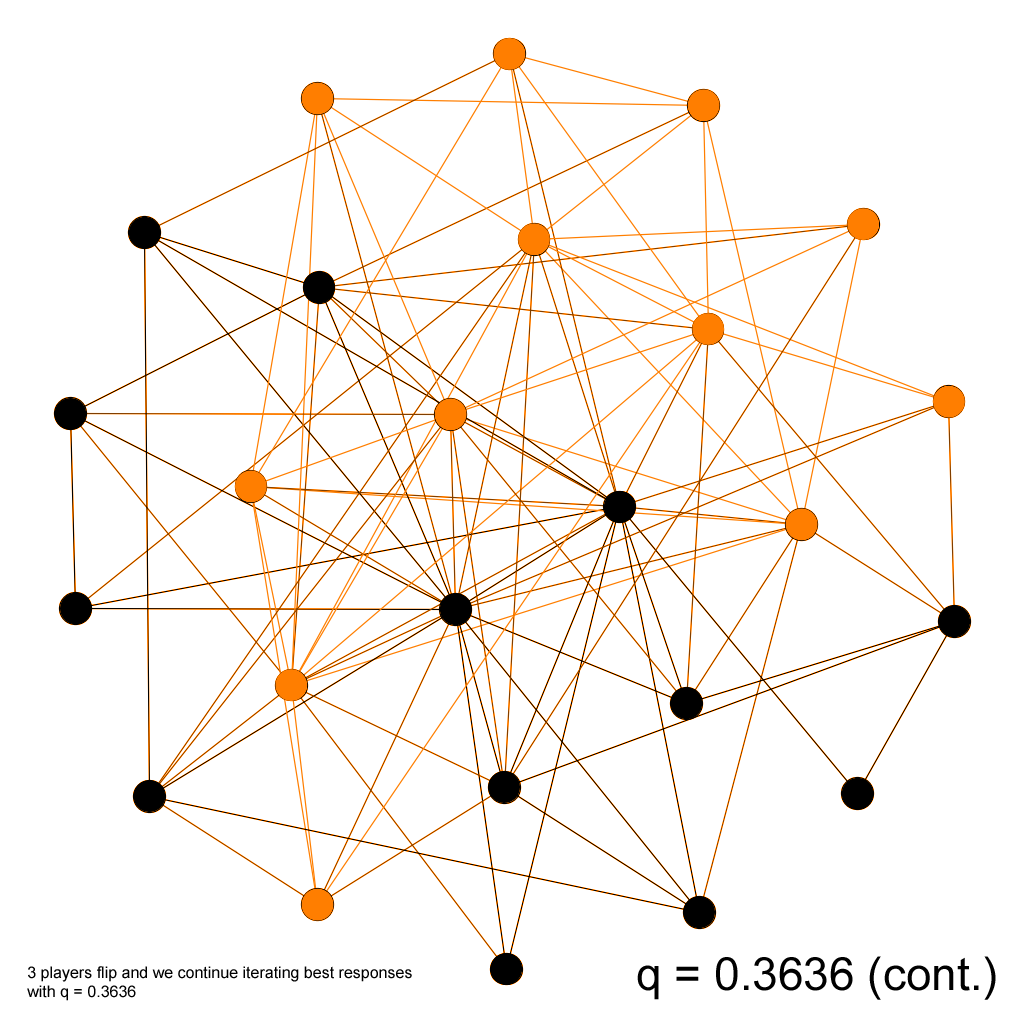}
    \end{subfigure}
    \begin{subfigure}{0.24\textwidth}
    \includegraphics[width=\linewidth]{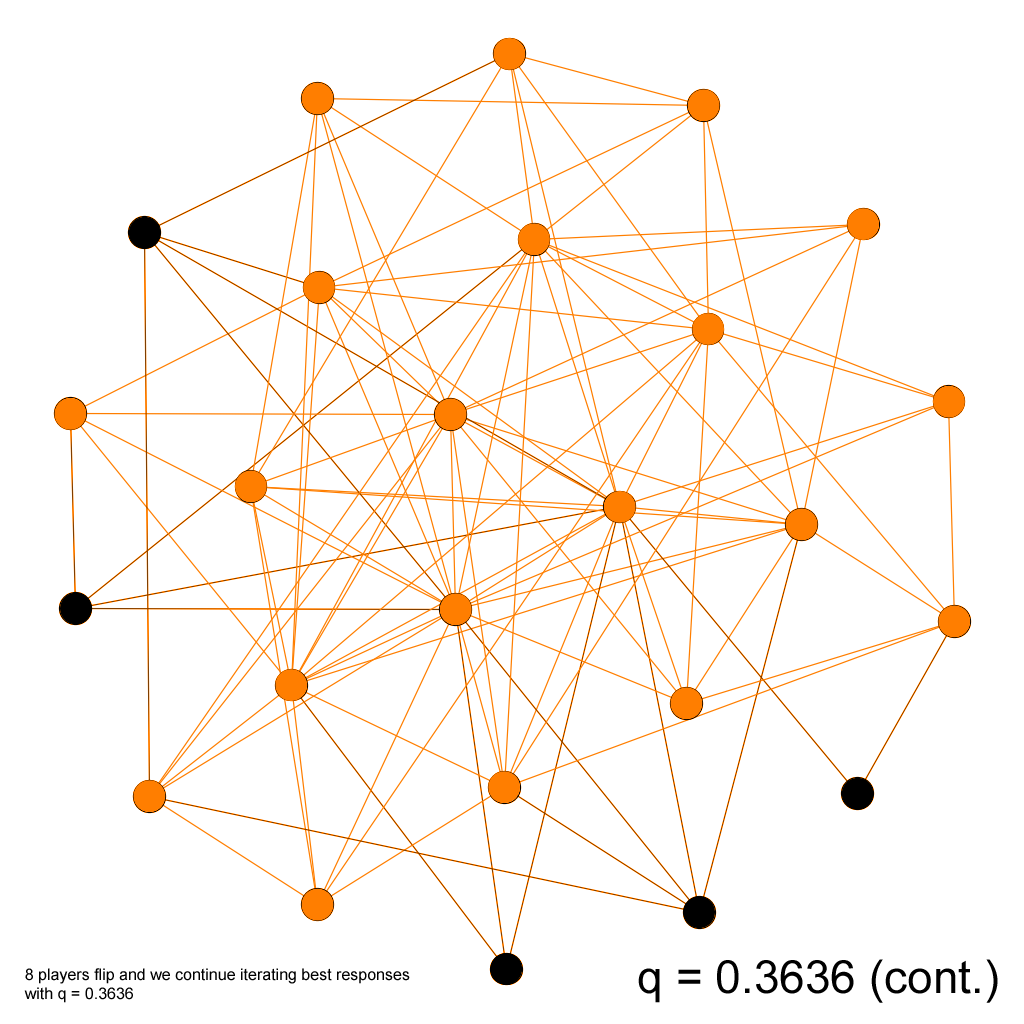}
    \end{subfigure}
    \begin{subfigure}{0.24\textwidth}
    \includegraphics[width=\linewidth]{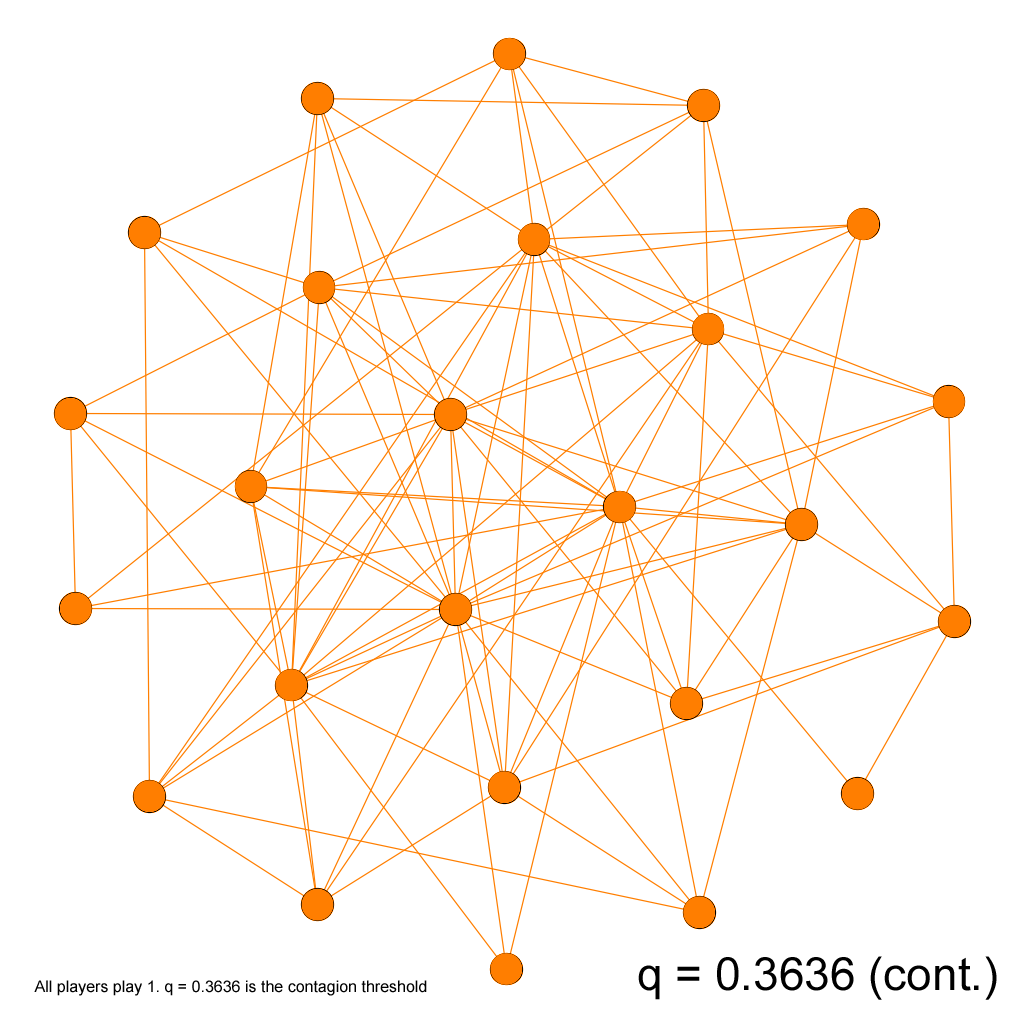}
    \end{subfigure}
    \caption{Contagion threshold with local effects only is $q^* = 0.3636$ ($I = 25$, $m=5$ $\alpha = 0$)} \label{fig:alg-ex-a}
    \end{subfigure}
    \medskip \vspace{0.1in} 
    \begin{subfigure}{\textwidth}
    \begin{subfigure}{0.24\textwidth}
    \includegraphics[width=\linewidth]{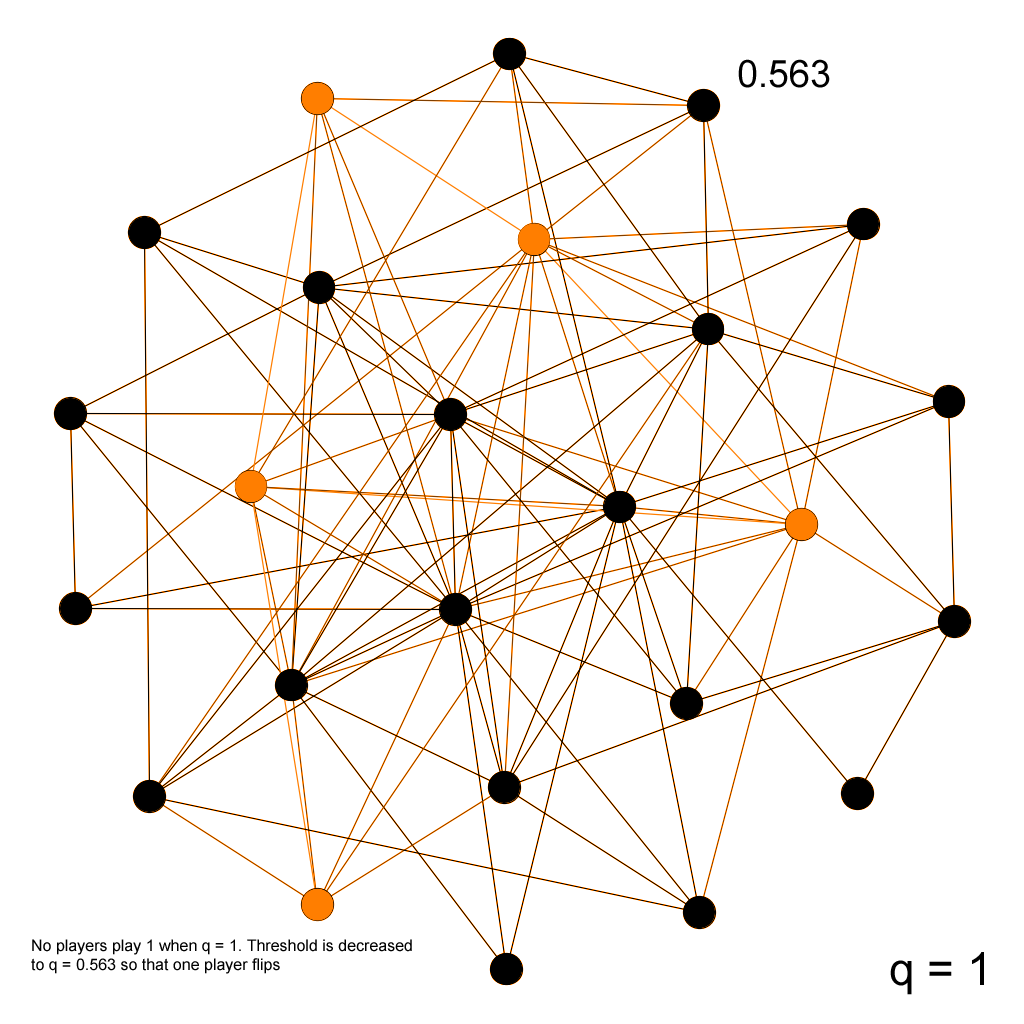}
    \end{subfigure}
    \begin{subfigure}{0.24\textwidth}
    \includegraphics[width=\linewidth]{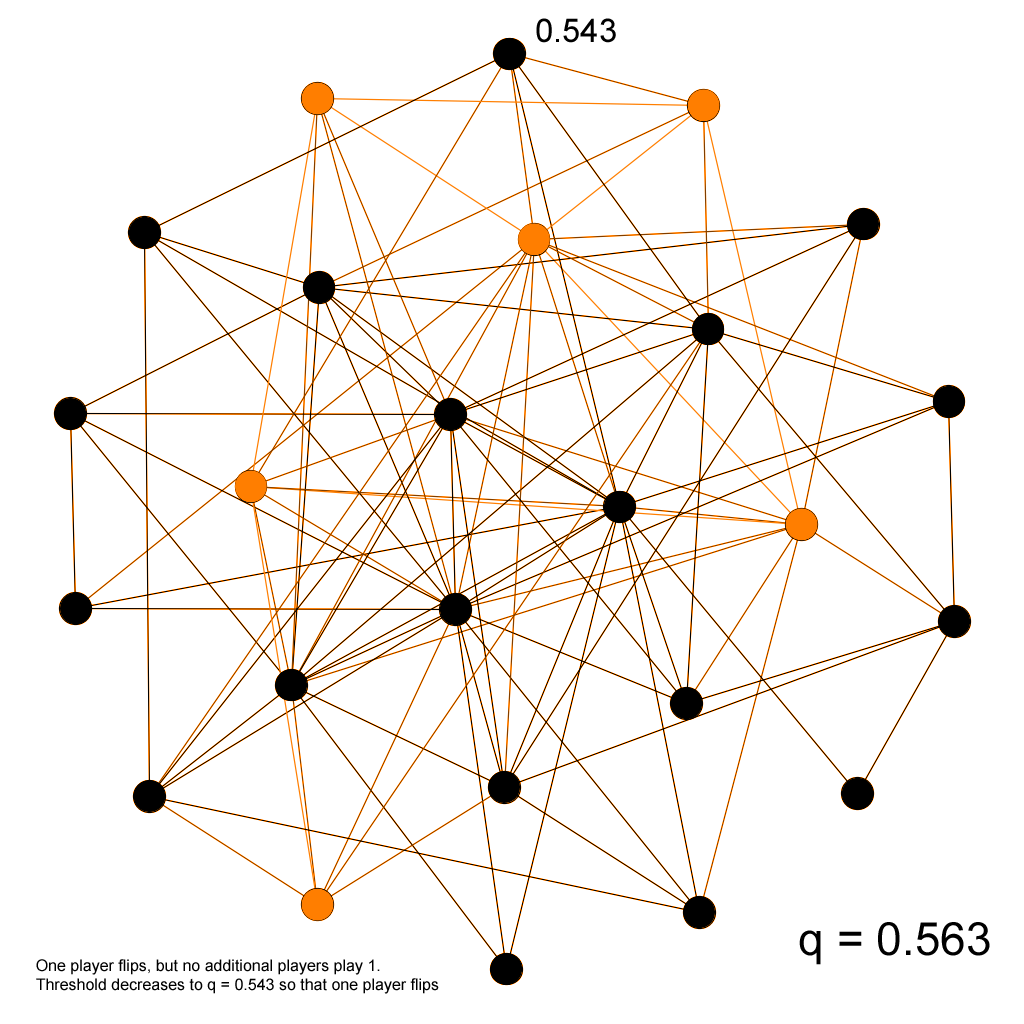}
    \end{subfigure}
    \begin{subfigure}{0.24\textwidth}
    \includegraphics[width=\linewidth]{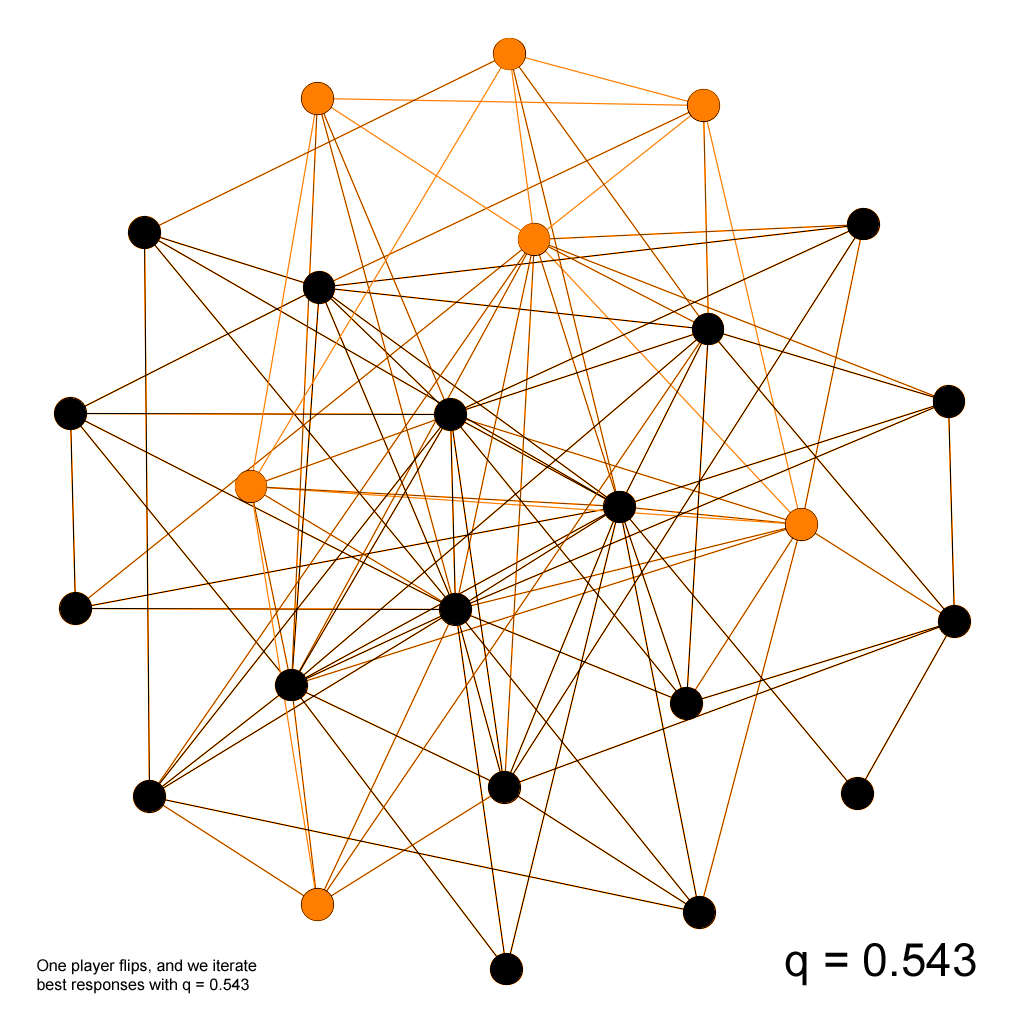}
    \end{subfigure}
    \begin{subfigure}{0.24\textwidth}
    \includegraphics[width=\linewidth]{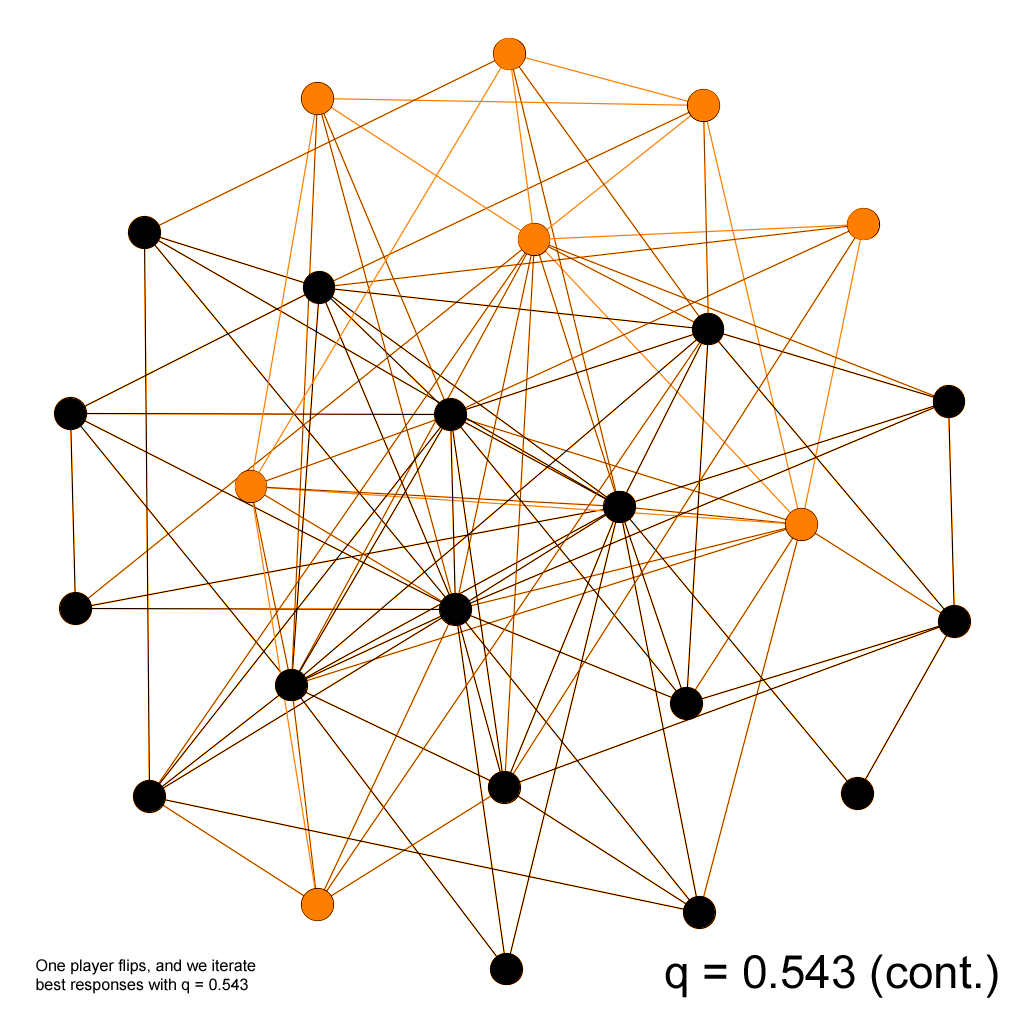}
    \end{subfigure}
    \medskip
    \centering
    \begin{subfigure}{0.24\textwidth}
    \includegraphics[width=\linewidth]{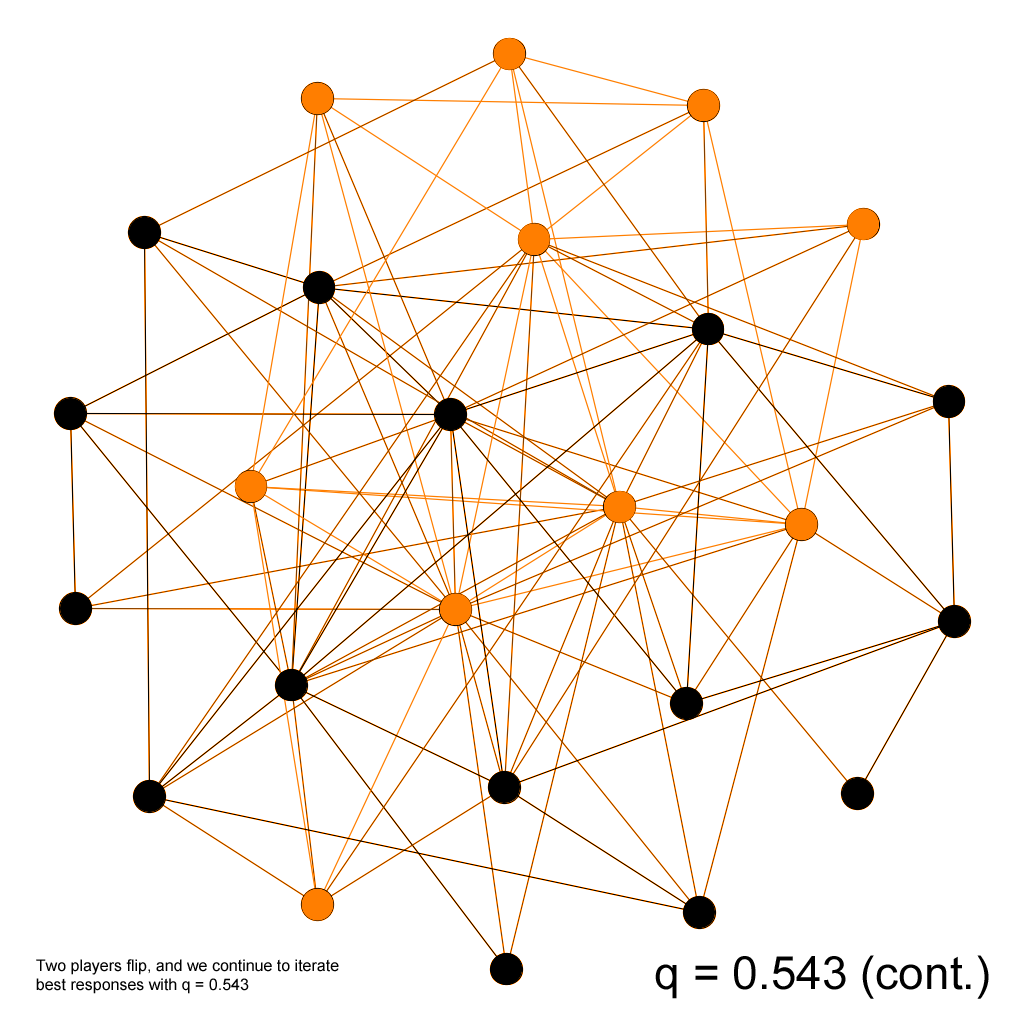}
    \end{subfigure}
    \begin{subfigure}{0.24\textwidth}
    \includegraphics[width=\linewidth]{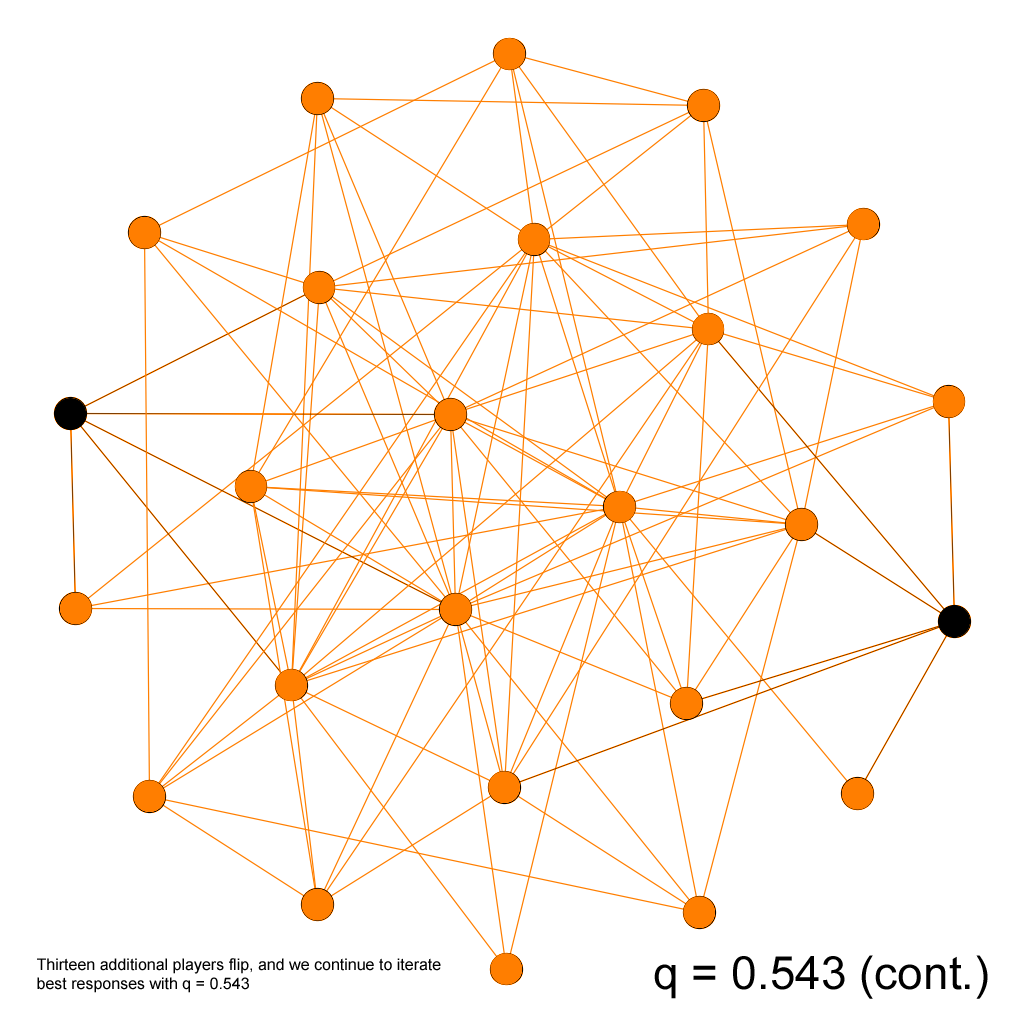}
    \end{subfigure}
    \begin{subfigure}{0.24\textwidth}
    \includegraphics[width=\linewidth]{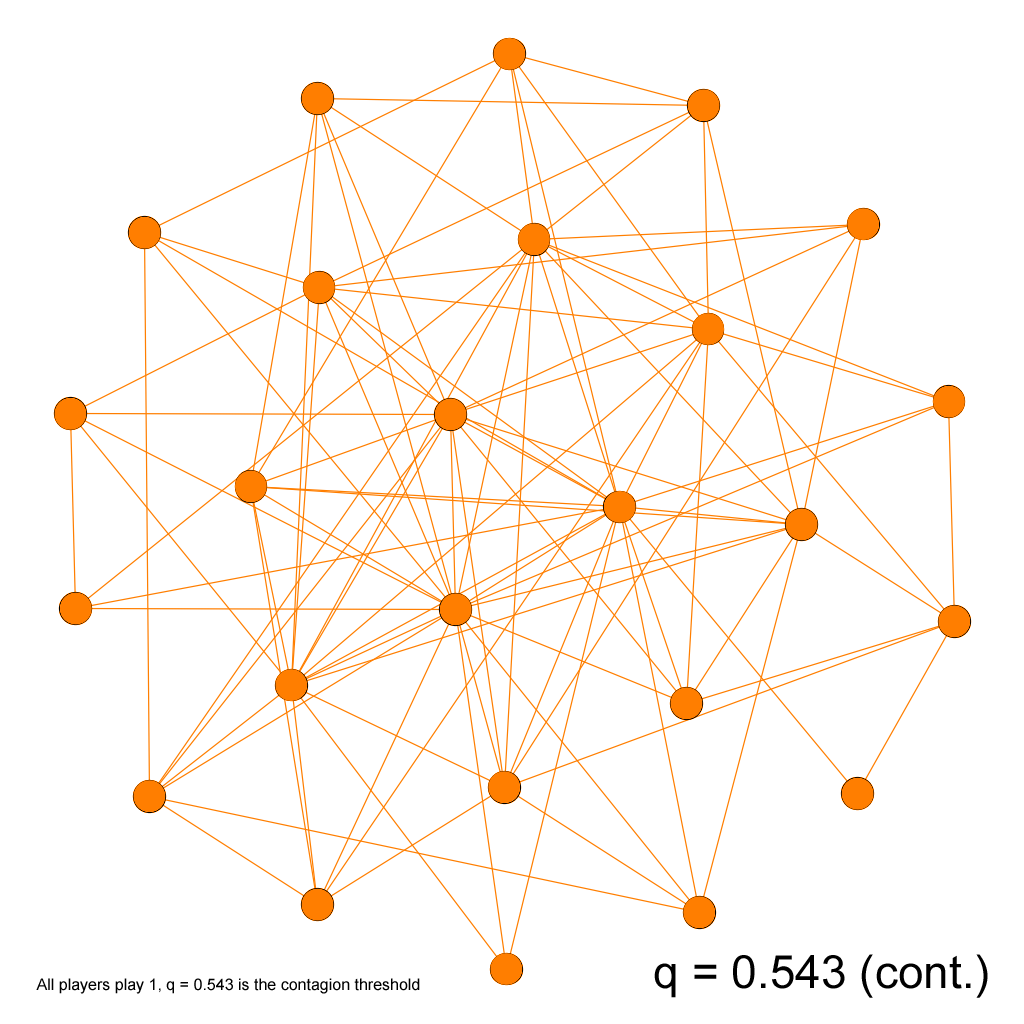}
    \end{subfigure}
    \caption{Contagion threshold with local and global effects is $q^* = 0.543$ ($I = 25$, $m=5$ $\alpha = 1$)} \label{fig:alg-ex-b}
    \end{subfigure}
    \caption{Illustration of network contagion algorithm}  \label{fig:alg-ex}
\end{figure}

The following example shows an explicit illustration of Algorithms 1 and 2. We use the Barab\'{a}si-Albert procedure to generate a scale-free network of 25 players with attachment parameter $m=5$. The generated network is presented in \Cref{fig:alg-ex} (each panel shows the same network). Global effect is modeled by $\phi_i(p_i) = \alpha c d_i p_i$ where cost of miscoordination is normalized to $c=1$. When $\alpha = 0$ there is no global network effect, when $\alpha = 1$ there is a global effect, and intermediate values of $\alpha$ correspond to differential impact of global impact on payoffs. For convenience, we focus on the cases when $\alpha = 0$ and $\alpha = 1$ in this example. 

Suppose there are $5$ players who have exogenously decided to play $1$ or who have been exogenously infected to play $1$. These comprise the initial set $S$, identified by the orange nodes in the first panel of the first row in both \Cref{fig:alg-ex-a} and \Cref{fig:alg-ex-b}. 

First consider the case of no global network effects, $\alpha = 0$, presented in \Cref{fig:alg-ex-a}. The orange nodes in the first panel in the first row are the initial set of 5 players with an incentive to play $1$. We start Algorithm 2 with $q_0 = 1$ and checking the remaining 20 players, find that no  other player becomes infected at step 1 ($A_1 = A_0 = S$). The next step down is $q_1 = 0.5$ (labeled next to the player who will switch at this threshold). Running Algorithm 1 with $(A_1, q_1)$ shows that in the first iteration (with 20 players remaining) only 1 player switches at step 1 of Algorithm 1 (shown in the second panel in top row of \Cref{fig:alg-ex-a}), leaving 19 players, and in the second iteration 1 more player flips (third panel in top row), leaving 18 players, and then Algorithm 1 stops. Thus contagion spreads to 7 players when $q=0.5$. It is the smallest Nash equilibrium of players playing $1$ if we start from $(S, q_1)$. 

Going further, Algorithm 2 denotes this set $A_2$, checks the remaining 18 players, and computes the next level for $q$ to be $q_2 = 0.4$ (labeled next to the player who will switch at this threshold). Running Algorithm 1 with $(A_2, q_2)$ flips 1 player only and stops after one iteration (last panel in the first row), leaving 17 players. The resulting set of 8 players is $A_3$, the next larger Nash equilibrium of players playing $1$. The next threshold is $q_3 = 0.3636$. Using $(A_3, q_3)$ as input to Algorithm 1 shows that 1 player flips in iteration one (first panel of row 2 in \Cref{fig:alg-ex-a}), leaving 16 players, 3 more players flip in iteration two (second panel), leaving 13 players, 8 more players flip in iteration three (third panel), leaving 5 players, and the remaining 5 players flip in iteration four (last panel). The contagion threshold for this network and starting set $S$ is $q^* = 0.3636$. Algorithm 2 also yields depth of contagion for different values of $q$. Depth (in terms of proportion of the network infected) is $1$  when $q \le 0.3636$, is $\frac{7}{25}$ when $q \in (0.3636, 0.4]$, is $\frac{6}{25}$ when $q \in (0.4, 0.5]$, and is $\frac{5}{25}$ when $q \in (0.5, 1]$. 

The entire process checks 8 subsets of players in the complement of $S$ (first with 20 players, then 19 players, then 18, 17, 16, 13, 5, and the last with 0 players), as compared to the total number of subsets of players in the complement of $S$, which is $2^{20} = 1,048,576$. 

Now consider the case with global effects, $\alpha = 1$, presented in \Cref{fig:alg-ex-b}. We start with the same initial set of 5 players in $S$ (first panel in first row of \Cref{fig:alg-ex-b}) and $q_0=1$ and find that no other player gets infected. The next level is $q_1 = 0.563$ at which the next marginal player will flip. (Notice that this is the same player as in the earlier case, but the threshold at which this player flips is higher due to impact of global effect which contributes $0.063$ to the earlier threshold of $0.5$ without global effect.) Only this player flips and the next threshold is $q_2=0.543$. In this case, 1 player flips in the first iteration of Algorithm 1, 1 more in the second iteration, 2 in the third, thirteen in the fourth, and the remaining two players in the fifth iteration. The contagion threshold with global effects for starting set $S$ is $q^* = 0.543$. Depth (in terms of proportion of the network infected) is $1$ when $q \le 0.543$, is $\frac{6}{25}$ when $q \in (0.543, 0.563]$, and is $\frac{5}{25}$ when $q \in (0.563, 1]$. The entire process checks 7 subsets of players in the complement of $S$. 

The contagion threshold with global effects is considerably higher than the one without local effect only. In other words, for every $q \in (0.3636, 0.543]$ there is full contagion with global effects but not without it. More generally, for every $q \in [0,1]$, depth of contagion at $q$ is (weakly) higher with global effects than without.

\section{Monte Carlo Simulation}

Our algorithms provide a computational tool to study equilibrium properties of contagion in networks. We use it for Monte Carlo simulation of scale-free networks following the Barab\'{a}si-Albert algorithm. Scale-free networks are used widely to study social networks and they naturally capture the idea of preferential attachment. 
We present results for a large number of combinations of network configurations,  attachment parameters, starting set sizes $S$, local network effects and global network effects. As mentioned in section 3, global network effects in simulations are formalized by $\phi_i(p_i) = \alpha cd_ip_i$, where $cd_i$ is the upper bound on $\phi_i$. The term $c$ is a constant that we set equal to $1$. The parameter $\alpha$ indexes impact of global network effects and is termed the global virality parameter. The \textbf{\textit{data generating process}} is as follows.

\begin{enumerate} 

\item Fix a global virality parameter $\alpha \in \{ 0, 0.5, 1 \}$.

\item Fix an attachment parameter $m \in \{ 5, 10, 20 \}$ and randomly generate 40 scale-free networks of size 1,000 with attachment parameter $m$ using Barab\'{a}si-Albert algorithm in NetworkX Python. 

\item In each generated network, for each starting set size $n \in \{ 10\lambda \mid \lambda = 1,\ldots, 99 \}$, randomly select 50 sets $S$ of nodes in the generated network, each set of size $\lvert S \rvert = n$. Players in $S$ are assumed to have an incentive to play $1$ at $q=1$. (We may consider these players to be exogenously infected to play $1$.) 

\item For each $S$, use the combined algorithm to compute the contagion threshold $q^*$ and the depth of contagion function $\delta(S,q)$. 

\item Repeat for each $\alpha$ and $m$.  

\end{enumerate}

For each $m$ and $\alpha$, this process generates 198,000 ($40 \times 50 \times 99$) complete runs of \Cref{alg:thresh}. Varying $m \in \{ 5, 10, 20 \}$ and $\alpha \in \{ 0, 0.5, 1 \}$ yields 1,782,000 complete runs of \Cref{alg:thresh} (9 scenarios, each with 198,000 runs). We report results based on these simulations. The parameter values facilitate analysis for three levels, intuitively corresponding to low, medium, and high parameter values. For the attachment parameter $m$, the value $m=5$ captures the idea of low attachment (5 connections as  each new player is added to form a network with 1,000 players), $m=10$ is medium attachment (10 connections for each new player), and $m=20$ is high attachment (20 connections for each new player). The value $\alpha=0$ corresponds to the model with local effects only, $\alpha=1$ is the model with local and global effects, and $\alpha = 0.5$ indexes intermediate levels of global effects. Similarly, we'll consider $q=0.25$ to be low relative cost of miscoordination (or low network resilience to switch to $1$), $q=0.5$ to be medium relative cost of miscoordination (or medium network resilience), and $q=0.75$ to be high relative cost of miscoordination (or high network resilience). A large number of robustness checks conducted with additional variation in the values of $m$, $q$, and $\alpha$ yield similar results.

\subsection{Contagion Thresholds} 

\begin{figure}[!htb]
 
 \centering
\begin{subfigure}{0.3\textwidth}
  \includegraphics[width=\linewidth]{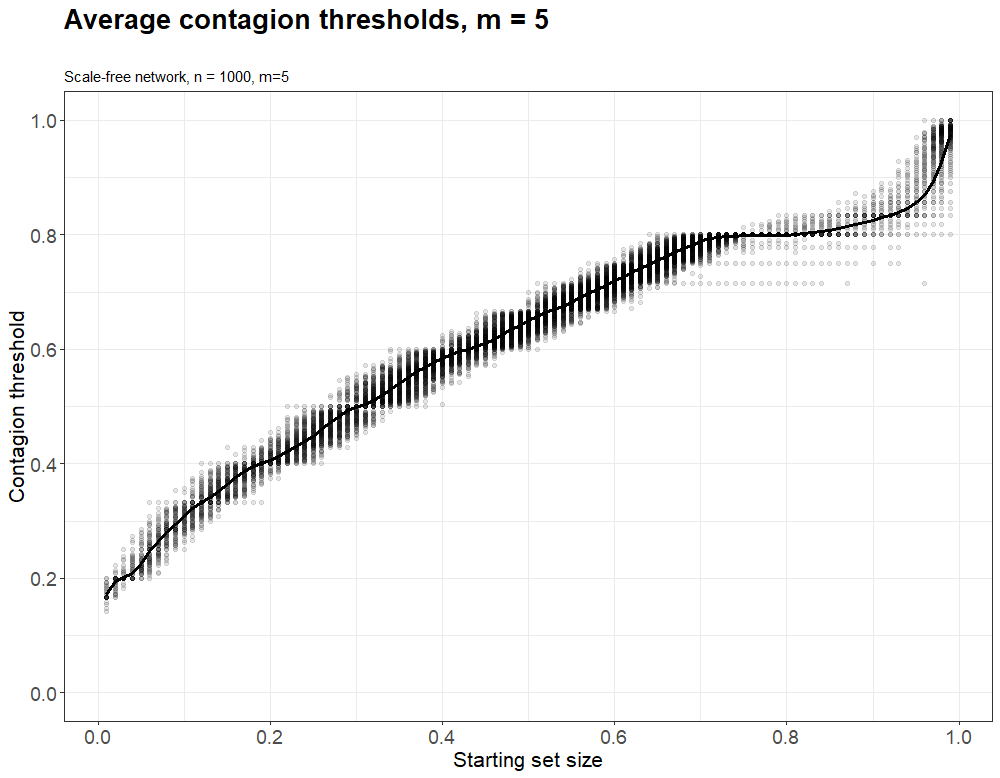}
  \caption{$m = 5, \alpha = 0$}
  \label{fig:m5thresh}
\end{subfigure}\hfil
\begin{subfigure}{0.3\textwidth}
  \includegraphics[width=\linewidth]{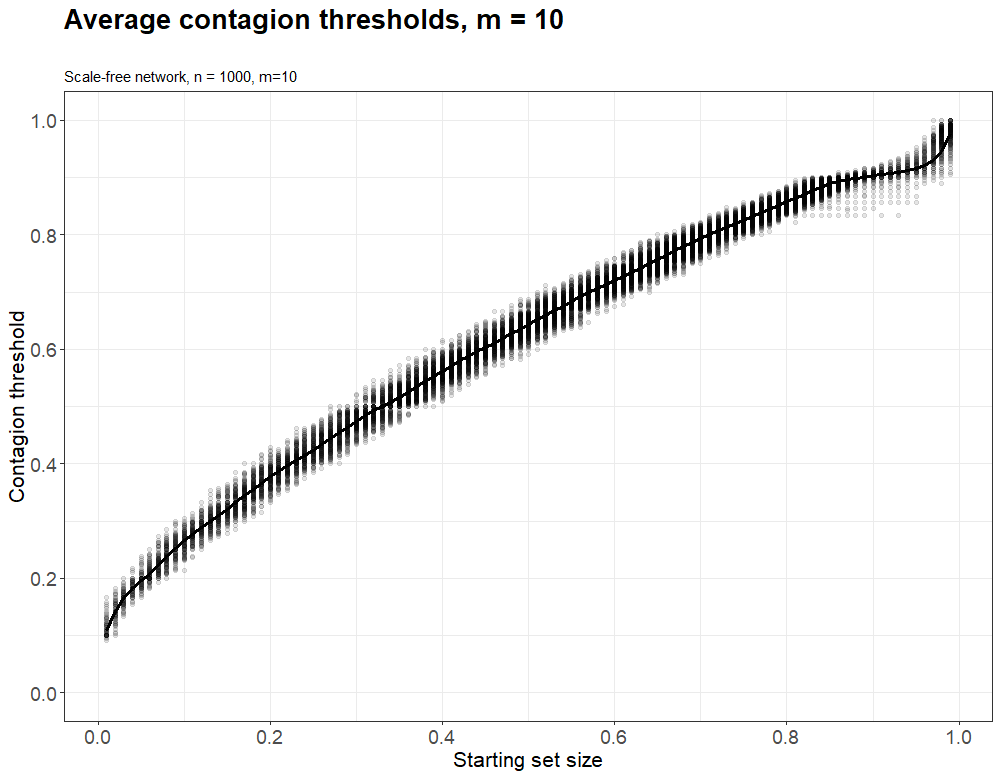}
  \caption{$m = 10, \alpha = 0$}
  \label{fig:m10thresh}
\end{subfigure}\hfil
\begin{subfigure}{0.3\textwidth}
  \includegraphics[width=\linewidth]{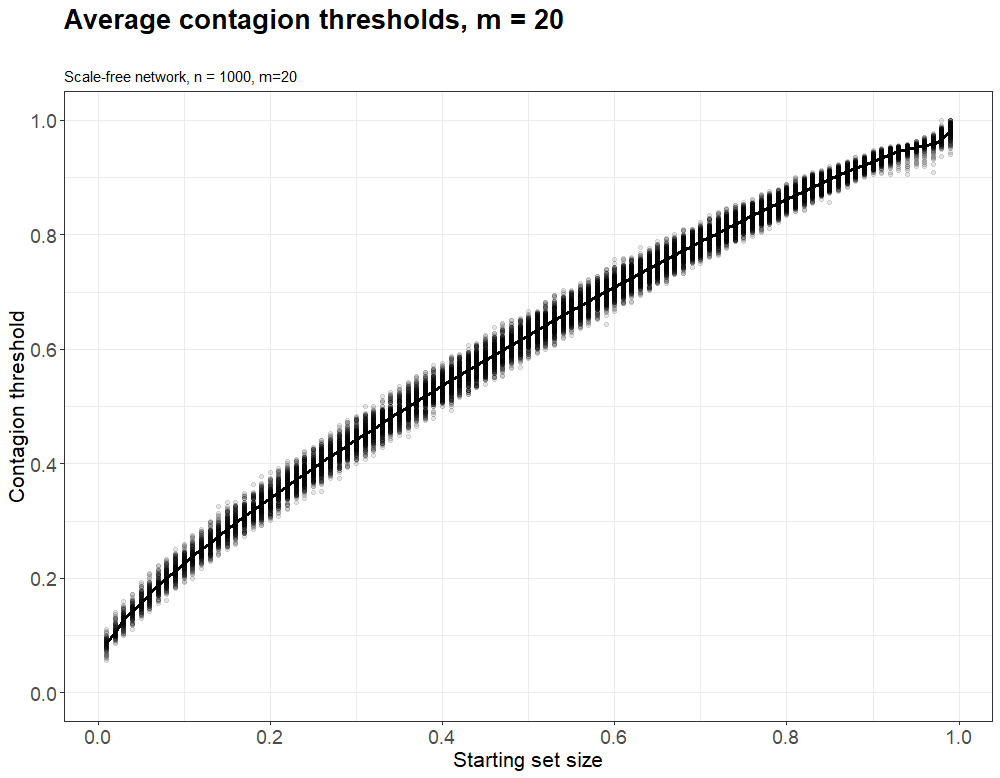}
  \caption{$m = 20, \alpha = 0$}
  \label{fig:m20thresh}
\end{subfigure} 

\medskip
\centering
\begin{subfigure}{0.3\textwidth}
  \includegraphics[width=\linewidth]{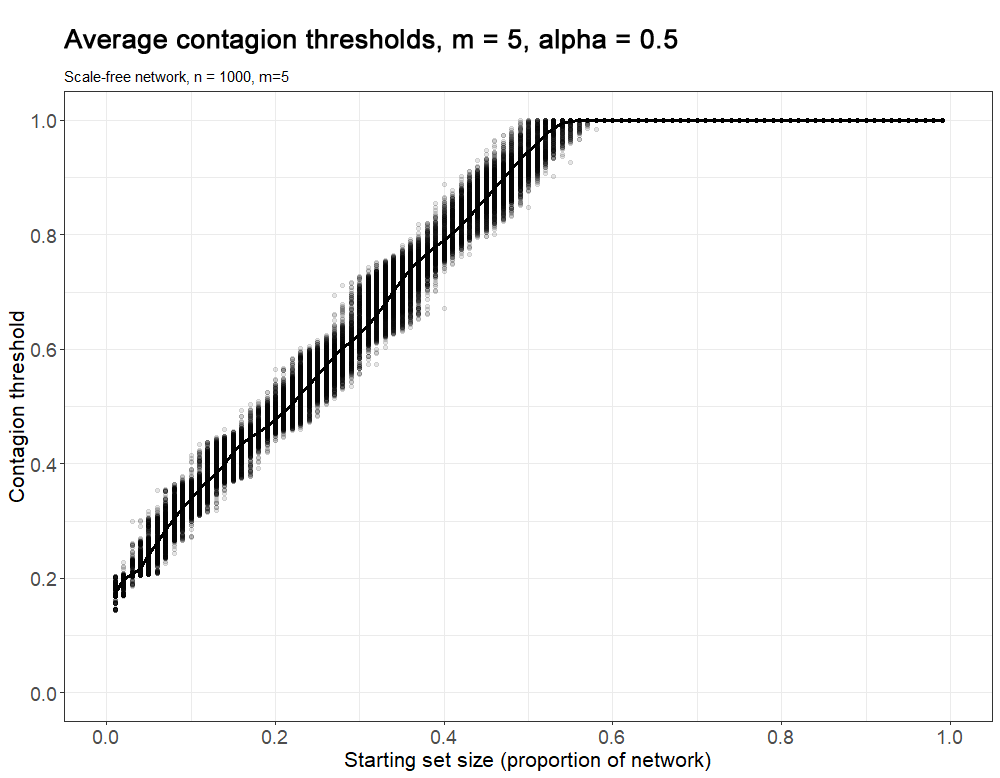}
  \caption{$m = 5, \alpha = 0.5$}
  \label{fig:m5threshv}
\end{subfigure}\hfil 
\begin{subfigure}{0.3\textwidth}
  \includegraphics[width=\linewidth]{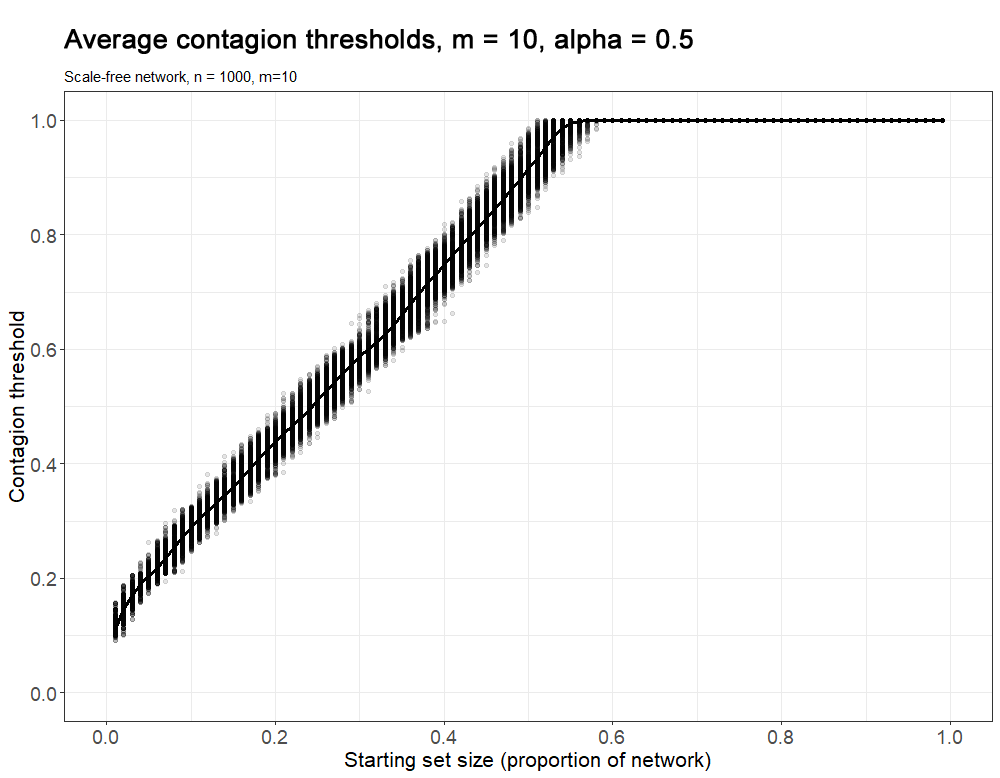}
  \caption{$m=10, \alpha = 0.5$}
  \label{fig:m10threshv}
\end{subfigure}\hfil
\begin{subfigure}{0.3\textwidth}
  \includegraphics[width=\linewidth]{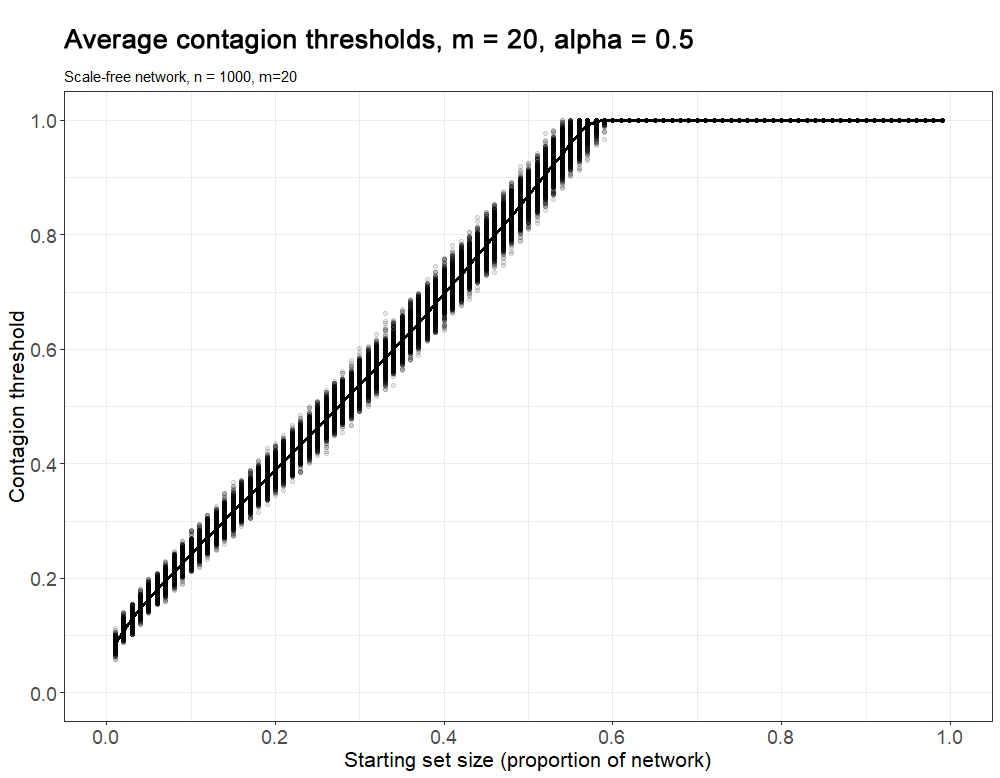}
  \caption{$m=20, \alpha = 0.5$}
  \label{fig:m20threshv}
\end{subfigure}

\medskip
\centering
\begin{subfigure}{0.3\textwidth}
  \includegraphics[width=\linewidth]{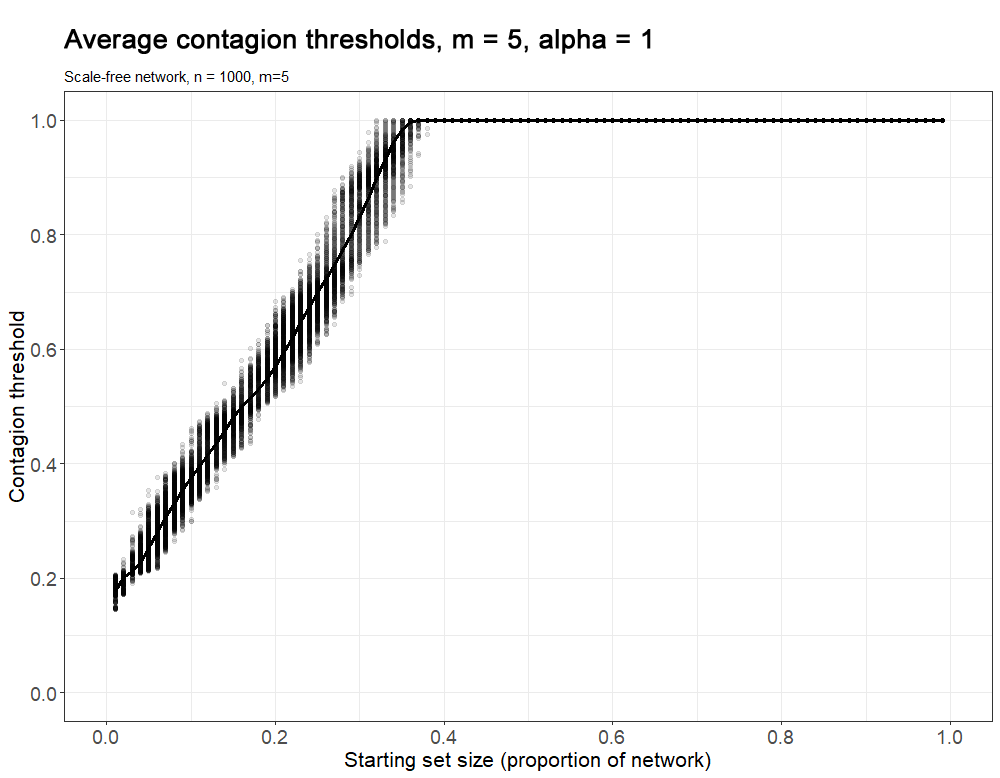}
  \caption{$m = 5, \alpha = 1$}
  \label{fig:m5threshv}
\end{subfigure}\hfil 
\begin{subfigure}{0.3\textwidth}
  \includegraphics[width=\linewidth]{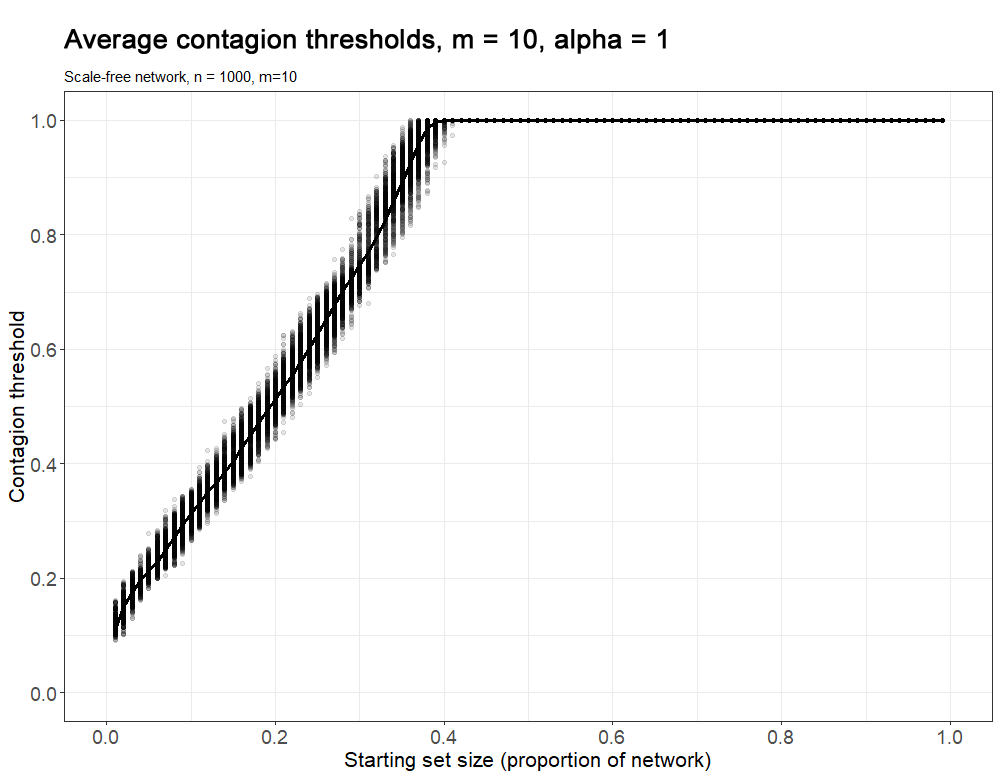}
  \caption{$m=10, \alpha = 1$}
  \label{fig:m10threshv}
\end{subfigure}\hfil
\begin{subfigure}{0.3\textwidth}
  \includegraphics[width=\linewidth]{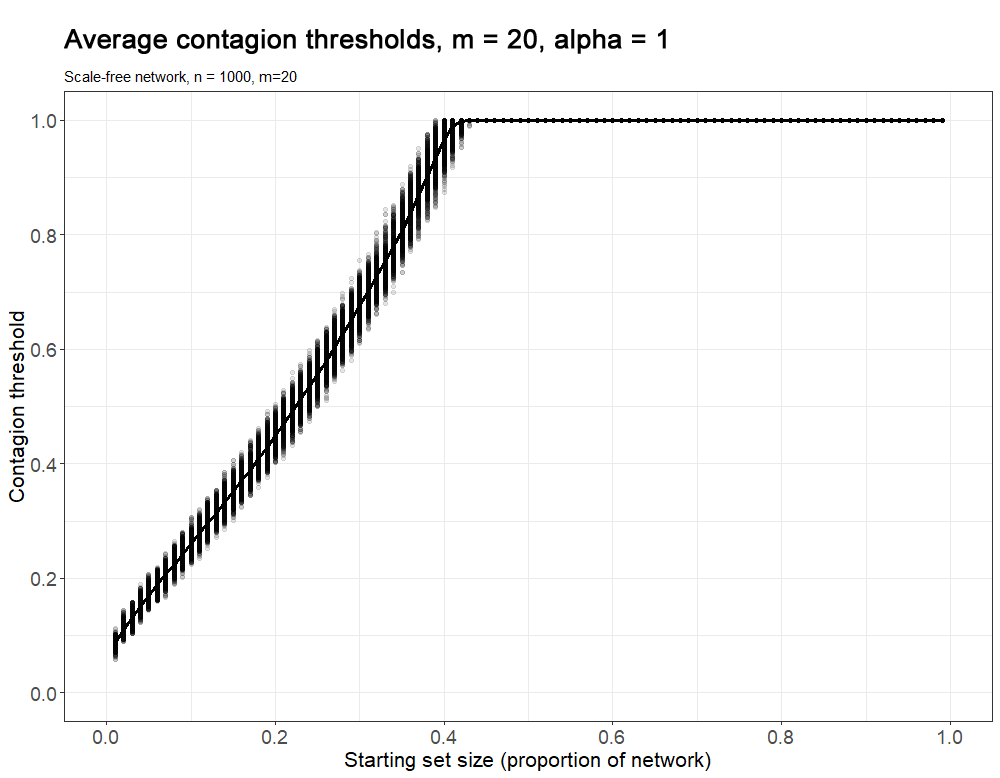}
  \caption{$m=20, \alpha = 1$}
  \label{fig:m20threshv}
\end{subfigure}
\caption{Contagion thresholds for scale-free networks}
\label{fig:threshsumm}
\end{figure}

\Cref{fig:threshsumm} shows the distribution of contagion thresholds ($q^*$) for each of the nine scenarios. A summary of selected data is included in \Cref{tab:thresh}. 

In each panel in \Cref{fig:threshsumm}, starting set size (as proportion of the network) is on the $x$-axis and network contagion threshold on the $y$-axis. Each dot is one run of the algorithm for the corresponding starting set to determine contagion threshold $q^*$. For each starting set size, there are 2,000 outcomes and there are 99 starting set sizes to yield a total of 198,000 outcomes in each panel. The bold line plots the average contagion threshold for the 2,000 outcomes for each starting set size. Graphically, the area under the curve signifies starting sets for which contagion occurs from $S$ to the entire network ($q \leq q^*$) and the area above the curve signifies starting sets for which contagion does not occur from $S$ to the entire network ($q>q^*$).

\begin{table}
\ra{1.1}
\centering
\begin{tabular}{@{\hspace*{\leftmargin}}lrrrcrrrcrrr@{}}\toprule
& \multicolumn{3}{c}{$m=5$} & \phantom{abc}& \multicolumn{3}{c}{$m=10$} &
\phantom{abc} & \multicolumn{3}{c}{$m=20$}\\
\cmidrule{2-4} \cmidrule{6-8} \cmidrule{10-12}
Size
&$\alpha = 0$& $\alpha = 0.5$ & $\alpha = 1$ && $\alpha = 0$& $\alpha = 0.5$ & $\alpha = 1$  && $\alpha = 0$& $\alpha = 0.5$ & $\alpha = 1$ \\ \midrule
$0.01$  & 0.171& 0.174&0.177 && 0.109& 0.110 &0.111 && 0.085&0.086&0.087 \\
$0.02$ & 0.194& 0.198&0.201&& 0.143&0.146& 0.149&& 0.105& 0.107&0.108\\
$0.03$ &0.201 &0.206& 0.211&& 0.166 & 0.170&0.175&& 0.126&0.129&0.132\\
$0.04$ & 0.209& 0.217&0.227 && 0.182&0.189&0.196&& 0.142&0.147&0.151\\
$0.05$ & 0.226& 0.239& 0.253&& 0.195&0.203&0.212&& 0.157 &0.162&0.168\\
$0.06$ & 0.246& 0.262&0.279  && 0.210&0.217&0.228 && 0.173&0.180&0.188\\
$0.07$ &0.263& 0.284&0.306&&0.222&0.235&0.249&& 0.187&0.196&0.206\\
$0.08$& 0.280&0.304& 0.331&& 0.238&0.254&0.272&& 0.199&0.211&0.223\\
$0.09$&0.294 & 0.322&0.354&& 0.253&0.271&0.293&& 0.212&0.226&0.241\\
$0.1$  & 0.308&0.340& 0.376 && 0.265&0.288&0.313  && 0.225&0.242&0.261 \\
$0.2$ &0.405& 0.477&0.571&& 0.377& 0.437&0.512&& 0.340&0.386& 0.449\\
$0.3$ &0.498&0.627& 0.831&& 0.474 & 0.587&0.746&& 0.442&0.538&0.676\\
$0.4$ &0.584& 0.790&1 && 0.562&0.748&1&& 0.536&0.698&0.966\\
$0.5$ &0.650&0.946& 1&& 0.643&0.916&1&& 0.624 &0.870&1\\
$0.6$ & 0.720&1& 1  && 0.720&1&1 && 0.709&1&1 \\
$0.7$ &0.788& 1&1&&0.794 &1&1&& 0.788&1&1\\
$0.8$& 0.800& 1&1&& 0.858&1&1&& 0.862&1&1\\
$0.9$&0.825 & 1&1&& 0.903&1&1&& 0.928&1&1\\
\bottomrule
\end{tabular}
\caption{Average thresholds for full network contagion for different starting set sizes (proportion of network)}
\label{tab:thresh}
\end{table}
\newcommand{\rowgroup}[1]{\hspace{-1em}#1}

Several features of contagion on networks emerge in \Cref{fig:threshsumm}. In all 9 cases, the average contagion threshold is increasing in starting set size (area under the curve increases as we move toward higher starting set sizes) signifying a greater range of $q$ for which contagion occurs from $S$ to the entire network, as expected. Also, the contagion thresholds are distributed tightly around the average, for the most part. 

Increasing global impact (an increase in $\alpha$, or going downward in a column through the panels) increases the contagion threshold, making the network more susceptible to contagion. Indeed, the same increase in starting set size causes a larger increase in the contagion threshold when global impact is higher, as shown in the higher slope of the average contagion threshold line when $\alpha$ goes up. 

This shows a complementarity in the effect of starting set size and global effects on contagion thresholds. The greater is starting set size, the greater is the impact of global effects in raising the contagion threshold. For example, in a low connectivity network ($m = 5$), with no global effects ($\alpha =0)$, when starting set size increases from 5 percent (50 players) to 10 percent, the contagion threshold $q^*$ increases by 36.3 percent (from 0.226 to 0.308), but with global effects ($\alpha =1$), the comparable increase is 48.6 percent (from 0.253 to 0.376). An increase of starting set size from 5 percent to 20 percent leads to a 79.2 percent increase in the contagion threshold (from 0.226 to 0.405) without global effects, but a 125.7 percent (from 0.253 to 0.571) with global effects, on average. 

This complementarity captures a snowball effect of contagion resulting from increasing returns. A marginal increase in the starting set size causes a more than proportional increase in the contagion threshold in the presence of global effects. 

Indeed, after a point, impact of global effects overrides local obstacles to switch to $1$. In low or moderate connectivity networks ($m=5, 10$) with global effects ($\alpha =1$), the contagion threshold is $q^*=1$ with a starting set size of 40 percent (see \Cref{tab:thresh}). In other words, in low or moderate connectivity networks with global effects, if about 40 percent of the network is infected (has an incentive to play $1$ at $q=1$), there is no hope of stopping contagion to the entire network, that is, even if for every remaining player, the cost of miscoordination with a neighbor ($c>0$) is unboundedly higher than the benefit of coordination ($b$ is arbitrarily close to $0$). But if we can stop impact of global effects ($\alpha = 0$), contagion from 40 percent of the network will not spread to the whole network whenever relative miscoordination cost $q$ is above 0.584, on average. 

Increasing connectivity (an increase in $m$, or going across in a row through the panels) makes the average contagion threshold curve less steep making it harder to propagate an action across the network. Other things equal, higher connectivity implies a higher degree, which lowers the incentive for a player to play $1$, as shown by the threshold calculation $\frac{s_i}{d_i} + \frac{\phi_i(p_i)}{d_i(b+c)} \geq q$. For example, suppose $q=0.25$ and consider a network with global effects ($\alpha = 1$). As shown in \Cref{tab:thresh}, for a starting set size of 5 percent, full contagion occurs with low attachment parameter ($m=5$) but does not occur with moderate or high attachment parameter ($m=10, 20$).  Intuitively, when a player has few neighbors, flipping one of them is more likely to flip a given player, and the opposite is true when a player has more neighbors.

\subsection{Depth of contagion} 

\begin{figure}[!htb]
    \centering 
\begin{subfigure}{0.25\textwidth}
  \includegraphics[width=\linewidth]{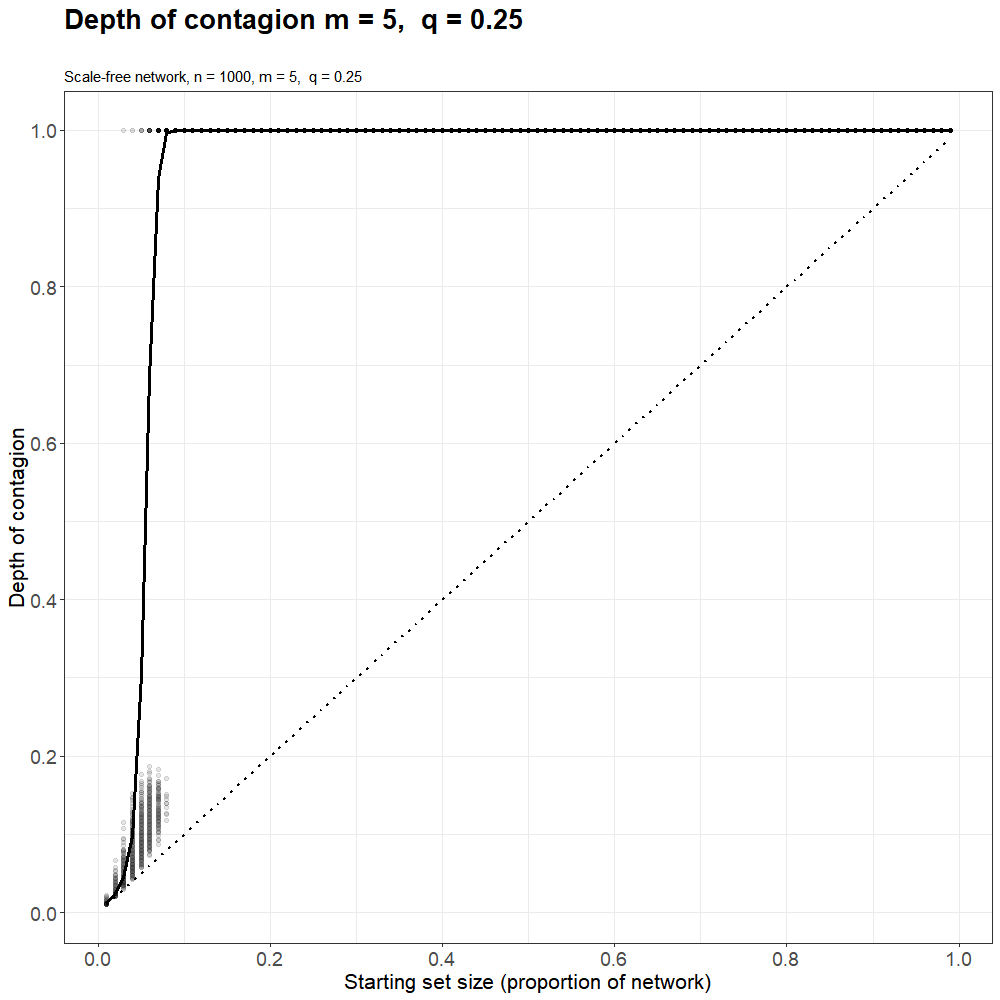}
  \caption{$m = 5, q = 0.25, \alpha = 0$}
  \label{fig:m5q25v0}
\end{subfigure}\hfil 
\begin{subfigure}{0.25\textwidth}
  \includegraphics[width=\linewidth]{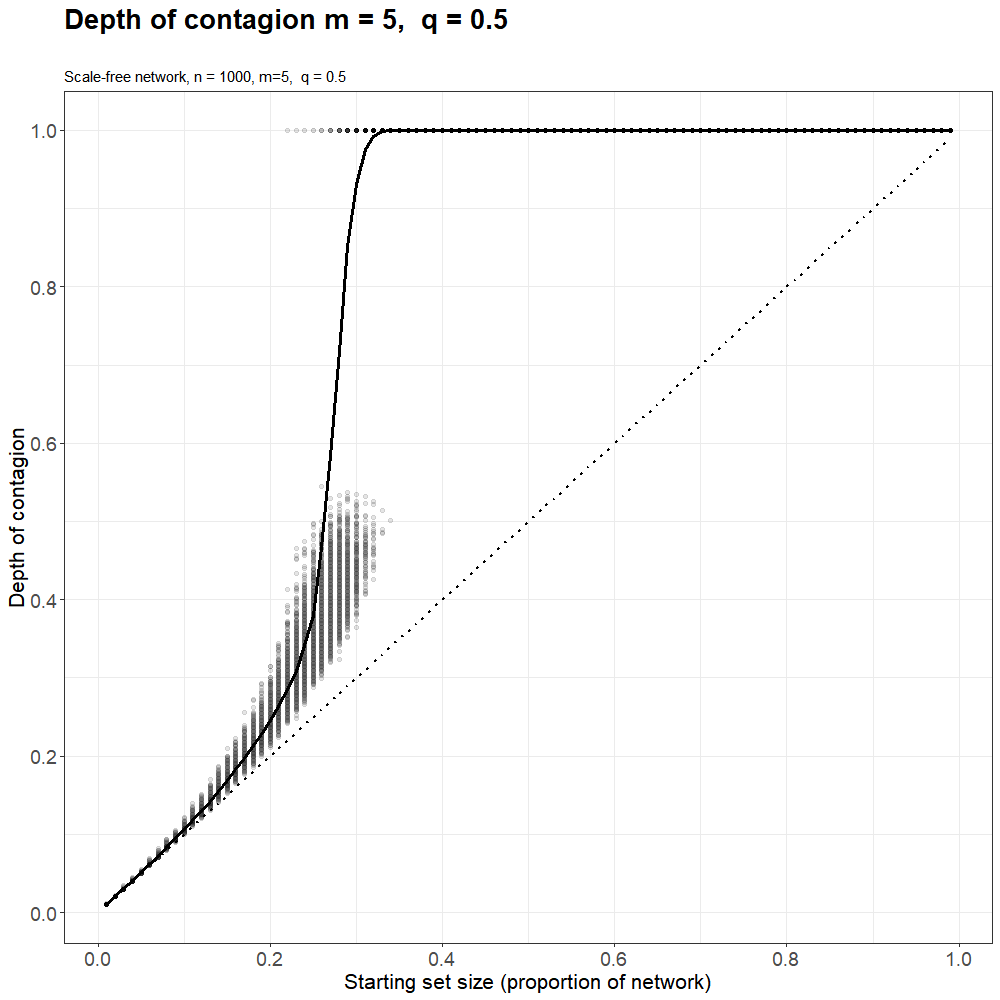}
  \caption{$m = 5, q = 0.5, \alpha = 0$}
  \label{fig:m5q5v0}
\end{subfigure}\hfil 
\begin{subfigure}{0.25\textwidth}
  \includegraphics[width=\linewidth]{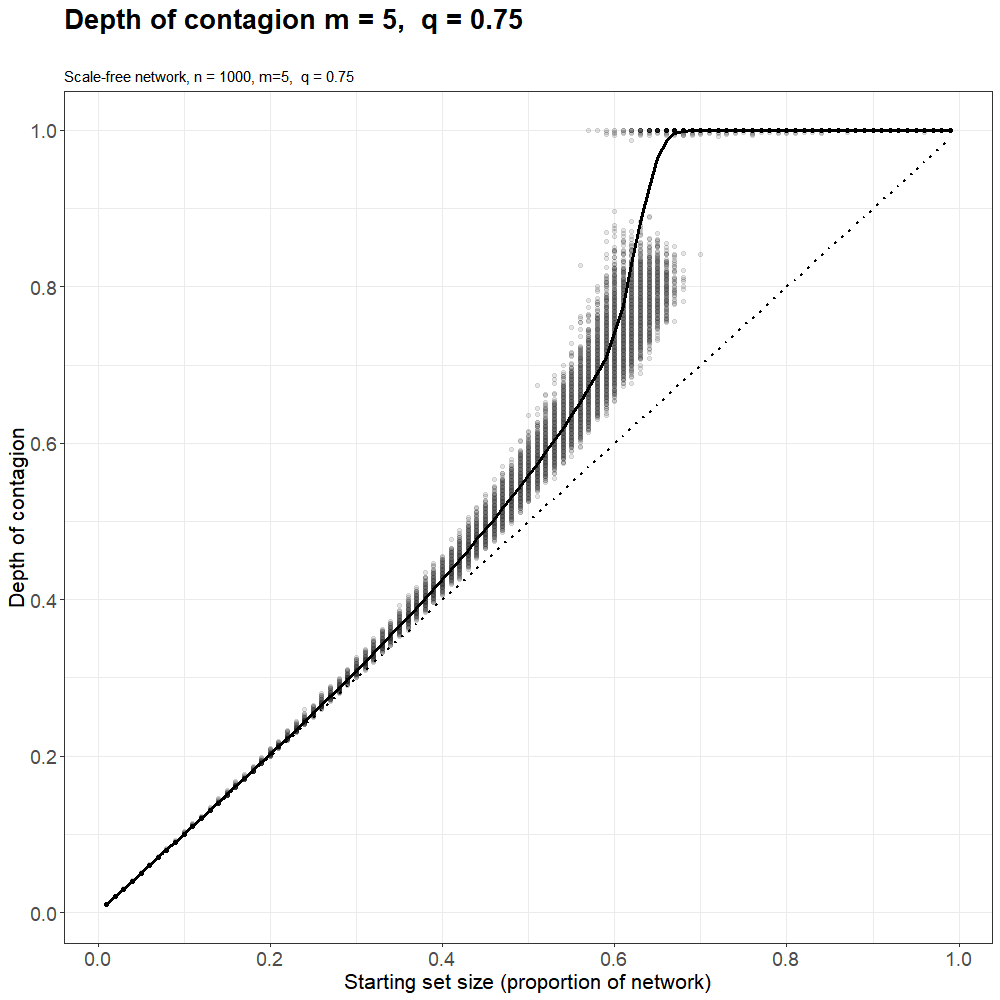}
  \caption{$m = 5, q = 0.75, \alpha = 0$}
  \label{fig:m5q75v0}
\end{subfigure}

\medskip

\begin{subfigure}{0.25\textwidth}
  \includegraphics[width=\linewidth]{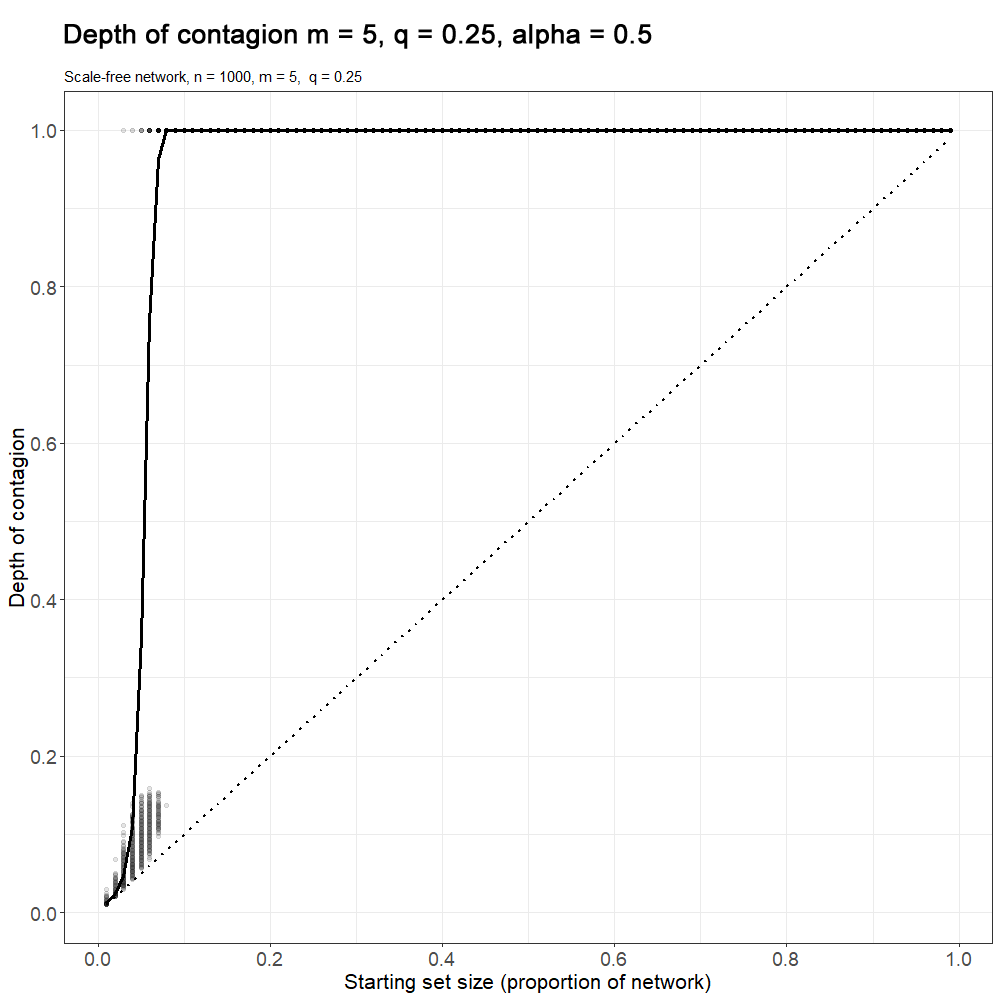}
  \caption{$m = 5, q = 0.25, \alpha = 0.5$}
  \label{fig:m5q25v5}
\end{subfigure}\hfil 
\begin{subfigure}{0.25\textwidth}
  \includegraphics[width=\linewidth]{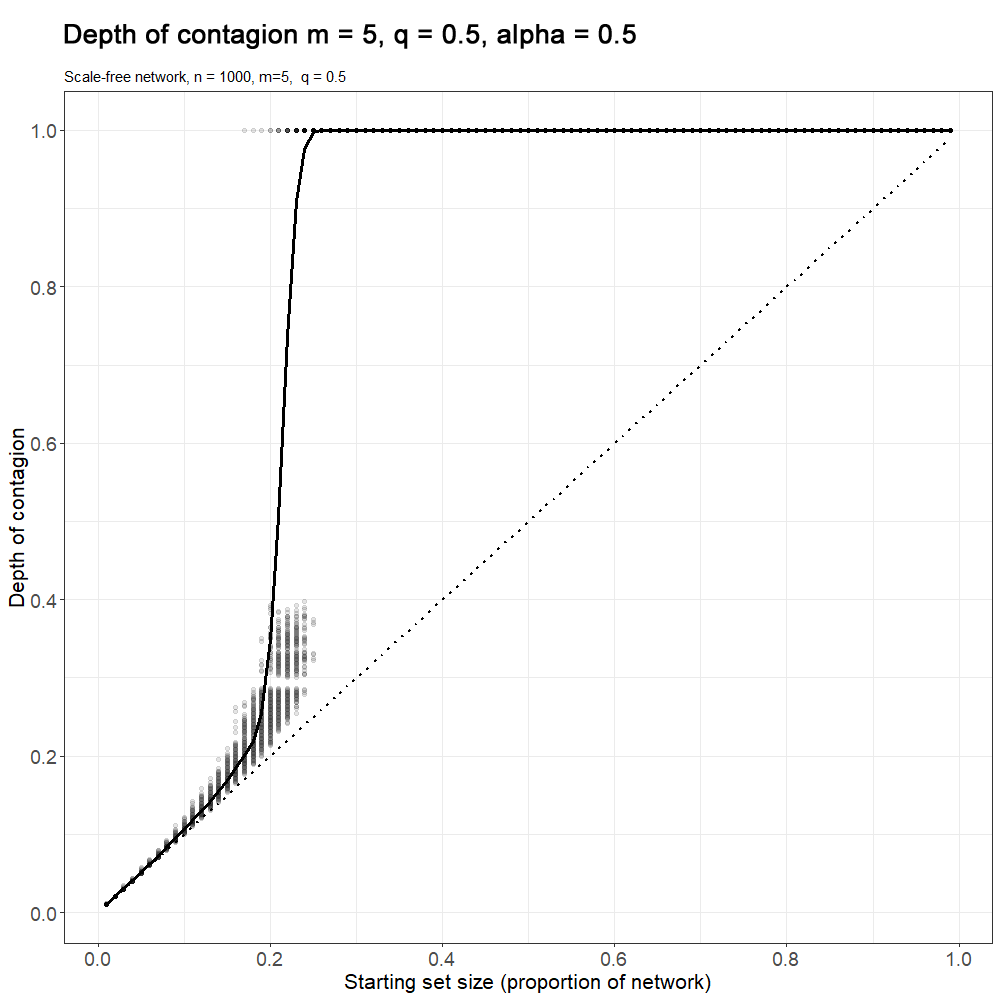}
  \caption{$m = 5, q = 0.5, \alpha = 0.5$}
  \label{fig:m5q5v5}
\end{subfigure}\hfil 
\begin{subfigure}{0.25\textwidth}
  \includegraphics[width=\linewidth]{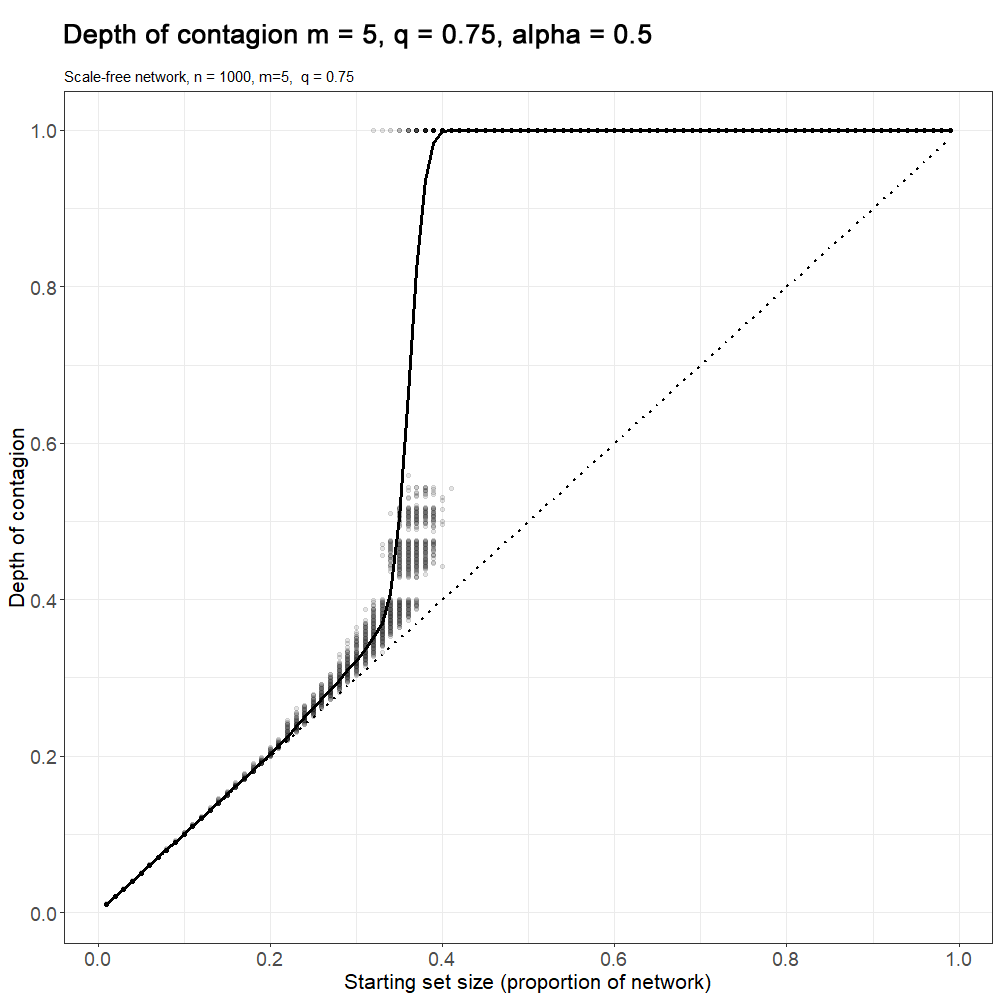}
  \caption{$m = 5, q = 0.75, \alpha = 0.5$}
  \label{fig:m5q75v5}
\end{subfigure}

\medskip

\begin{subfigure}{0.25\textwidth}
  \includegraphics[width=\linewidth]{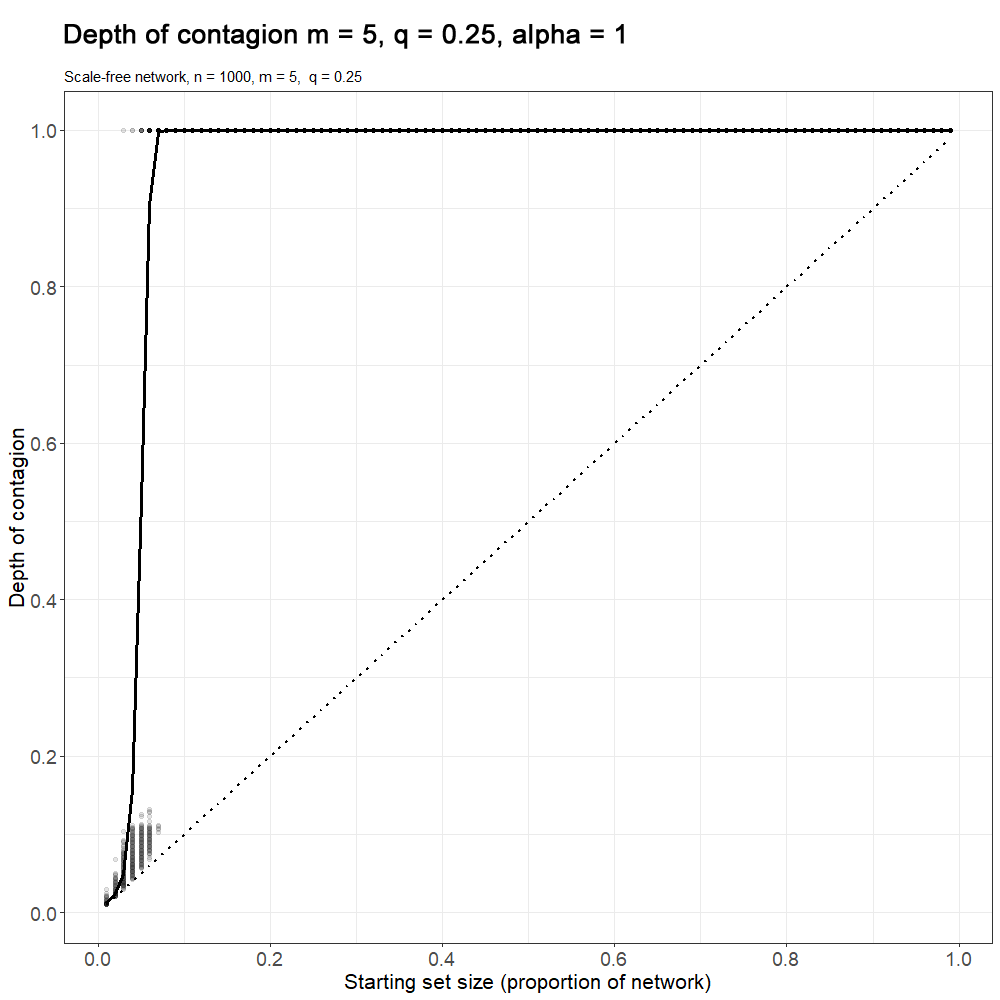}
  \caption{$m = 5, q = 0.25, \alpha = 1$}
  \label{fig:m5q25v1}
\end{subfigure}\hfil
\begin{subfigure}{0.25\textwidth}
  \includegraphics[width=\linewidth]{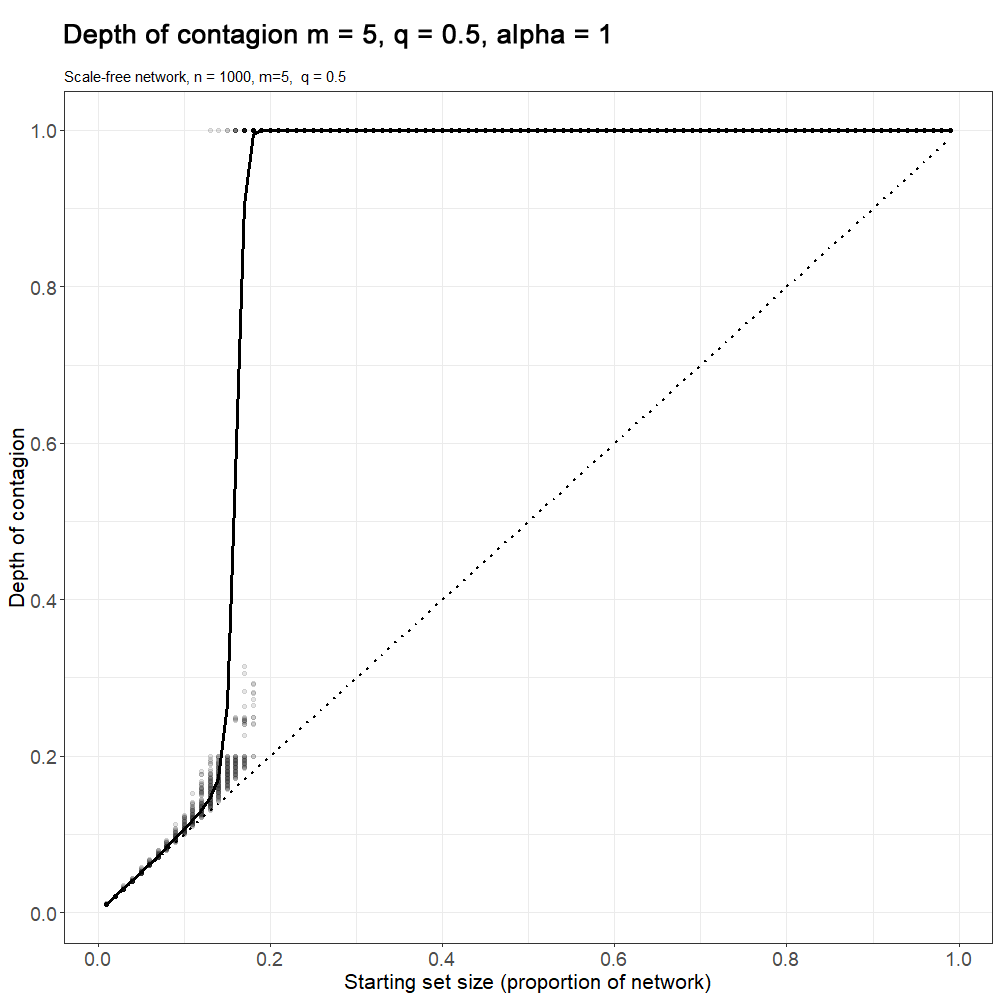}
  \caption{$m = 5, q = 0.5, \alpha = 1$}
  \label{fig:m5q5v1}
\end{subfigure}\hfil 
\begin{subfigure}{0.25\textwidth}
  \includegraphics[width=\linewidth]{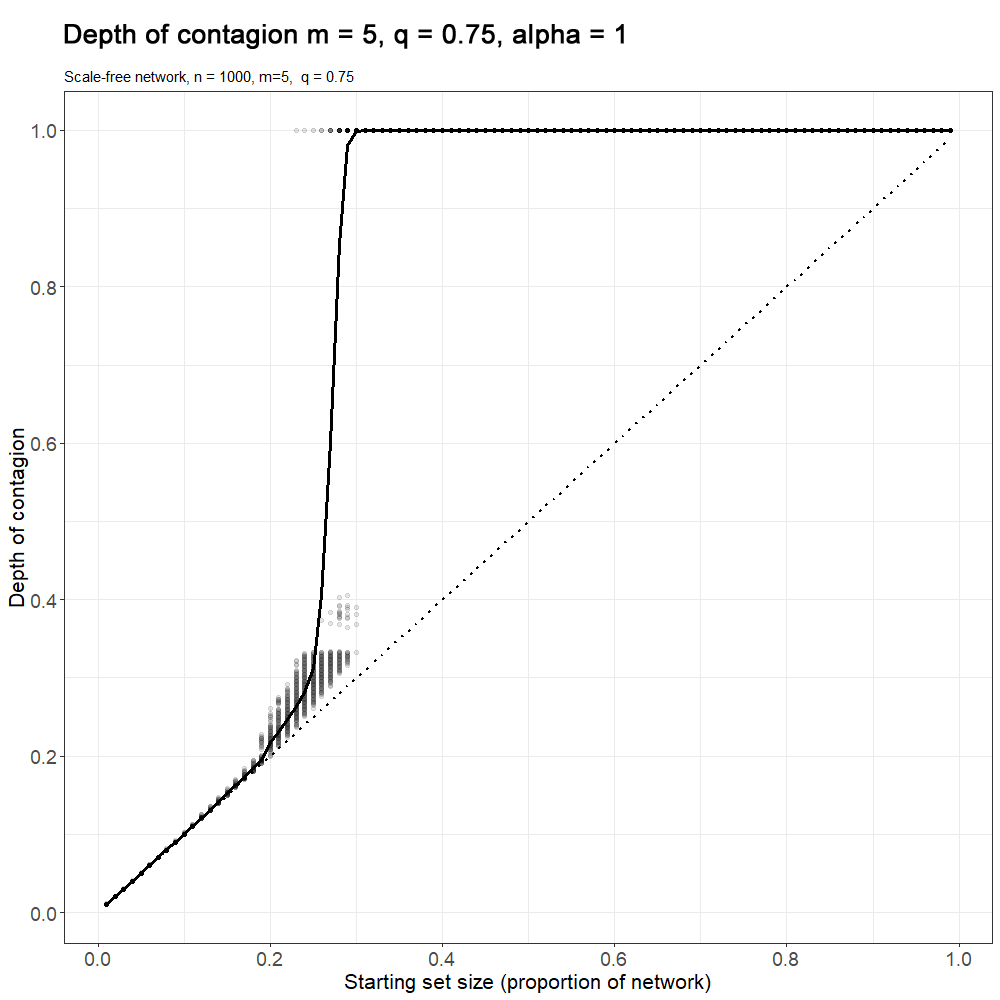}
  \caption{$m = 5, q = 0.75, \alpha = 1$}
  \label{fig:m5q75v1}
\end{subfigure} 

\caption{Depth of contagion for scale-free networks with $m=5$}
\label{fig:depths5}
\end{figure}

\begin{figure}[!htb]
    \centering 
\begin{subfigure}{0.25\textwidth}
  \includegraphics[width=\linewidth]{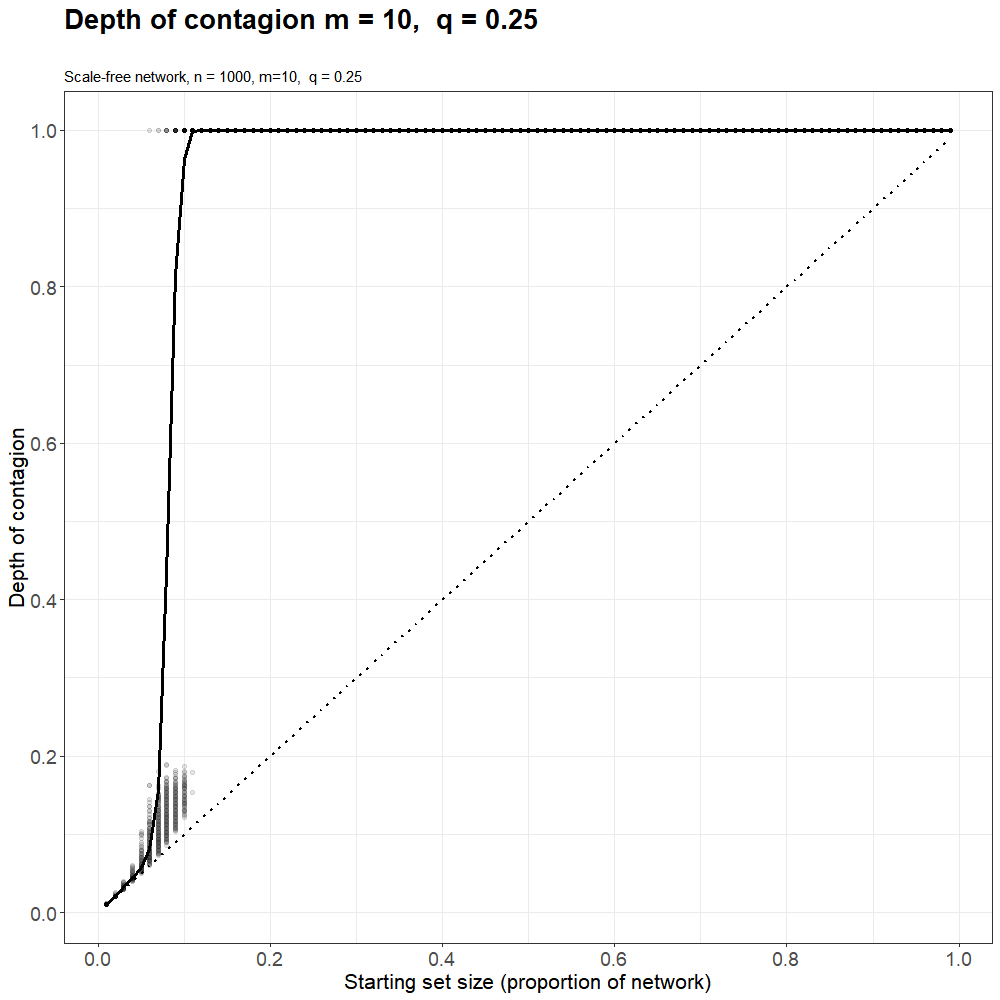}
  \caption{$m = 10, q = 0.25, \alpha = 0$}
  \label{fig:m10q25}
\end{subfigure}\hfil 
\begin{subfigure}{0.25\textwidth}
  \includegraphics[width=\linewidth]{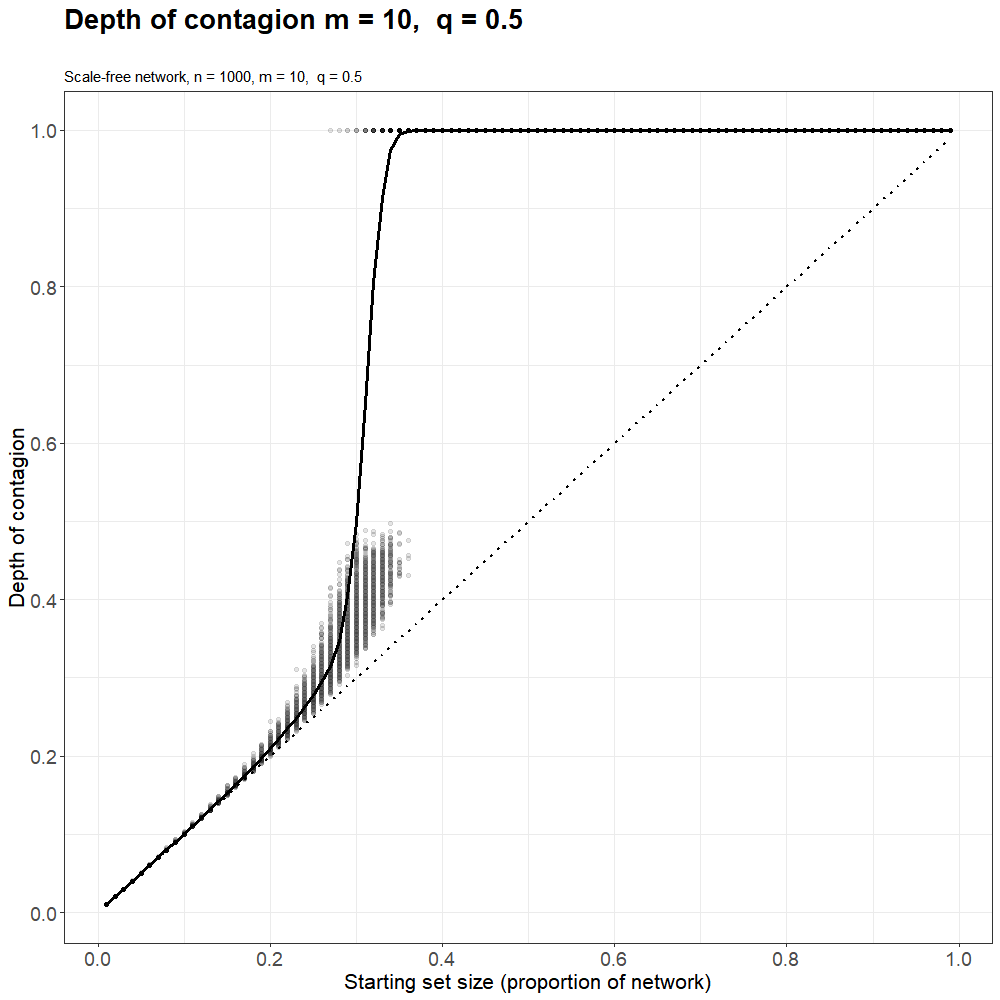}
  \caption{$m = 10, q = 0.5, \alpha = 0$}
  \label{fig:m10q5}
\end{subfigure}\hfil 
\begin{subfigure}{0.25\textwidth}
  \includegraphics[width=\linewidth]{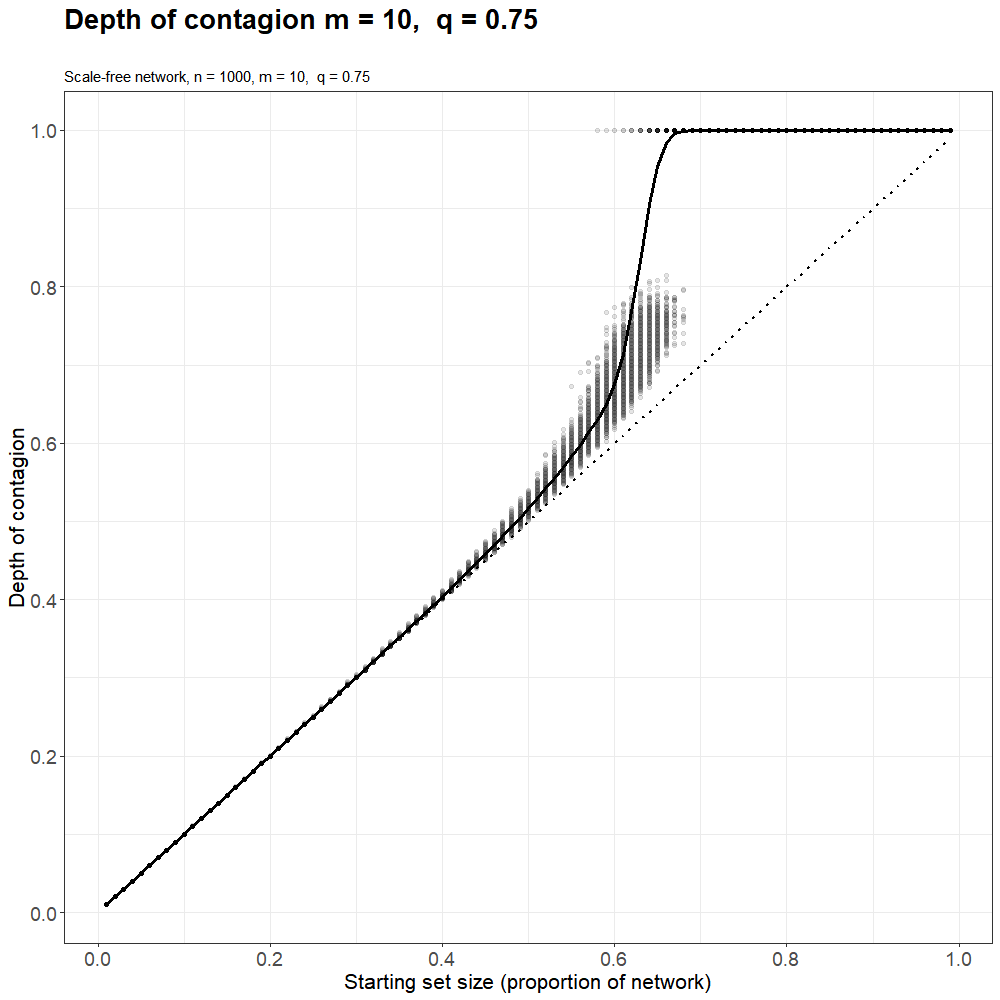}
  \caption{$m = 10, q = 0.75, \alpha = 0$}
  \label{fig:m10q75}
\end{subfigure}

\medskip
\begin{subfigure}{0.25\textwidth}
  \includegraphics[width=\linewidth]{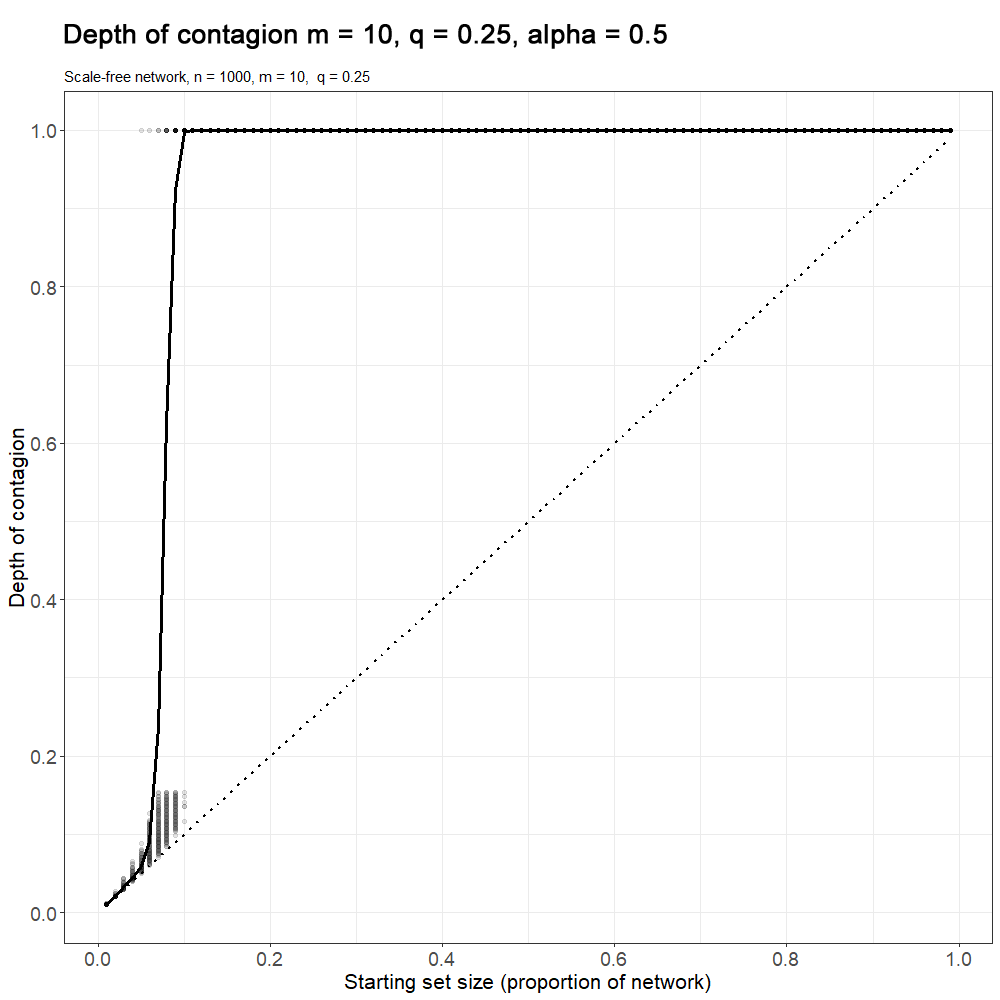}
  \caption{$m = 10, q = 0.25, \alpha = 0.5$}

\end{subfigure}\hfil 
\begin{subfigure}{0.25\textwidth}
  \includegraphics[width=\linewidth]{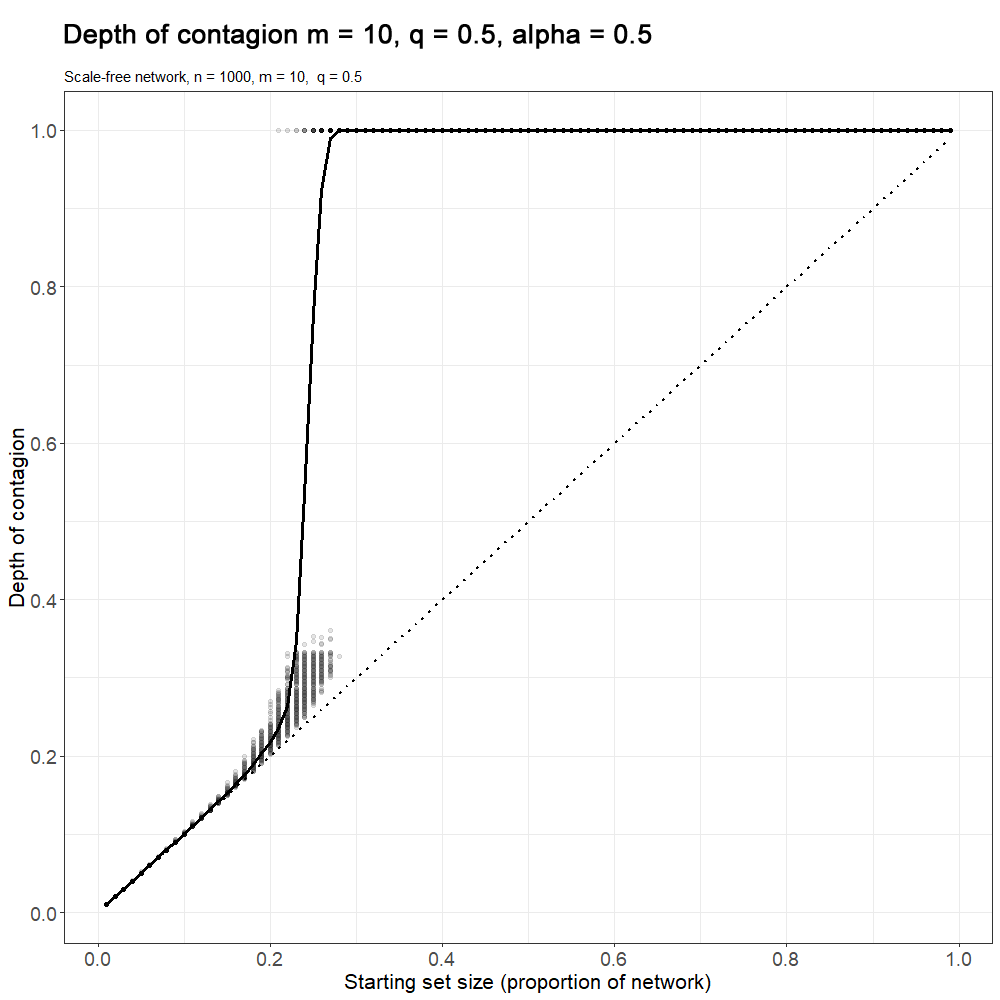}
  \caption{$m = 10, q = 0.5, \alpha = 0.5$}

\end{subfigure}\hfil 
\begin{subfigure}{0.25\textwidth}
  \includegraphics[width=\linewidth]{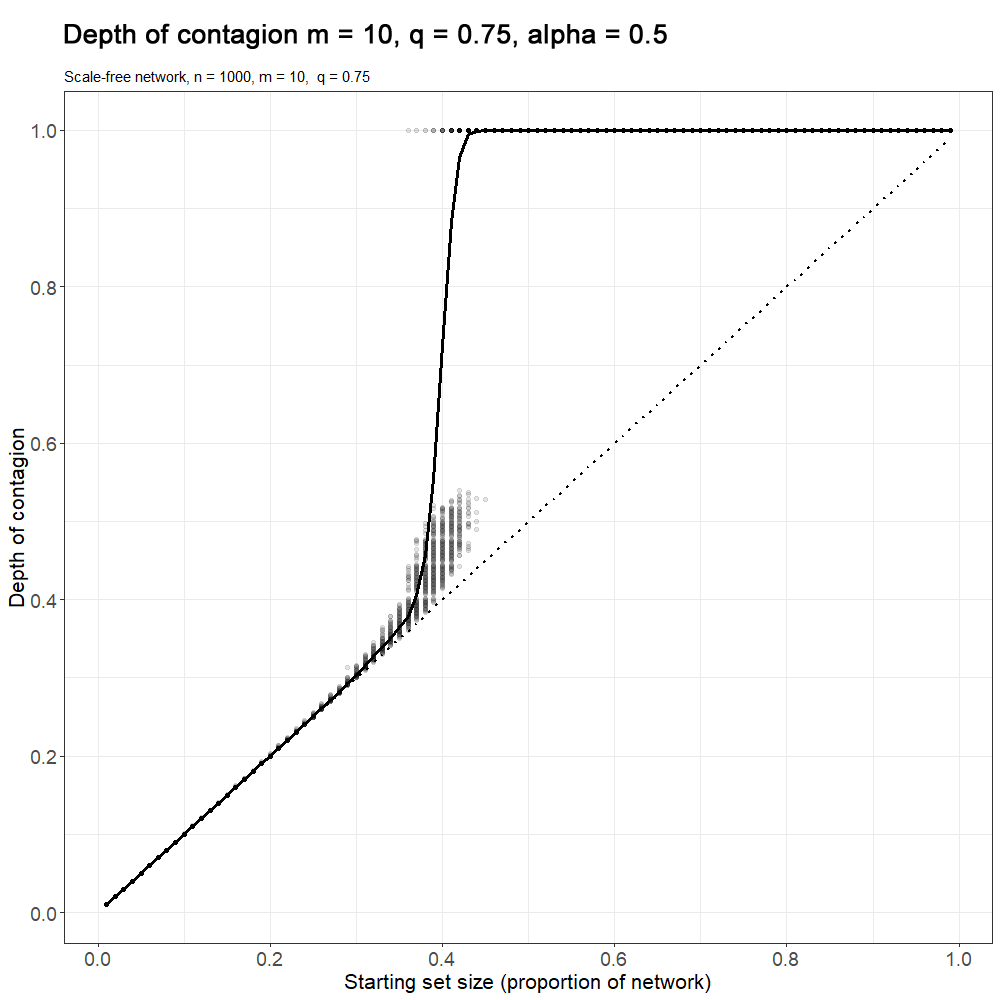}
  \caption{$m = 10, q = 0.75, \alpha = 0.5$}

\end{subfigure}

\medskip
\begin{subfigure}{0.25\textwidth}
  \includegraphics[width=\linewidth]{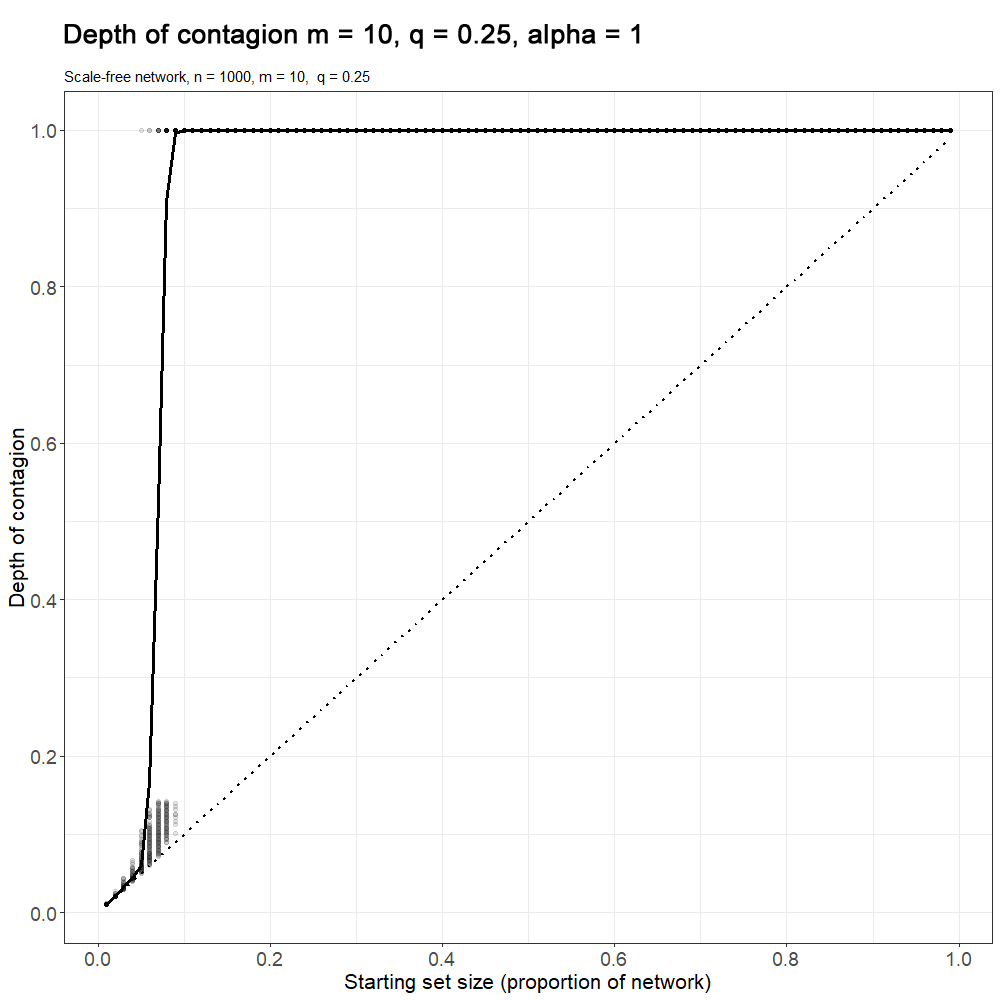}
  \caption{$m = 10, q = 0.25, \alpha = 1$}
  \label{fig:m20q25v}
\end{subfigure}\hfil 
\begin{subfigure}{0.25\textwidth}
  \includegraphics[width=\linewidth]{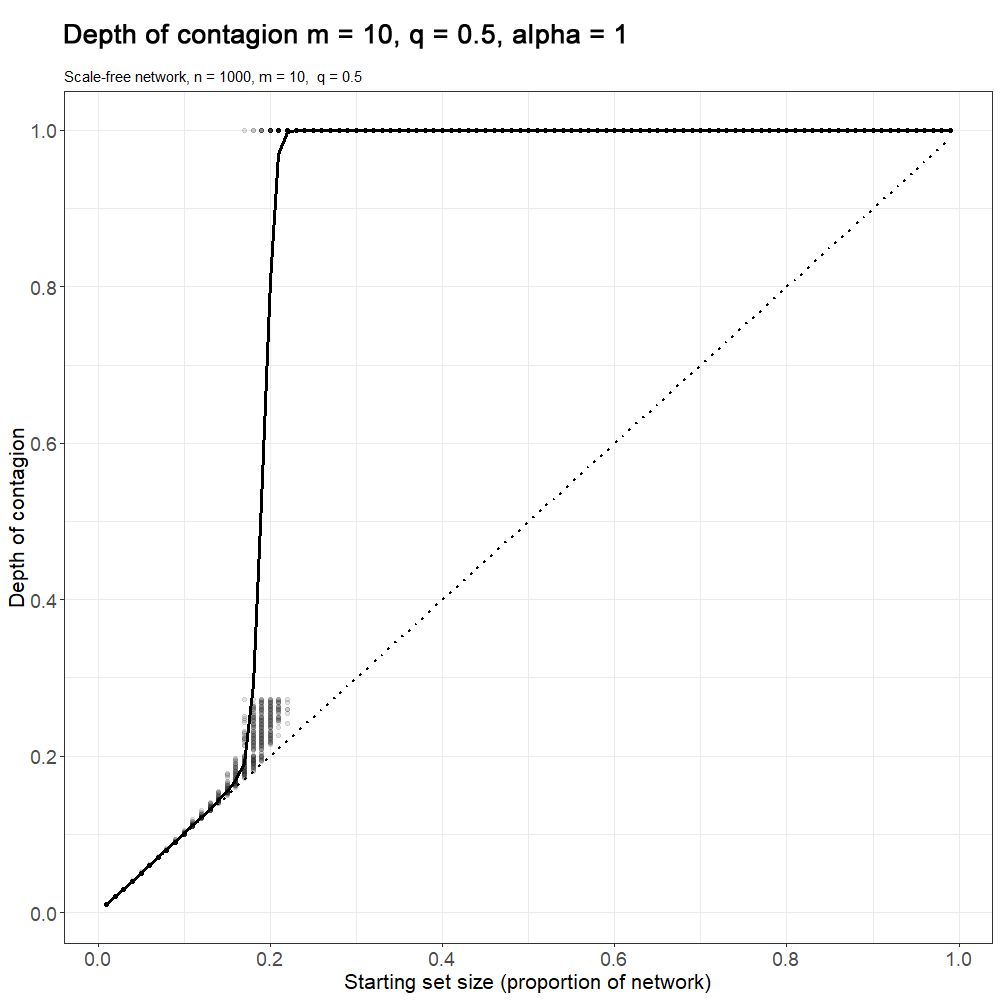}
  \caption{$m = 10, q = 0.5, \alpha = 1$}
  \label{fig:m20q5v}
\end{subfigure}\hfil 
\begin{subfigure}{0.25\textwidth}
  \includegraphics[width=\linewidth]{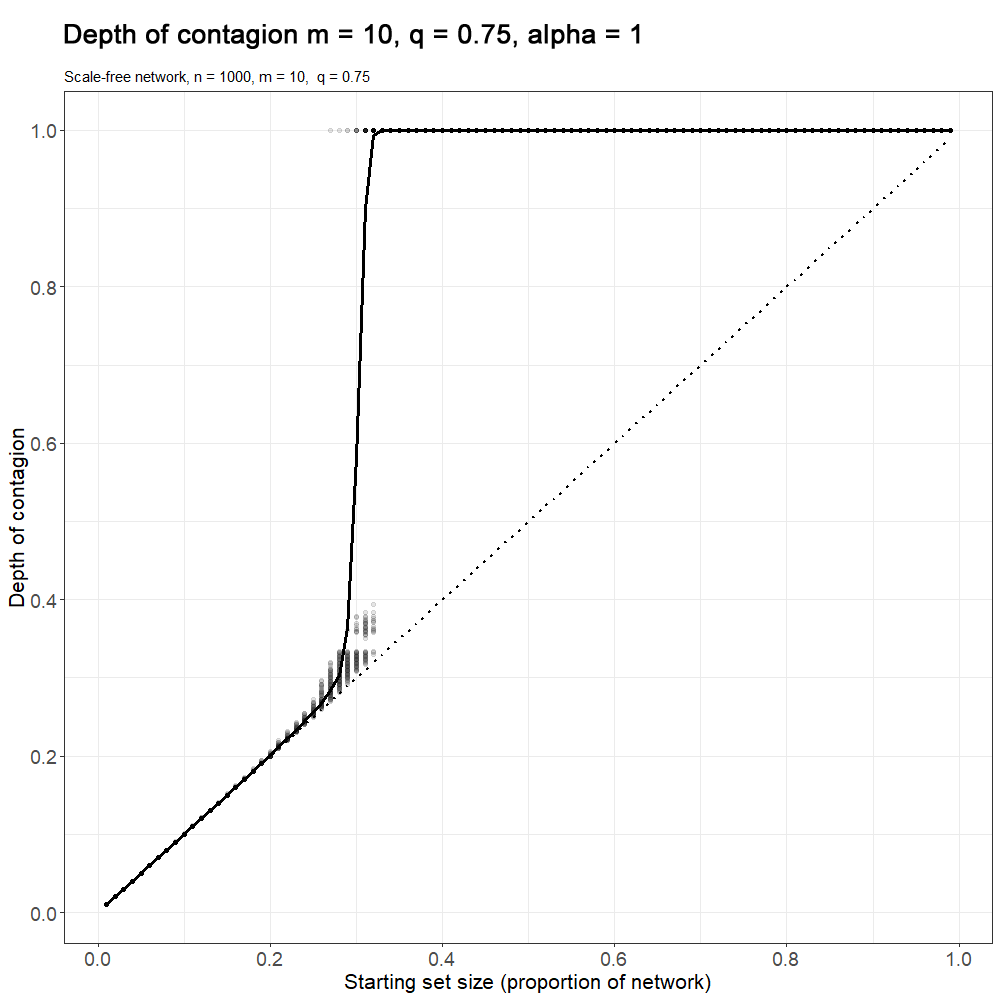}
  \caption{$m = 10, q = 0.75, \alpha = 1$}
  \label{fig:m20q75v}
\end{subfigure}

\caption{Depth of contagion for scale-free networks with $m=10$}
\label{fig:depths10}
\end{figure}

\begin{figure}[!htb]
    \centering 
\begin{subfigure}{0.25\textwidth}
  \includegraphics[width=\linewidth]{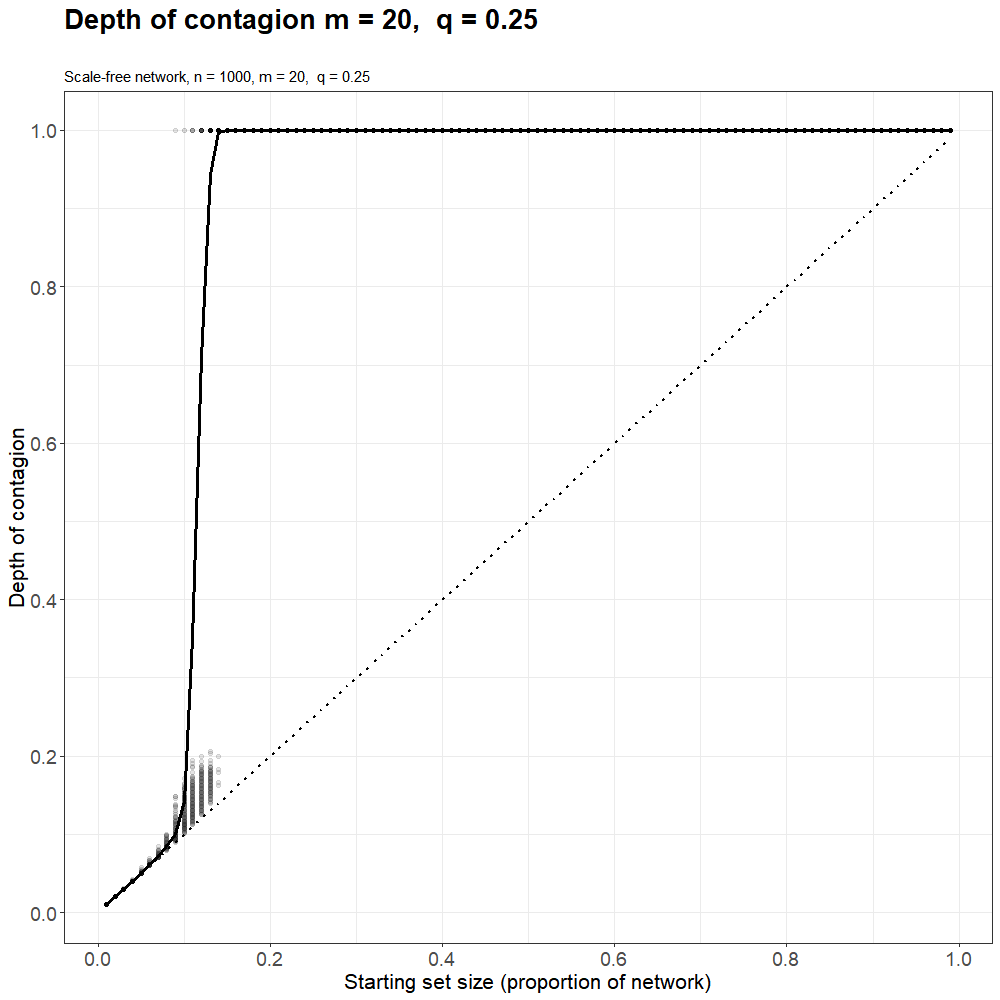}
  \caption{$m = 20, q = 0.25, \alpha = 0$}
  \label{fig:m20q25}
\end{subfigure}\hfil 
\begin{subfigure}{0.25\textwidth}
  \includegraphics[width=\linewidth]{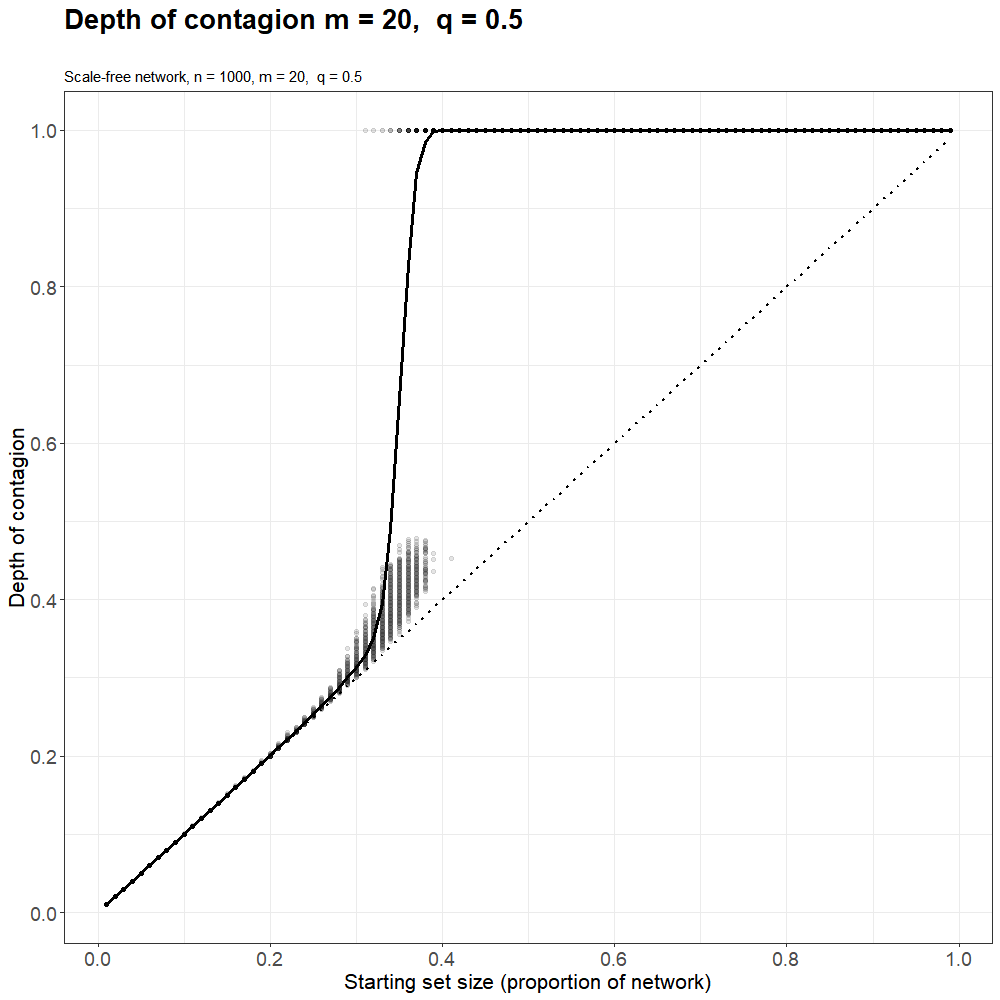}
  \caption{$m = 20, q = 0.5, \alpha = 0$}
  \label{fig:m20q5}
\end{subfigure}\hfil 
\begin{subfigure}{0.25\textwidth}
  \includegraphics[width=\linewidth]{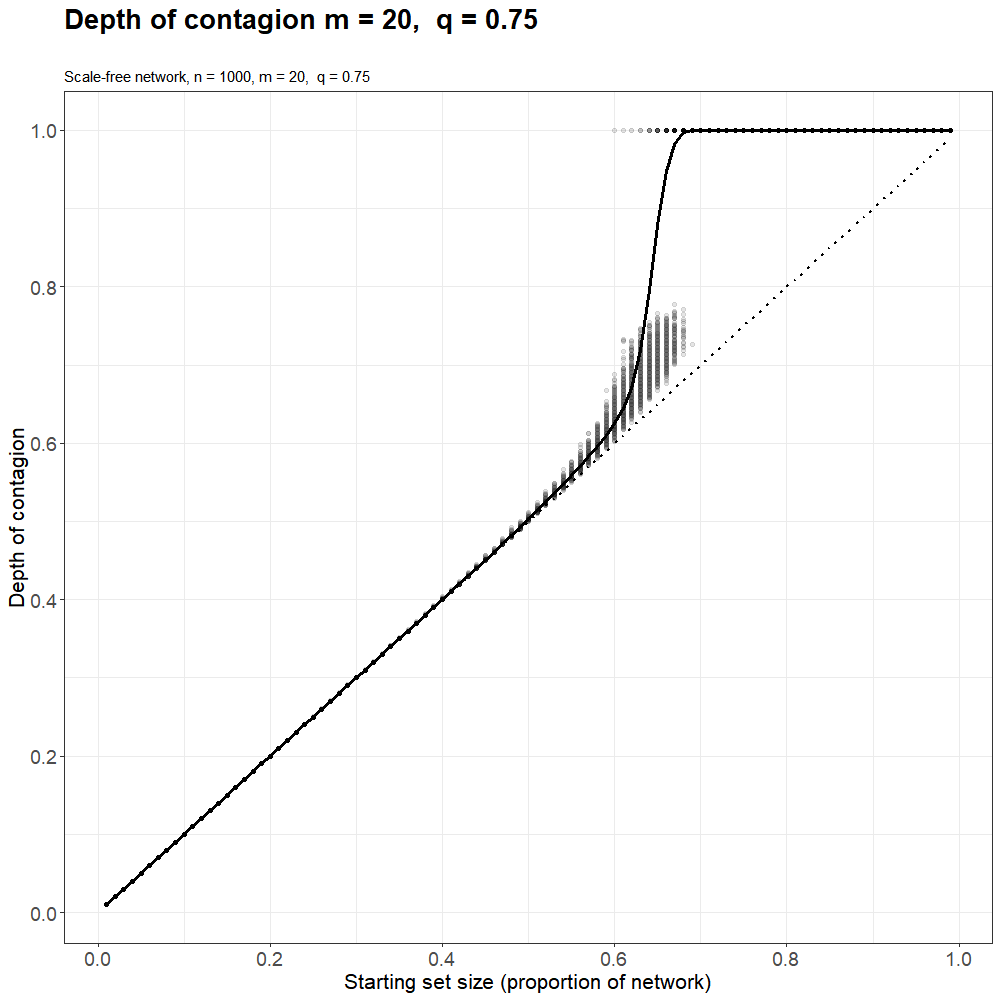}
  \caption{$m = 20, q = 0.75, \alpha = 0$}
  \label{fig:m20q75}
\end{subfigure}

\medskip
\begin{subfigure}{0.25\textwidth}
  \includegraphics[width=\linewidth]{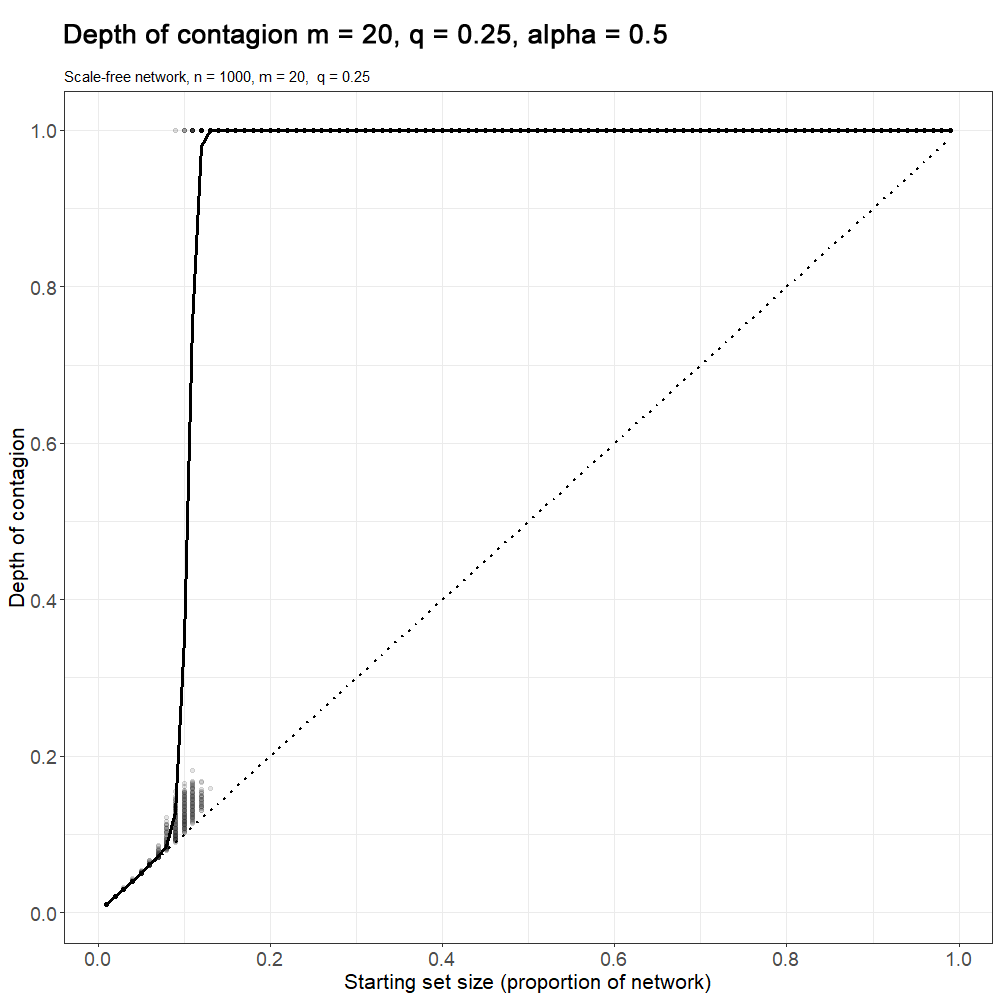}
  \caption{$m = 20, q = 0.25, \alpha = 0.5$}

\end{subfigure}\hfil 
\begin{subfigure}{0.25\textwidth}
  \includegraphics[width=\linewidth]{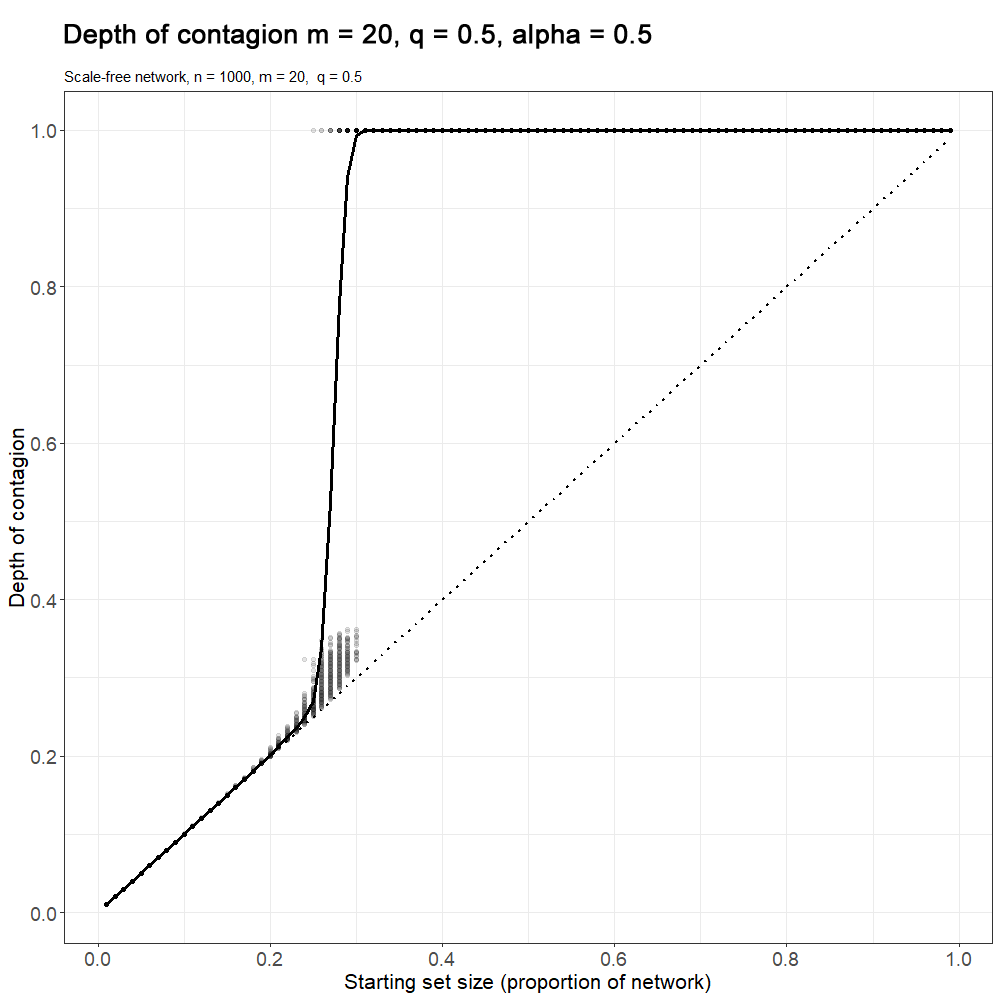}
  \caption{$m = 20, q = 0.5, \alpha = 0.5$}

\end{subfigure}\hfil 
\begin{subfigure}{0.25\textwidth}
  \includegraphics[width=\linewidth]{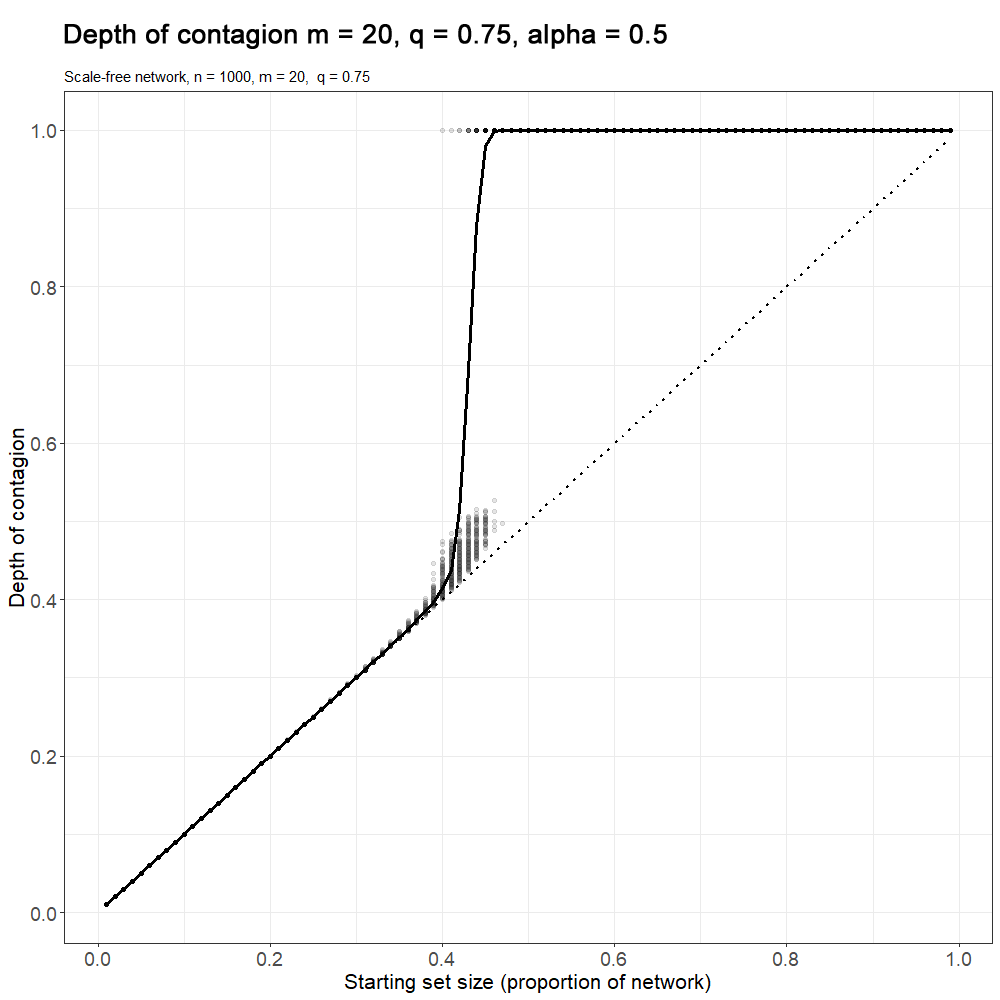}
  \caption{$m = 20, q = 0.75, \alpha = 0.5$}

\end{subfigure}

\medskip
\begin{subfigure}{0.25\textwidth}
  \includegraphics[width=\linewidth]{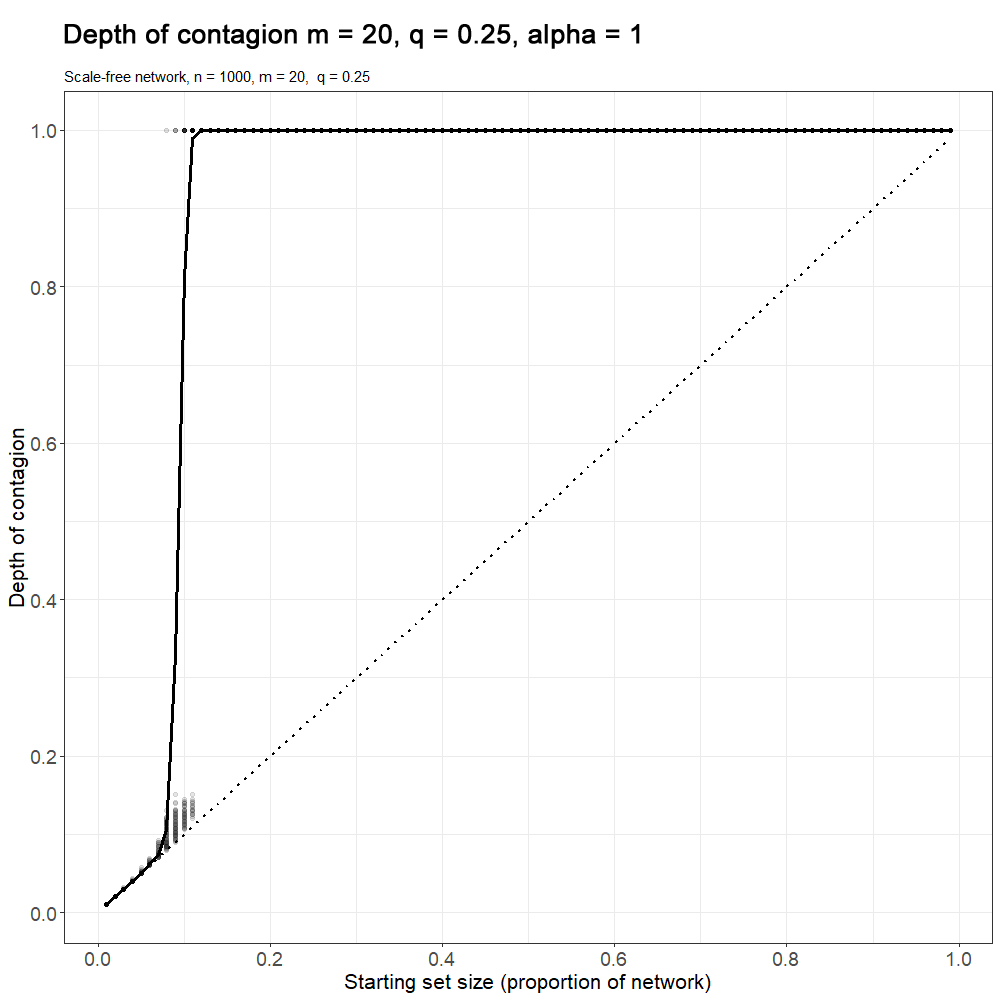}
  \caption{$m = 20, q = 0.25, \alpha = 1$}
  \label{fig:m20q25v}
\end{subfigure}\hfil 
\begin{subfigure}{0.25\textwidth}
  \includegraphics[width=\linewidth]{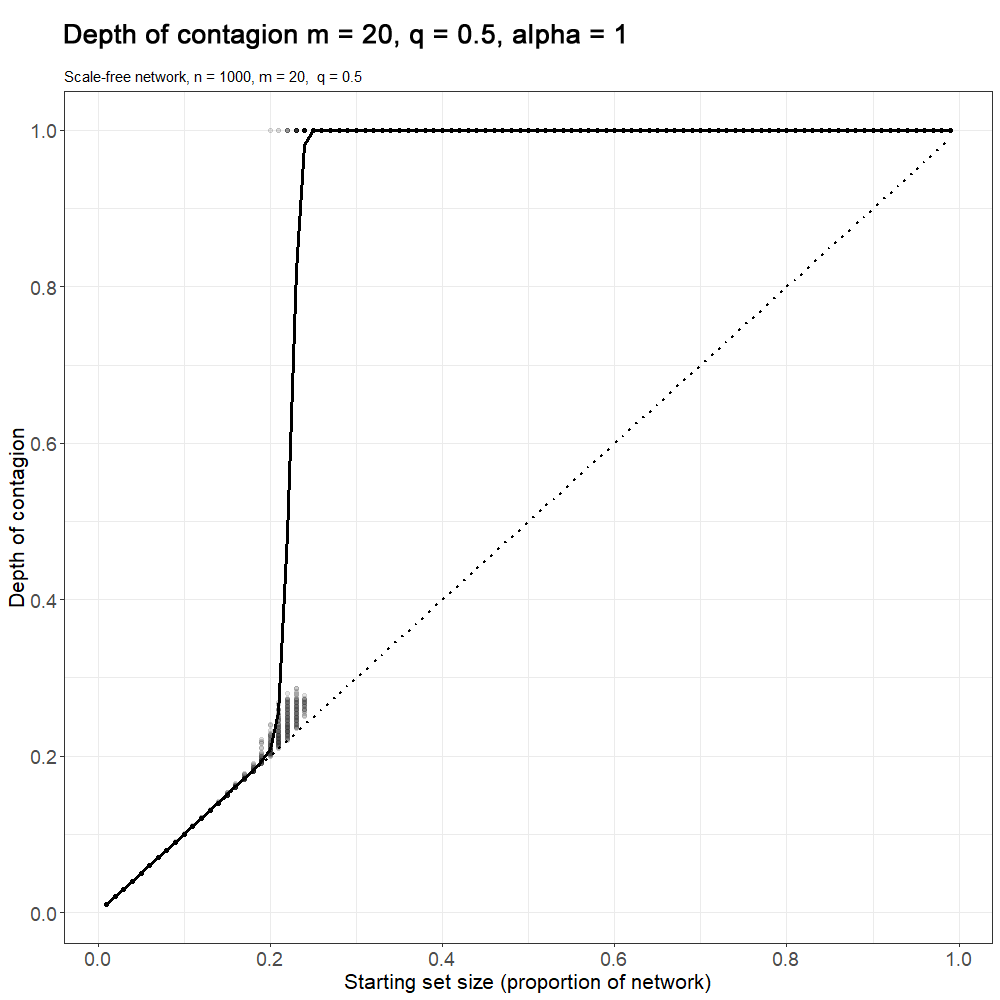}
  \caption{$m = 20, q = 0.5, \alpha = 1$}
  \label{fig:m20q5v}
\end{subfigure}\hfil 
\begin{subfigure}{0.25\textwidth}
  \includegraphics[width=\linewidth]{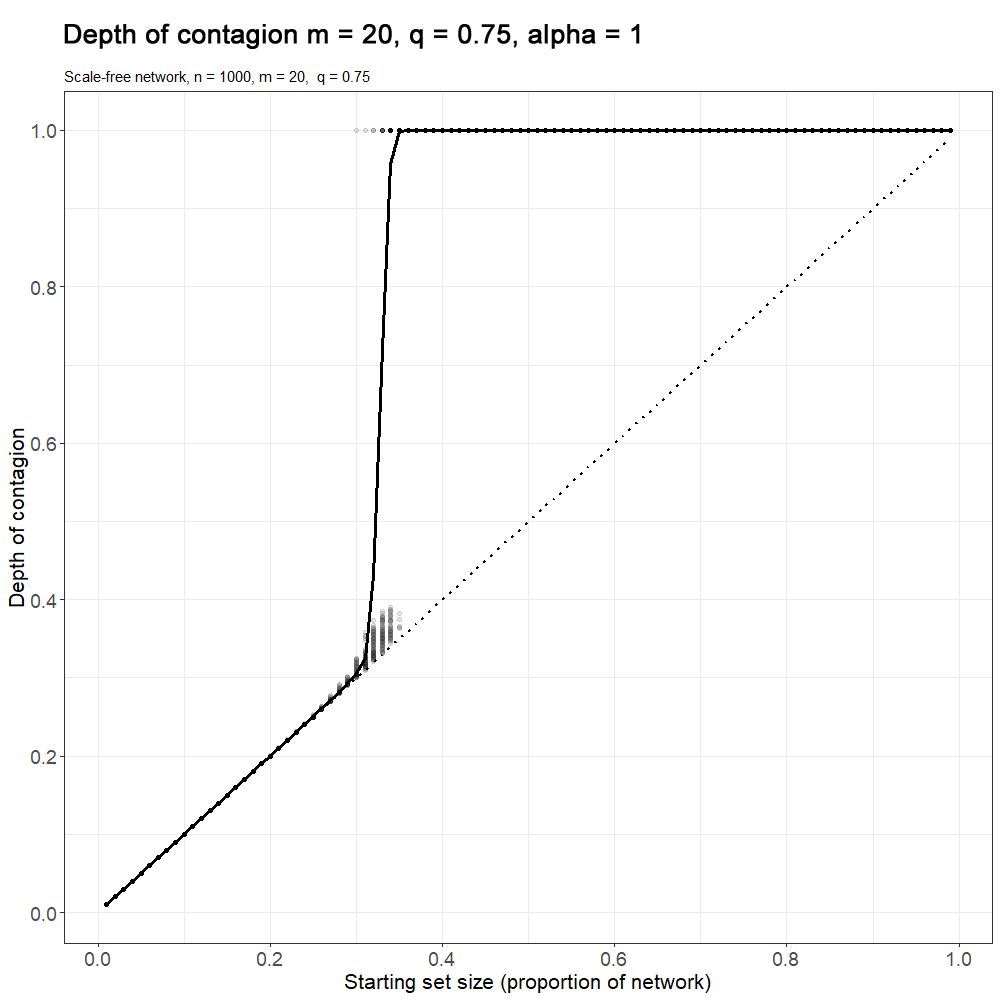}
  \caption{$m = 20, q = 0.75, \alpha = 1$}
  \label{fig:m20q75v}
\end{subfigure}

\caption{Depth of contagion for scale-free networks with $m=20$}
\label{fig:depths20}
\end{figure}

The depth of contagion function, $\delta(S,q)$, measures how far contagion spreads in equilibrium starting from set $S$ and network resilience parameter $q=\frac{c}{b+c} \in [0,1]$. This is reported in \Cref{fig:depths5} (for scale-free networks with attachment parameter $m=5$), in \Cref{fig:depths10} (for $m=10$), and in \Cref{fig:depths20} (for $m=20$). In each panel of each figure, the starting set size (as proportion of the network) is on the horizontal axis and equilibrium depth of contagion is on the vertical axis. In each panel, for each starting set size, there are 2,000 computations of depth of contagion. Each dot represents one such computation. As depth of contagion uses equilibrium sets of the form $C(S,q)$, each dot corresponds to a Nash equilibrium in the corresponding network for starting set $S$ at $q$. As $S \subseteq C(S,q)$, each dot is above the diagonal. The difference between $\delta(S,q)$ and the diagonal measures how far contagion spreads in equilibrium from a given initial set, formalizing an equilibrium notion of virality. Points on the diagonal show that there is no further contagion beyond the corresponding starting set size ($S = C(S,q)$). The average of these depths is overlaid as a bold line across different starting set sizes.

In each panel, an increase in starting set size increases depth of contagion (both due to greater person-to-person spread and due to greater impact of global effects). 

In the absence of global effects, it is harder to spread contagion at higher $q$. This can be seen by going across the top row in each figure. For example, in \Cref{fig:depths5}, first row, when network resilience is low ($q=0.25$), depth of contagion rises steeply in a narrow range of small starting set sizes (about 5-10 percent of the network) and spreads quickly to the entire network beyond that. When resilience increases to $q=0.5$, contagion spreads more slowly for smaller starting set sizes, speeds up if about 20-25 percent of the network has an incentive to play $1$, and spreads quickly to the entire network shortly thereafter. When resilience is high ($q=0.75$), there is almost no contagion beyond the starting set till about 35 percent of the network is infected, contagion spreads gradually in the range of 35-65 percent of the network and then spreads to the entire network.

With global effects, contagion spreads deeper and using smaller starting sets. This can be seen by going down in a column in each figure. For example, consider \Cref{fig:depths5}, third column. Without global effects ($\alpha=0$), starting set size needs to go beyond about 35 percent of the network for contagion to start spreading and it takes about 68 percent of the network to be affected before contagion spreads to the entire network. With $\alpha=0.5$, contagion spreads earlier, with starting set size around 25 percent, and spreads to the entire network once about 40 percent of the network is affected. With $\alpha=1$, contagion starts even earlier, at around 20 percent of the network and spreads to the entire network once about 30 percent of the network is infected. 

Complementarity between global effects and network resilience (local effects) implies that the equilibrium impact of global effects is higher when network resilience is higher. When $q$ is low (see \Cref{fig:depths5}, first column), contagion spreads easily without global effects ($\alpha=0$) at starting set sizes that are about 5-8 percent of the network and adding global effects ($\alpha = 0.5, 1$) contributes little to further increase spread of contagion. When $q$ is high (third column), without global effects, contagion starts to spread only after about 35 percent of the network is already infected (with full network contagion after about 68 percent of the network is infected) but with global effects, this occurs at about 20 percent of the network (with full network contagion guaranteed if about 30 percent of the network is infected).

Impact of global effects increases the range of starting set sizes that lead to full network contagion and brings contagion depth closer to a type of singularity (a narrow interval of starting set sizes below which there is no contagion and above which there is full contagion). 

As shown by going across the figures for the panel with $\alpha=1$, when $q=0.25$, for $m=5$, this interval is about [4,9] percent of the network, for $m=10$, it is about [6,11] percent of the network, and for $m=20$, it is about [9,15] percent of the network. For starting sets below this interval, contagion to any additional depth rarely occurs and beyond this interval, contagion occurs to the entire network.

When $q=0.5$, this interval increases and complementarity implies that global effects play a more visible role in shrinking this interval. For $m=5$, this interval is about [13,35] percent of the network without global effects ($\alpha=0$), but [12,19] percent with global effects ($\alpha=1$). For $m=10$, it is about [20,37] percent with $\alpha=0$, and shrinks to about [17,23] percent with $\alpha=1$. For $m=20$, it is about [29,40] percent of the network with $\alpha=0$, but [20,26] percent with $\alpha=1$.

When $q=0.75$, this interval increases even more with local effects only and global effects plays an even stronger role in shrinking this range. For $m=5$, this interval is about [30,68] percent of the network with $\alpha=0$, but [20,30] percent with $\alpha=1$. For $m=10$, it is about [47,69] percent with $\alpha=0$ but about [26,33] percent with $\alpha=1$. For $m=20$, it is about [56,69] percent with $\alpha=0$ and [30,36] percent with $\alpha=1$. 

In most cases, global effects shrink the singularity interval to about 6-7 percentage point difference in starting set sizes; below this there is no contagion and beyond this contagion spreads to the entire network. In this sense, global effects reduce the variability of equilibrium outcomes, and once contagion starts, its spread to the entire network is more inevitable. The impact of global effects to shrink this interval and move it toward smaller set sizes is more pronounced in networks with higher resilience $q$. In other words, it is especially in networks where we believe it is hard to spread contagion because local network resilience is high that global effects can cause large-scale contagion.

\subsection{Inverse depth of contagion} 

The equilibrium contagion depth function $\delta(S,q)$ can be used to answer the reverse question as well, that is, given a depth of contagion, what is the minimum starting set size to achieve this depth, on average? As average depth of contagion is a (weakly) increasing function of starting set size, we can invert it numerically to derive the (smallest) starting set size needed to achieve a desired depth of contagion. We term this \textit{inverse depth of contagion}. These numbers are shown in \Cref{tab:depthsummary} for depths of contagion from 0.1 (10 percent of network) to 1 (100 percent of network) at steps of 0.1. 

The broad pattern that emerges is that in order to spread contagion to a given depth of the network (on average), smaller starting sets are needed if relative miscoordination cost is lower ($q$ is low; local network resilience is low), or global effect is higher ($\alpha$ is high; global effect is strong), or connectivity is lower ($m$ is low), or a combination of these. Different combinations of these parameters have differing impacts on the starting set size needed to achieve a given depth of contagion.

\begin{table*}
\begin{subfigure}{\textwidth}
\ra{1.1}
\centering
\begin{tabular}{@{\hspace*{\leftmargin}}lrrrrrrrrrrr@{}}\toprule
Depth
& $0.1$ & $0.2$ & $0.3$ & $0.4$ & $0.5$ & $0.6$ & $0.7$ & $0.8$ & $0.9$ & $1$\\ \midrule
\rowgroup{$m=5, q=0.25$}\\
\hspace{0.2cm}$\alpha=0$ & 0.05& 0.05&0.05 &0.06 &0.06 &0.06  & 0.07&0.07 &0.07 &0.09 \\
\hspace{0.2cm}$\alpha=0.5$ & 0.04 &0.05  &0.05  & 0.06&0.06   & 0.06&0.06 &  0.07& 0.07 &0.08 \\
\hspace{0.2cm}$\alpha=1$ & 0.04&0.05&0.05 &0.05 &0.05&0.06  &0.06 &0.06 &0.06 &0.08\\
\rowgroup{$m=5, q=0.5$}\\
\hspace{0.2cm}$\alpha=0$ & 0.10 & 0.18 & 0.23 & 0.26 & 0.27 & 0.28 & 0.28 & 0.29 & 0.30 &0.35\\
\hspace{0.2cm}$\alpha=0.5$ & 0.10&0.17& 0.20& 0.21& 0.21& 0.22& 0.22& 0.23& 0.23& 0.26\\
\hspace{0.2cm}$\alpha=1$ & 0.10& 0.15& 0.16& 0.16& 0.16& 0.17& 0.17&0.17& 0.17&0.19\\
\rowgroup{$m=5, q=0.75$}\\
\hspace{0.2cm}$\alpha=0$ & 0.10& 0.20&0.30 &0.38 &0.46 &0.53  & 0.59&0.62 &0.64 &0.68\\
\hspace{0.2cm}$\alpha=0.5$ &0.10 &0.20&0.29 & 0.34&0.35 & 0.36 & 0.37& 0.37&0.38 &0.41\\
\hspace{0.2cm}$\alpha=1$ &0.10 &0.20&0.25 &0.26 & 0.27&0.27  & 0.28&0.28 & 0.29&0.31\\
\midrule
\rowgroup{$m=10, q=0.25$}\\
\hspace{0.2cm}$\alpha=0$ & 0.07 & 0.08 &0.08 &0.08  &0.09   &0.09 &0.09 &0.09  &0.10  &0.11 \\
\hspace{0.2cm}$\alpha=0.5$ &0.07  & 0.07 & 0.08&  0.08&  0.08 & 0.08& 0.09& 0.09 &0.09  &0.11 \\
\hspace{0.2cm}$\alpha=1$ & 0.06& 0.07 &0.07  &0.07 &0.07  &0.08   &0.08 &0.08 &0.08  &0.10\\
\rowgroup{$m=10, q=0.5$}\\
\hspace{0.2cm}$\alpha=0$ & 0.10 & 0.20 & 0.27& 0.29 &0.31   &0.31 &0.32 &0.32  &0.33  &0.37\\
\hspace{0.2cm}$\alpha=0.5$ &0.10  & 0.19 & 0.23&0.24  &0.24   & 0.25& 0.25& 0.26 &0.26  &0.28\\
\hspace{0.2cm}$\alpha=1$ & 0.10 &0.18  &0.19 &0.19  &0.19   &0.20 &0.20 &0.20  &0.21  &0.23\\
\rowgroup{$m=10, q=0.75$}\\
\hspace{0.2cm}$\alpha=0$ &0.10  & 0.20 & 0.30& 0.40 & 0.49  &0.57 &0.61 & 0.63 &0.64  &0.69\\
\hspace{0.2cm}$\alpha=0.5$ &0.10  & 0.20 &0.30 &0.37  & 0.39  &0.40 & 0.40& 0.41 &0.42  &0.45\\
\hspace{0.2cm}$\alpha=1$ &0.10  &0.20  &0.28 &0.30  &0.30   &0.31 &0.31 & 0.31 &0.31  &0.33\\
\midrule
\rowgroup{$m=20, q=0.25$}\\
\hspace{0.2cm}$\alpha=0$ & 0.09 &0.11  &0.11 &0.12  &0.12   &0.12 &0.12 &0.13  &0.13  &0.15\\
\hspace{0.2cm}$\alpha=0.5$ & 0.09 &0.10  &0.10 &0.11  &0.11   &0.11 & 0.11& 0.12 &0.12  &0.13 \\
\hspace{0.2cm}$\alpha=1$& 0.08 &0.09  &0.09 &0.10  &0.10   &0.10 &0.10 &0.10  &0.11  &0.12\\
\rowgroup{$m=20, q=0.5$}\\
\hspace{0.2cm}$\alpha=0$ &0.10  &0.20  &0.29 &0.34  &0.35   &0.35 &0.36 &0.36  &0.37  &0.40\\
\hspace{0.2cm}$\alpha=0.5$ & 0.10 &0.20  &0.26 &0.27  &0.27   &0.28 &0.28 &0.29  &0.29  &0.31\\
\hspace{0.2cm}$\alpha=1$ &0.10  &0.20  &0.22 &0.22  &0.23   &0.23 &0.23 & 0.23 & 0.24 &0.25\\
\rowgroup{$m=20, q=0.75$}\\
\hspace{0.2cm}$\alpha=0$ & 0.10 &0.20  &0.30 &0.40  &0.50   &0.59 &0.63 &0.65  &0.66  &0.69\\
\hspace{0.2cm}$\alpha=0.5$ & 0.10 &0.20  &0.30 &0.40  &0.42   &0.43 &0.44 &0.44  &0.45  &0.47\\
\hspace{0.2cm}$\alpha=1$ & 0.10 &0.20  &0.30 &0.32  &0.33   &0.33 &0.33 &0.34  &0.34  &0.36\\
\bottomrule
\end{tabular}

\end{subfigure}
\caption{Starting set size needed to achieve a given depth of contagion}
\label{tab:depthsummary}
\end{table*}

As shown in \Cref{tab:depthsummary}, when $m=5$ and $q=0.25$, in order to spread contagion to one-half of the network (depth = 0.5), a starting set size of 6 percent (60 players) of the network is required with local effects only ($\alpha =0$) and 5 percent is required with global effects ($\alpha =1$). When $q$ is low, person-to-person spread is high and global effect has a small additional impact. This can be explained by examining the incentive to play $1$, given by $\frac{s_i}{d_i} \ge q(1-\alpha p_i)$. The partial derivative of the right-hand side with respect to $\alpha$ is $-qp_i$, which is small in absolute value when $q$ is low. When $q$ is low, an increase in $\alpha$ has a smaller contribution to lowering the threshold to play $1$. 

When $m=5$ and $q=0.5$, in order to spread contagion to one-half of the network on average, a starting set size of 27 percent of the network is required if $\alpha=0$ and 16 percent if $\alpha=1$. That is, instead of needing 270 players in the initial starting set, including a global impact enables us to achieve the same expected depth with 160 players. This is a 40.7 percent reduction in required starting set size. For full contagion, we need 35 percent starting set size when $\alpha = 0$ and 19 percent when $\alpha = 1$, implying that global impact yields a 45.7 percent reduction in starting set size. 

When $m=5$ and $q=0.75$, full contagion requires 68 percent of the network with $\alpha = 0$ and only 31 percent when $\alpha = 1$. This is a 54.4 percent reduction in required starting set size. We go from needing more than $\frac{2}{3}$ of the network in the starting set when $\alpha=0$ to requiring less than $\frac{1}{3}$ when $\alpha = 1$. 

The differential impact of global effects is especially pronounced when $q$ increases from 0.5 to 0.75. Without global effect ($\alpha=0$), the required starting set increases from 35 percent of the network to 68 percent (a 94.3 percent increase), whereas with global effect ($\alpha=1$), the starting set size increases from 19 percent to 31 percent (a 63.2 percent increase). This can also be explained in terms of the threshold to play $1$: $\frac{s_i}{d_i} \ge q(1-\alpha p_i)$. The partial derivative of the right-hand side with respect to $\alpha$ is $-qp_i$, which is large in absolute value when $q$ is high. When $q$ is high, an increase in $\alpha$ has a larger contribution to lowering the threshold to play $1$. 

The implication is that to achieve a given depth of contagion, opening the channel of global effects has the largest impact when larger starting sets are required in the absence of global effects. In particular, this is the case when the network is more resilient to contagion ($q$ is high). In these cases, global effects can have an outsized impact in reducing the required starting set size to achieve a given depth of contagion.

\section{Policy implications}

Our analysis provides insight into the design of policies to control or spread contagion in networks. Limiting starting set size, making the network more resilient, and reducing cross-network impact (global effects) each reduces contagion by itself and taking the reverse actions increases spread of contagion. Complementarities among these factors have important effects on depth of contagion.  

Complementarity between starting set sizes and global effects amplifies the contagion effect of a given starting set making the network even more susceptible to contagion. Indeed, as shown somewhat starkly in \Cref{tab:thresh} and \Cref{fig:threshsumm}, with global effects ($\alpha=1$), for moderately connected networks ($m=5, 10$), if 40 percent of the network has an initial incentive to play $1$ at $q=1$, then in equilibrium, 100 percent of the network has an incentive to play $1$ \textit{for every $q \in [0,1]$}, making full network contagion inevitable. This is not the case in the absence of global effects, even with much larger starting sets. Therefore, controlling starting sets limits proliferation through multiple channels. 
In other words, in networks where global interaction is a common feature (or promoted by network owners), it is easier to curtail spread of misinformation by nipping it in the bud. Waiting for things to play out will add to the likelihood of considerably larger spread of contagion. This provides theoretical motivation for more aggressive policies to identify and limit accounts designed mainly to spread misinformation. 

Complementarity between global effects and network resilience (local effects) implies that the equilibrium impact of global effects is higher when network resilience is higher. In other words, it is especially in networks where we believe it is hard to spread contagion because local network resilience is high that we need greater vigilance against global effects (as they have a greater impact on virality). As incendiary news is more likely to cause greater engagement in social networks, this complementarity provides additional theoretical impetus for curtailing cross-network proliferation of unverified incendiary news. 

Complementarities among starting set size, network resilience, and global effects bring the spread of contagion closer to a singularity or tipping point. Indeed, in more resilient networks, an ability to control global effects or global interaction may be the difference between localized contagion around the starting set or widespread contagion throughout the network. In other words, when global effects are present, simply having strong local effects (in terms of high relative cost of miscoordination $q$ or high connectivity $m$) does not necessarily prevent deeper contagion. Therefore, whether it is spread of vaccine misinformation or an attack on election integrity, identifying the phenomenon early and implementing measures to prevent its spread when the problem is small is more likely to prevent widespread damage. Conversely, delaying action is more likely to result in widespread contagion and a potentially large and systemic impact.

Partial proliferation of an action may be sufficient to achieve the objective of promoters of a particular action. Our results show that in several classes of situations, contagion can occur from a relatively small starting set to a relatively large set. These include situations in which network resilience is low, or connectivity is low, or global impact is high. Combinations of these may amplify or dampen the spread of contagion. This has numerous implications for social networks that may affect societal cohesion and democratic institutions. 
For example, with lower voter turnout, convincing a small subset of the population to vote in a particular manner is sufficient to alter the election outcome. Under several scenarios, this can be achieved with a considerably smaller initial number of people promoting a particular outcome. Similarly, if a group of people seeks to sow discord within a country by spreading misinformation, then under several scenarios they only need to convince a relatively small minority of citizens to believe the misinformation, and this can be sufficient to influence a large proportion of society. 
As contagion spreads to a fraction of the network more easily than it does to the entire network, policies to control or spread contagion take on greater immediacy in these situations. 

More broadly, our research provides a framework to analyze the effects of different policies in a situation-dependent manner. The scope of application is enlarged given the many other situations across different fields and disciplines that may be modeled using this framework.

\renewcommand{\baselinestretch}{1} \small \normalsize

\bibliographystyle{apalike}
\bibliography{ref}

\begin{thebibliography}{}

\bibitem[Adler, 1991]{adler1991bootstrap}
Adler, J. (1991).
\newblock Bootstrap percolation.
\newblock {\em Physica A: Statistical Mechanics and its Applications},
  171(3):453--470.

\bibitem[Akbarpour et~al., 2020]{akbarpour2020}
Akbarpour, M., Malladi, S., and Saberi, A. (2020).
\newblock Just a few seeds more: Value of network information for diffusion.
\newblock {\em Working Paper}.

\bibitem[Angeletos et~al., 2006]{angeletos2006signaling}
Angeletos, G.-M., Hellwig, C., and Pavan, A. (2006).
\newblock Signaling in a global game: {Coordination} and policy traps.
\newblock {\em Journal of Political Economy}, 114(3):452--484.

\bibitem[Angeletos et~al., 2007]{angeletos2007dynamic}
Angeletos, G.-M., Hellwig, C., and Pavan, A. (2007).
\newblock Dynamic global games of regime change: {Learning}, multiplicity, and
  the timing of attacks.
\newblock {\em Econometrica}, 75(3):711--756.

\bibitem[Barab{\'a}si and Albert, 1999]{barabasi1999emergence}
Barab{\'a}si, A.-L. and Albert, R. (1999).
\newblock Emergence of scaling in random networks.
\newblock {\em Science}, 286(5439):509--512.

\bibitem[Basak and Zhou, 2020]{basak2020diffusing}
Basak, D. and Zhou, Z. (2020).
\newblock Diffusing coordination risk.
\newblock {\em American Economic Review}, 110(1):271--97.

\bibitem[Bich and Morhaim, 2020]{bich2020}
Bich, P. and Morhaim, L. (2020).
\newblock On the existence of pairwise stable weighted networks.
\newblock {\em Mathematics of Operations Research}, 45(4):1393--1404.

\bibitem[Blume et~al., 2011]{blume2011}
Blume, L., Easley, D., Kleinberg, J., Kleinberg, R., and Tardos, {\'E}. (2011).
\newblock Which networks are least susceptible to cascading failures?
\newblock In {\em 2011 IEEE 52nd Annual Symposium on Foundations of Computer
  Science}, pages 393--402. IEEE.

\bibitem[Board and Meyer-ter Vehn, 2021]{board2021}
Board, S. and Meyer-ter Vehn, M. (2021).
\newblock Learning dynamics in social networks.
\newblock {\em Econometrica}, 89(6):2601--2635.

\bibitem[Bramoull{\'e} and Kranton, 2016]{bramoulle15}
Bramoull{\'e}, Y. and Kranton, R. (2016).
\newblock Games played on networks.
\newblock In Bramoull{\'e}, Y., Galeotti, A., and Rogers, B., editors, {\em The
  Oxford Handbook of the Economics of Networks}. Oxford University Press.

\bibitem[Brennen et~al., 2020]{brennen2020types}
Brennen, J.~S., Simon, F.~M., Howard, P.~N., and Nielsen, R.~K. (2020).
\newblock Types, sources, and claims of {COVID}-19 misinformation.
\newblock {\em Reuters Institute}.

\bibitem[Broniatowski et~al., 2018]{broniatowski}
Broniatowski, D.~A., Jamison, A.~M., Qi, S., AlKulaib, L., Chen, T., Benton,
  A., Quinn, S.~C., and Dredze, M. (2018).
\newblock Weaponized health communication: Twitter bots and russian trolls
  amplify the vaccine debate.
\newblock {\em American Journal of Public Health}, 108(10):1378--1384.

\bibitem[Carlsson and Van~Damme, 1993]{carlsson1993global}
Carlsson, H. and Van~Damme, E. (1993).
\newblock Global games and equilibrium selection.
\newblock {\em Econometrica}, pages 989--1018.

\bibitem[Centola, 2010]{centola2010spread}
Centola, D. (2010).
\newblock The spread of behavior in an online social network experiment.
\newblock {\em Science}, 329(5996):1194--1197.

\bibitem[Centola and Macy, 2007]{centola2007complex}
Centola, D. and Macy, M. (2007).
\newblock Complex contagions and the weakness of long ties.
\newblock {\em American Journal of Sociology}, 113(3):702--734.

\bibitem[Chwe, 2000]{chwe}
Chwe, M. S.-Y. (2000).
\newblock Communication and coordination in social networks.
\newblock {\em The Review of Economic Studies}, 67(1):1--16.

\bibitem[Dasaratha, 2020]{dasaratha2019innovation}
Dasaratha, K. (2020).
\newblock Innovation and strategic network formation. {Working paper}.

\bibitem[Dodds and Watts, 2004]{dodds2004universal}
Dodds, P.~S. and Watts, D.~J. (2004).
\newblock Universal behavior in a generalized model of contagion.
\newblock {\em Physical Review Letters}, 92(21):218701.

\bibitem[Elliott and Golub, 2019]{elliott2019network}
Elliott, M. and Golub, B. (2019).
\newblock A network approach to public goods.
\newblock {\em Journal of Political Economy}, 127(2):730--776.

\bibitem[Elliott et~al., 2014]{elliott2014financial}
Elliott, M., Golub, B., and Jackson, M.~O. (2014).
\newblock Financial networks and contagion.
\newblock {\em American Economic Review}, 104(10):3115--53.

\bibitem[Frankel et~al., 2003]{frankel2003equilibrium}
Frankel, D.~M., Morris, S., and Pauzner, A. (2003).
\newblock Equilibrium selection in global games with strategic
  complementarities.
\newblock {\em Journal of Economic Theory}, 108(1):1--44.

\bibitem[Freiberger et~al., 2022]{freiberger2022chasing}
Freiberger, M., Grass, D., Kuhn, M., Seidl, A., and Wrzaczek, S. (2022).
\newblock Chasing up and locking down the virus: Optimal pandemic interventions
  within a network.
\newblock {\em Journal of Public Economic Theory}, 24(5):1182--1217.

\bibitem[Frenkel, 2020]{frenkel_2020}
Frenkel, S. (2020).
\newblock Meet the top election misinformation ‘superspreaders.’ {Nov} 23,
  2020.
\newblock {\em New York Times}.

\bibitem[Galeotti et~al., 2010]{galeotti2010network}
Galeotti, A., Goyal, S., Jackson, M.~O., Vega-Redondo, F., and Yariv, L.
  (2010).
\newblock Network games.
\newblock {\em The Review of Economic Studies}, 77(1):218--244.

\bibitem[Gallic et~al., 2022]{gallic2022optimal}
Gallic, E., Lubrano, M., and Michel, P. (2022).
\newblock Optimal lockdowns for covid-19 pandemics: Analyzing the efficiency of
  sanitary policies in europe.
\newblock {\em Journal of Public Economic Theory}, 24(5):944--967.

\bibitem[Golub and Jackson, 2010]{golub2010naive}
Golub, B. and Jackson, M.~O. (2010).
\newblock Naive learning in social networks and the wisdom of crowds.
\newblock {\em American Economic Journal: Microeconomics}, 2(1):112--49.

\bibitem[Golub and Sadler, 2016]{golublearning}
Golub, B. and Sadler, E. (2016).
\newblock Learning in social networks.
\newblock In {\em The Oxford Handbook of the Economics of Networks}. Oxford
  University Press.

\bibitem[Goyal, 2009]{goyal2009connections}
Goyal, S. (2009).
\newblock {\em Connections: An introduction to the economics of networks}.
\newblock Princeton University Press.

\bibitem[Granovetter, 1978]{granovetter1978threshold}
Granovetter, M. (1978).
\newblock Threshold models of collective behavior.
\newblock {\em American Journal of Sociology}, 83(6):1420--1443.

\bibitem[Grinberg et~al., 2019]{grinberg}
Grinberg, N., Joseph, K., Friedland, L., Swire-Thompson, B., and Lazer, D.
  (2019).
\newblock Fake news on {Twitter} during the 2016 {US} {Presidential Election}.
\newblock {\em Science}, 363(6425):374--378.

\bibitem[Hoffmann and Sabarwal, 2019]{hoffmann2019global}
Hoffmann, E.~J. and Sabarwal, T. (2019).
\newblock Global games with strategic complements and substitutes.
\newblock {\em Games and Economic Behavior}, 118:72--93.

\bibitem[Hritonenko et~al., 2022]{hritonenko2022model}
Hritonenko, N., Yatsenko, O., and Yatsenko, Y. (2022).
\newblock Model with transmission delays for covid-19 control: Theory and
  empirical assessment.
\newblock {\em Journal of Public Economic Theory}, 24(5):1218--1244.

\bibitem[Jackson, 2008]{jackson2008}
Jackson, M.~O. (2008).
\newblock {\em Social and Economic Networks}.
\newblock Princeton University Press.

\bibitem[Jackson et~al., 2017]{jackson2017}
Jackson, M.~O., Rogers, B.~W., and Zenou, Y. (2017).
\newblock The economic consequences of social-network structure.
\newblock {\em Journal of Economic Literature}, 55(1):49--95.

\bibitem[Jackson and Storms, 2019]{jackson2019behavioral}
Jackson, M.~O. and Storms, E.~C. (2019).
\newblock Behavioral communities and the atomic structure of networks. {Working
  Paper}.

\bibitem[Jackson and Yariv, 2006]{jackson2006diffusion}
Jackson, M.~O. and Yariv, L. (2006).
\newblock Diffusion on social networks.
\newblock {\em Economie publique/Public economics}, (16).

\bibitem[Jackson and Yariv, 2007]{jackson2007diffusion}
Jackson, M.~O. and Yariv, L. (2007).
\newblock Diffusion of behavior and equilibrium properties in network games.
\newblock {\em American Economic Review}, 97(2):92--98.

\bibitem[Jackson and Zenou, 2015]{jackson2015}
Jackson, M.~O. and Zenou, Y. (2015).
\newblock Games on networks.
\newblock In {\em Handbook of Game Theory with Economic Applications},
  volume~4, pages 95--163. Elsevier.

\bibitem[Kempe et~al., 2003]{kempe2003}
Kempe, D., Kleinberg, J., and Tardos, {\'E}. (2003).
\newblock Maximizing the spread of influence through a social network.
\newblock In {\em Proceedings of the ninth ACM SIGKDD international conference
  on Knowledge discovery and data mining}, pages 137--146.

\bibitem[Kobayashi and Onaga, 2021]{kobayashi2021dynamics}
Kobayashi, T. and Onaga, T. (2021).
\newblock Dynamics of diffusion on monoplex and multiplex networks: A
  message-passing approach.
\newblock {\em Working Paper}.

\bibitem[Leister et~al., 2021]{leister2021}
Leister, M., Zenou, Y., and Zhou, J. (2021).
\newblock Social connectedness and local contagion.
\newblock {\em Review of Economic Studies, forthcoming}.

\bibitem[Morris, 2000]{morris2000}
Morris, S. (2000).
\newblock Contagion.
\newblock {\em The Review of Economic Studies}, 67(1):57--78.

\bibitem[Morris and Shin, 2003]{morris2003global}
Morris, S. and Shin, H.~S. (2003).
\newblock Global games: Theory and applications.
\newblock In {\em Advances in Economics and Econometrics: Theory and
  Applications, Eighth World Congress, Volume 1}, pages 56--114. Cambridge
  University Press.

\bibitem[Mossel and Roch, 2010]{mossel2010}
Mossel, E. and Roch, S. (2010).
\newblock Submodularity of influence in social networks: From local to global.
\newblock {\em SIAM Journal on Computing}, 39(6):2176--2188.

\bibitem[Oyama and Takahashi, 2015]{oyama2015contagion}
Oyama, D. and Takahashi, S. (2015).
\newblock Contagion and uninvadability in local interaction games: The
  bilingual game and general supermodular games.
\newblock {\em Journal of Economic Theory}, 157:100--127.

\bibitem[Ressa, 2021]{ressa_2021}
Ressa, M. (2021).
\newblock {Maria Ressa Nobel Lecture.
  \\https://www.nobelprize.org/prizes/peace/2021/ressa/lecture/ }.

\bibitem[Sadler, 2020]{sadler2020diffusion}
Sadler, E. (2020).
\newblock Diffusion games.
\newblock {\em American Economic Review}, 110(1):225--70.

\bibitem[Schelling, 1978]{schelling1978sorting}
Schelling, T.~C. (1978).
\newblock Sorting and mixing.
\newblock {\em Micromotives and {Macrobehavior}}.

\bibitem[Shao et~al., 2018]{shao2018}
Shao, C., Ciampaglia, G.~L., Varol, O., Yang, K.-C., Flammini, A., and Menczer,
  F. (2018).
\newblock The spread of low-credibility content by social bots.
\newblock {\em Nature Communications}, 9(1):4787.

\bibitem[Varol et~al., 2017]{varol2017}
Varol, O., Ferrara, E., Davis, C.~A., Menczer, F., and Flammini, A. (2017).
\newblock Online human-bot interactions: Detection, estimation, and
  characterization.
\newblock In {\em Eleventh international AAAI conference on web and social
  media}.

\bibitem[Vosoughi et~al., 2018]{vosoughi2018}
Vosoughi, S., Roy, D., and Aral, S. (2018).
\newblock The spread of true and false news online.
\newblock {\em Science}, 359(6380):1146--1151.

\bibitem[Watts, 2002]{watts2002simple}
Watts, D.~J. (2002).
\newblock A simple model of global cascades on random networks.
\newblock {\em Proceedings of the National Academy of Sciences},
  99(9):5766--5771.

\bibitem[Wells et~al., 2021]{wells2021}
Wells, G., Horwitz, J., and Seetharaman, D. (2021).
\newblock {Facebook Knows Instagram is Toxic for Teen Girls, Company Documents
  Show}.

\bibitem[Wiedermann et~al., 2020]{wiedermann2020network}
Wiedermann, M., Smith, E.~K., Heitzig, J., and Donges, J.~F. (2020).
\newblock A network-based microfoundation of {Granovetter’s} threshold model
  for social tipping.
\newblock {\em Scientific Reports}, 10(1):1--10.

\bibitem[Xu, 2018]{xu2018social}
Xu, H. (2018).
\newblock Social interactions in large networks: A game theoretic approach.
\newblock {\em International Economic Review}, 59(1):257--284.

\bibitem[Young, 2011]{young2011}
Young, H.~P. (2011).
\newblock The dynamics of social innovation.
\newblock {\em Proceedings of the National Academy of Sciences}, 108(Supplement
  4):21285--21291.

\end{thebibliography}

\end{document}